# A SEARCH FOR $\Lambda_b$ IN $\pi^-$-A COLLISIONS AT 515 GeV/c

BY

FRANCISCO JAVIER VACA-ALVAREZ
B.S., Universidad Michoacana, Morelia, Michoacán, México, 1989
M.S., University of Illinois at Chicago, 1992

THESIS

Submitted as partial fulfillment of the requirements
for the degree of Doctor of Philosophy in Physics
in the Graduate College of the

University of Illinois at Chicago, 1995
Chicago, Illinois

This thesis is dedicated to my wife Maria Del Carmen Montes de Vaca and to my son Francisco Javier Vaca Jr. for their love and support, and to my parents, Mariano and Consuelo Vaca, without whom it would never have been accomplished.

Francisco Javier

## ACKNOWLEDGMENTS

I am pleased to have the opportunity to thank everyone who contributed to my education of more than twenty years. I begin by thanking God, who has always looked after me. Second, my parents, whom with many sacrifices put me through college and the beginning of graduate school. Also to my spouse Carmen, that has been very supportive, loving, and caring. Thank you Carmen. Next, my son Paquito, I guess if it had not been for him, writing this thesis would have not been half as fun. Every night, while I was in front of the computer typing or working on this thesis, he would approach me and say in Spanish: “quiero trabajar” (I want to work). So I would sit him in my lap and let him type. Thank you Paquito for helping me type my thesis and for making me type parts of it more than once. I also thank all my family for supporting me when I most needed them.

My formation as a physicist I owe to several Professors. The first one to thank is my advisor Prof. Seymour Margulies, who passed away May 4, 1995. He was not only the best physics teacher I have ever had, but a role model as a physicist. Thank you Sy, wherever you are. I thank Prof. Howard Goldberg for taking over what Prof. Margulies began. He has friendlily helped me throughout my thesis project. His support, advice, and encouragement in the last few months were much appreciated in a very difficult time. Thank you Howard. I also thank Prof. Julius Solomon, he has been very supportive and helpful during my training as graduate student. Thank you Julius. The members of the E672 collaboration deserve special thanks for teaching me much of what I know about experimental high energy physics. I thank Prof. Andrzej Zieminski, E672 spokesman, who supported my work as a member of the collaboration and always had confidence in me. To Dr. Hector Mendez and Dr. Richard Jesik, I owe more than thanks. Both of them have been very helpful throughout my years as a member of E672. Thank you Hector and Ricky. To the rest of the E672 collaboration, Prof. Loretta Dauwe, Dr. Andrei Gribushin, Dr. Victor Koreshev, Dr. Victor Abramov, Dr. Rui Li, and all the others whom I have not mentioned, thank you for being my colleagues. I also want to thank all of my colleagues from our sister experiment E706. In particular, I thank Dr. George Ginther for all his helpful comments about my studies in E672. "Mil gracias" (One thousand thanks) to the E672 and E706 collaborations for putting together such wonderful experiments.

I also thank all my Professors and fellow students at the UIC physics department. Prof. Wee Y. Kung, Prof. Clive Halliwell, Prof. Pagnamenta, Mark Oram, Salima Yala, Thomas Mckibben, and the list continues ....Thank you every one.

Francisco Javier

# TABLE OF CONTENTS

## LIST OF TABLES

## LIST OF FIGURES

## SUMMARY

An extensive analysis was performed on 8.0 $pb^{-1}$ of dimuon data produced in $\pi^-$ A collisions at 515 GeV/c to search for $\Lambda_b$ events in the decay channel $\Lambda_b \rightarrow J/\psi\ \Lambda^0$, with $J/\psi \rightarrow \mu^+ \mu^-$ and $\Lambda^0 \rightarrow p\ \pi^-$ (and for the conjugate reactions). The muon tracks from $J/\psi$ decays were refittedwith the mass constraint of two-body decay and with the constraint that both muon tracks intersect at a common point. The $\Lambda^0$ s were identified by their characteristic decay, giving the large fraction of their momenta to the protons. $\Lambda^0$ s were reconstructed in three regions of the E672/E706 spectrometer, in the SSD/target region, in the region between the SSDs and the dipole magnet, and inside the dipole magnet using an iterating algorithm. Several cuts were applied to the $\Lambda^0$ (and $\overline{\Lambda}^0$) to make a clean $\Lambda^0$ (and $\overline{\Lambda}^0$) data sample. Among the cuts there was a $K^0_s$ mass cut, in which if the $\Lambda^0$ (or $\overline{\Lambda}^0$) had the mass of the $K^0_s$ under the hypothesis of both tracks being $\pi^+ \pi^-$, the $\Lambda^0$ (or $\overline{\Lambda}^0$) candidates were rejected. These give a clean $\Lambda^0$ (and $\overline{\Lambda}^0$) data sample of $575 \pm 35$ $\Lambda^0$ (and $\overline{\Lambda}^0$) candidates.

To search for the $\Lambda_b \rightarrow J/\psi\ \Lambda^0$ (and charge conjugate reaction), $J/\psi$ s that passed the muon refit were combined with $\Lambda^0$ s (or $\overline{\Lambda}^0$s) when they existed in the same event. The results show 2 events in the $\Lambda_b$ mass region. Using the E672 measurement of the $b\bar{b}$ cross-section, and considering the 2 $\Lambda_b$ event candidates as signal with zero background, an upper limit to $F(\Lambda_b) * Br(\Lambda_b \rightarrow J/\psi\ \Lambda^0)$ was found to be less than 6.2 $x10^{-2}$ at 90 %

C.L. An upper limit was also calculate without using a $K^0_s$ mass cut for the $\Lambda^0$ s (and $\overline{\Lambda}^0$ s), and then, requiring that the J/ψ s originate from secondary vertices, giving $F(\Lambda_b) * Br(\Lambda_b \rightarrow J/\psi\ \Lambda^0) < 3.1\ x10^{-2}$ at 90 % C.L and $F(\Lambda_b) * Br(\Lambda_b \rightarrow J/\psi\ \Lambda^0) < 3.2\ x10^{-2}$ at 90 % C.L, respectively.

# CHAPTER 1

# INTRODUCTION

Since the discovery of the *b* quark in 1977 [1], much progress has been made in measuring and understanding the properties of beauty particles; this is primarily true for the $B_u$ and $B_d$ mesons, but not as much for the beauty baryons. The existence of $\Lambda_b$ (*bud*), the lightest of the baryons containing a *b* quark, has been somewhat controversial. In 1981, experiment R415, which used the Split Field Magnet (SFM) detector at the Intersecting Storage Ring (ISR) at CERN, reported an observation of the $\Lambda_b$ through the decay channel $\Lambda_b \to p\, D^0\, \pi^-$ [2] yielding a rest mass for the $\Lambda_b$ of $(5425^{+175}_{-75})$ MeV/c$^2$. Later, in 1982, experiment R416, using an upgraded version of the R415 detector, performed the same search for $\Lambda_b \to p\, D^0\, \pi^-$ with a negative result [3]. This led to some discussion by the R415 collaboration claiming that experiment R416 did not have the necessary rejection power against charged hadrons; therefore, it should not have been able to observe beauty baryons at the ISR [4]. Further evidence for the $\Lambda_b$ was reported in 1991 by the former R415 collaboration in a second upgraded experiment, R422, at the ISR. Two different $\Lambda_b$ decay modes were observed: $\Lambda_b \to p\, D^0\, \pi^-$ and $\Lambda_b \to \Lambda_c^+ \pi^+ \pi^- \pi^-$ [5]; the mass of the $\Lambda_b$ was found to be $(5640^{+150}_{-200})$ MeV/c$^2$ and $(5650^{+150}_{-200})$ MeV/c$^2$ for each mode, respectively. In 1986, an experiment using the Fermilab Multiparticle Spectrometer reported the observation of a heavy baryon decaying into $\Lambda^0 K_s^0 \pi^+ \pi^- \pi^+ \pi^-$ and having an invariant mass of 5750 MeV/$c^2$ [6]; however, this experiment did not claim to have observed the $\Lambda_b$. CERN experiment UA1 reported in 1991 the discovery of the $\Lambda_b$ decay channel $\Lambda_b \rightarrow J/\psi\ \Lambda^0$, claiming a signal of 16 ± 5 events above a background of 9 ± 1 events [7]; they measured the mass of the $\Lambda_b$ to be (5640 ± 50 (*stat*) ± 30 (*sys*)) MeV/ $c^2$. Since then, other experiments have searched for this decay mode such as ALEPH and OPAL at CERN [8], and CDF at Fermilab [9], however they have all failed to confirm UA1's observation. In 1992, LEP experiments ALEPH, DELPHI, and OPAL reported evidence for the existence of the $\Lambda_b$ [10] through observation of the semi-leptonic decay channels $\Lambda_b \rightarrow \Lambda^0\, l^- \bar{\nu}_l\, X$ and $\Lambda_b \rightarrow \Lambda_c^+\, l^- \bar{\nu}_l\, X$ in $Z^0$ decays.

In the fixed-target experiment E672/E706 at Fermilab, we have measured the $b\bar{b}$ total cross-section using our 1990 $\pi^-$ -A interactions at 515 GeV/c [11]. In addition we observed exclusive decays of B hadrons such as $B^\pm \rightarrow J/\psi\ K^\pm$ and $B^0 \rightarrow J/\psi\ K^{0*}$[11]. This suggests us to search for the $\Lambda_b \rightarrow J/\psi\ \Lambda^0$ decay mode, and to measure the product of the production fraction $F(\Lambda_b)$, times the branching ratio $Br(\Lambda_b \rightarrow J/\psi\ \Lambda^0)$.

The remainder of this chapter involves a review of the basic concepts in particle physics which are pertinent to the study of beauty baryons. It also presents detailed results of previous experimental searches for the $\Lambda_b$.

## 1.1 Theoretical overview

Present evidence indicates that matter is built from two types of fundamental particles called *quarks* and *leptons*, which are structureless and point-like on a scale of $10^{-17}$ m. *Quarks* carry fractional electric charge -*e*/3 or +2*e*/3, where *e* is the magnitude of the electron charge, and have spin 1/2. They come in several different *flavors* labeled *u* (*up*), *d* (*down*), *s* (*strange*), *c* (*charm*), *b* (*bottom*), and *t* (*top*). Each quark has it own internal quantum numbers. The *u* and *d* quarks are grouped in an isospin doublet with I = 1/2, and with the third component $I_3$ = +1/2 for the *u* and $I_3$= -1/2 for the *d*. The *s* quark is assigned an internal quantum number called *strangeness*, with value S = -1. The *c* quark is assigned an internal quantum number called *charm*, with value C= +1. The *b* quark is assigned the *bottom* quantum number B= -1, and the *t* quark is assigned the *top* quantum number T = +1. Quarks also have a *baryon number* assigned to them, $B^*$= 1/3 for quarks and $B^*$ = -1/3 for the antiquarks, which is conserved in any interaction. Because of their masses, the *u*, *d*, and *s* quarks are referred to as light quarks and the *c*, *b*, and *t* as heavy quarks. Table I shows the electric charges and masses for the six quarks along with their internal quantum numbers. Each of the six quarks has its antiquark partner, which has the opposite quantum numbers and the same rest mass.

Table I. The Quarks.

| Symbol | Internal Quantum Number | Electric Charge Q/*e* | Rest Mass [a] (MeV/$c^2$) |
|---|---|---|---|
| u | $I_3 = +\frac{1}{2}$ } $I = \frac{1}{2}$ doublet | $+\frac{2}{3}$ | 2 - 8 |
| d | $I_3 = -\frac{1}{2}$ } $I = \frac{1}{2}$ doublet | $-\frac{1}{3}$ | 5 -15 |
| s | S = -1 | $-\frac{1}{3}$ | 100 - 300 |
| c | C= +1 | $+\frac{2}{3}$ | 1000 - 1600 |
| b | B =-1 | $-\frac{1}{3}$ | 4100 - 4500 |
| t | T = +1 | $+\frac{2}{3}$ | 176,000; 199,000 |

[a] The quark masses are taken from Ref. [12], except the top quark mass which is taken from Ref.[13] and [14].

The *leptons* carry integral charge, 0 or $\pm e$, and have spin 1/2. The neutral leptons are called neutrinos, and have very small (perhaps zero) rest masses. The leptons appear to come in doublets, with each neutrino being assigned a subscript corresponding to its charged partner. The three different types of charged leptons are known as the electron ($e$), the muon ($\mu$), and the tau ($\tau$). Charged leptons are distinguished from antileptons by the sign of their charges. The neutrinos are longitudinally polarized: they have their spins opposite to their velocity vectors (left handed), while antineutrinos have spins in the same direction (right handed). A lepton number $L_e$, $L_\mu$, and $L_\tau$ of +1 is assigned to each type of lepton, respectively, and -1 to each type of antilepton. The lepton number is always conserved in any interaction. The properties of the leptons are summarized in Table II.

Table II. The Leptons

| Symbol | Electric Charge Q/*e* | Rest Mass[a] (MeV/c$^2$) | Antiparticle |
|---|---|---|---|
| $e^-$ | –1 | 0.511 | $e^+$ |
| $\nu_e$ | 0 | < 0.0051 | $\bar{\nu}_e$ |
| $\mu^-$ | –1 | 105.6 | $\mu^+$ |
| $\nu_\mu$ | 0 | < 0.27 | $\bar{\nu}_\mu$ |
| $\tau^-$ | –1 | 1777.1 | $\tau^+$ |
| $\nu_\tau$ | 0 | < 31 | $\bar{\nu}_\tau$ |

[a] The mass values are obtained from Ref. [12].

Quarks and leptons exist in three generations: The $u$, $d$, $e^-$, and $\nu_e$ are the first generation; the $s$, $c$, $\mu^-$, and $\nu_\mu$ are the second generation; and the $b$, $t$, $\tau^-$, and $\nu_\tau$ are the third generation. There are experimental measurements that indicate that there are only three generations[1].

Four fundamental forces (the gravitational, the electromagnetic, the weak, and the strong) govern the interactions between quarks and leptons. The gravitational force is by far the weakest of the four, and can be neglected for the study of interactions between elementary particles at typical distances of the order of one femtometer. Thus, ignoring the gravitational force, the charged leptons have electromagnetic and weak interactions, while the neutrinos have only weak interactions. The quarks are subject to the electromagnetic, weak, and strong interactions. The field quanta, or mediators, for the electromagnetic, weak, and strong forces are the photon ($\gamma$), the intermediate vector bosons $W^\pm$ and $Z^0$, and the gluon ($g$), respectively. All of the field quanta have spin 1 and are called, in general, *gauge bosons*. Table III summarizes the properties of the gauge bosons.

Table III. The Gauge Bosons

| Force | Field Quantum | Electric Charge Q/$e$ | Rest Mass[a] (GeV/$c^2$) |
|---|---|---|---|
| Electromagnetic | $\gamma$ | 0 | $< 3 \times 10^{-27}$ |
| Weak | $W^\pm$, $Z^0$ | ±1, 0 | 80.22, 91.18 |
| Strong | $g$ | 0 | 0 |

[a] The values for the boson masses are obtained from Ref.[12].

[1] See page 1333 of reference [12].

The electromagnetic force is described by the theory of Quantum Electrodynamics (QED). Here, the sources of the field are the electric charges of the interacting particles. The strength of the coupling constant of this force is called $\alpha$ and has a value of approximately 1/137. In the case of the strong force, which is described by the theory called Quantum Chromodynamics (QCD), quarks have an additional quantum number, whimsically called *color charge*, which is the source of this interaction. Each quark can have one of three primary color charges: red, green, or blue, denoted symbolically by R, G, and B, respectively. The antiquarks have, in similar fashion, color charges denoted by $\overline{\mathrm{R}}, \overline{\mathrm{G}}$, and $\overline{\mathrm{B}}$. Just as electric charge is conserved, so is color charge. Gluons also carry the strong-field color charge. They are bicolored objects and can be of eight different color/anti-color combinations[1]:

$$R\overline{B},\ R\overline{G},\ B\overline{G},\ B\overline{R},\ G\overline{R},\ G\overline{B},\ \frac{R\overline{R}-B\overline{B}}{\sqrt{2}},\ \text{and}\ \frac{R\overline{R}+B\overline{B}-2G\overline{G}}{\sqrt{6}}.$$

Quarks combine to form *colorless* particles (color singlets) called *hadron*s, which are strongly interacting particles. In practice, we cannot observe isolated quarks or gluons; instead, we observe hadrons. The quark and gluon constituents of hadrons are generally known as partons. Hadrons that contain $q\overline{q}$ valence quarks are called *mesons*, and those that contain *qqq* (or $\bar{q}\bar{q}\bar{q}$) valence quarks are called *baryons*. The coupling constant of the strong interactions is called $\alpha_s$ and it depends on the momentum transfer $Q^2$ in a given reaction. Thus, $\alpha_s$ is said to be a running coupling constant.

---

[1]With three colors and three anti-colors one expects $3^2$=9 combinations, but one of these is a color singlet and must be excluded.

$$\alpha_s\left(Q^2\right) = \frac{12\pi}{\left(33 - 2n_f\right)\log\left(\frac{Q^2}{\Lambda^2_{QCD}}\right)};$$

where $n_f$ is the number of flavors, $\Lambda_{QCD}$ is a parameter to be determined by experiment. Therefore, if the $n_f$ = 5 threshold is at $Q^2 = (2\ m_b)^2$ = 100 $GeV^2$, and with $\Lambda_{QCD} = \Lambda_5$ = 0.2 GeV, then $\alpha_s$ is equal to 0.48.

The weak force is described by the electroweak theory developed by Weinberg and Salam in the late sixties [15]. This gauge theory unifies the electromagnetic and weak interactions. The source of the weak force is a property called the weak charge, which gives quarks and leptons flavor-changing transitions. The corresponding effective coupling constant is called the Fermi constant, and has a value of $G_F$ =1.2x10$^{-5}$ GeV$^{-2}$. The fundamental electroweak coupling constant, called $g_W$, is related to $G_F$ through $g_W^2 / 8M_W = G_F / \sqrt{2}$, where $M_w$ is the mass of the $W$ boson.

The above brief introduction to quarks and leptons and the interactions among them is the modern view of fundamental particles and forces known as the *Standard Model*. Table IV illustrates Feynman diagrams[1] of typical interactions in the Standard Model.

Table IV. Feynman diagrams of Typical Interactions.

| Interaction | Charge | Quarks | Leptons |
|---|---|---|---|
| Strong | Color | 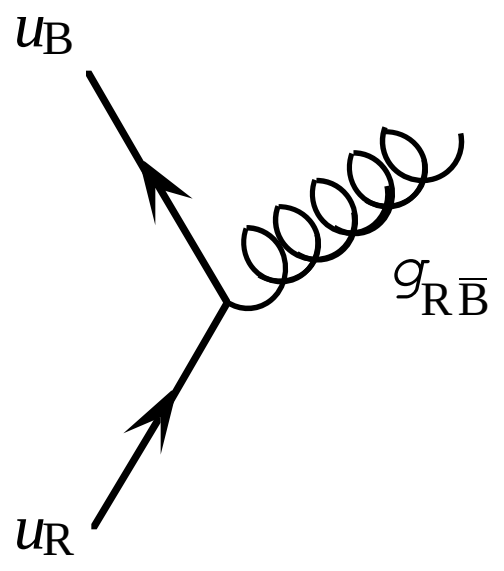 | |
| Electromagnetic | Electric charge (*e*) | 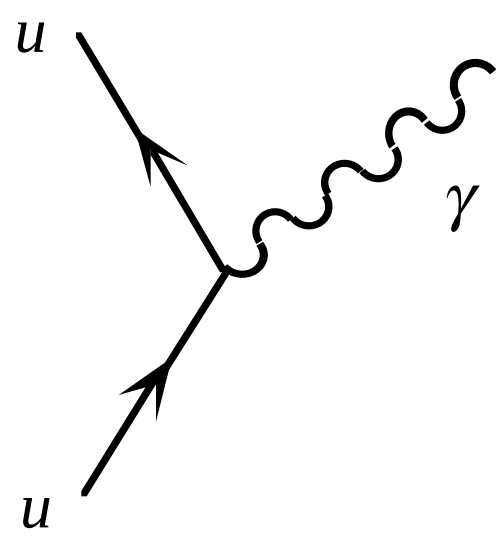 | 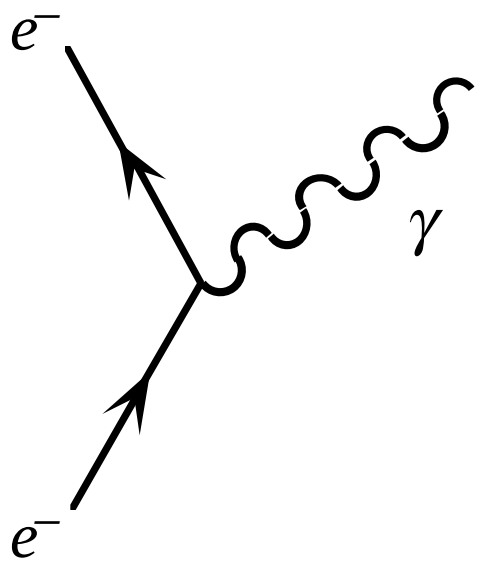 |
| Weak | Weak charge($g_W$), giving $u$ -> $d$ or $\nu_e$ -> $e^-$ flavor-changing transions | 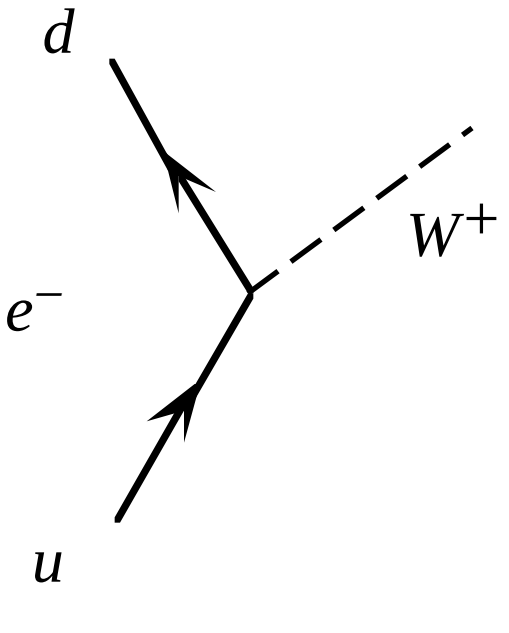 | 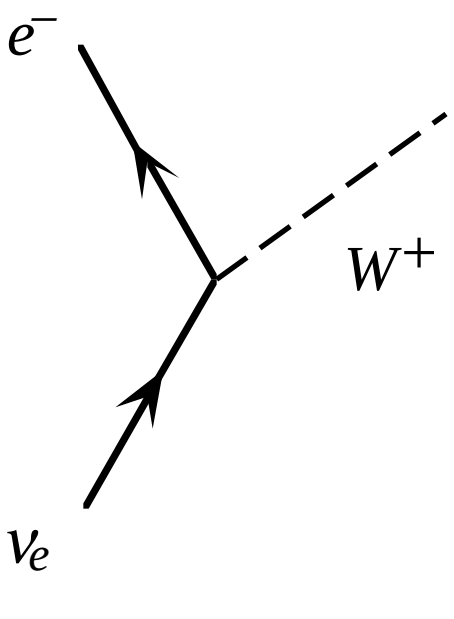 |

[1] For a description of how to read and construct Feynman diagram see Ref. [16].

### 1.1.1 Hadroproduction of b quarks

The hadronic production of *b* quarks is fundamental for the study of perturbative quantum chromodynamics (QCD) since results may be used to test and constrain the current form of the theory. The cross-sections for open-flavor (i.e., unbound) production of heavy quark pairs ($Q\overline{Q}$) are predicted by QCD. The lowest-order parton subprocesses are quark-antiquark fusion and gluon-gluon fusion, illustrated in Figure 1.1.

The hadron-hadron ($h_1 h_2 \rightarrow Q\overline{Q}X$) production cross-section for heavy quark pairs is expressed as convolution of the parton-parton scattering cross-sections, $\hat{\sigma}_{i,j}$, with the distribution function of the partons, summed over all partons:

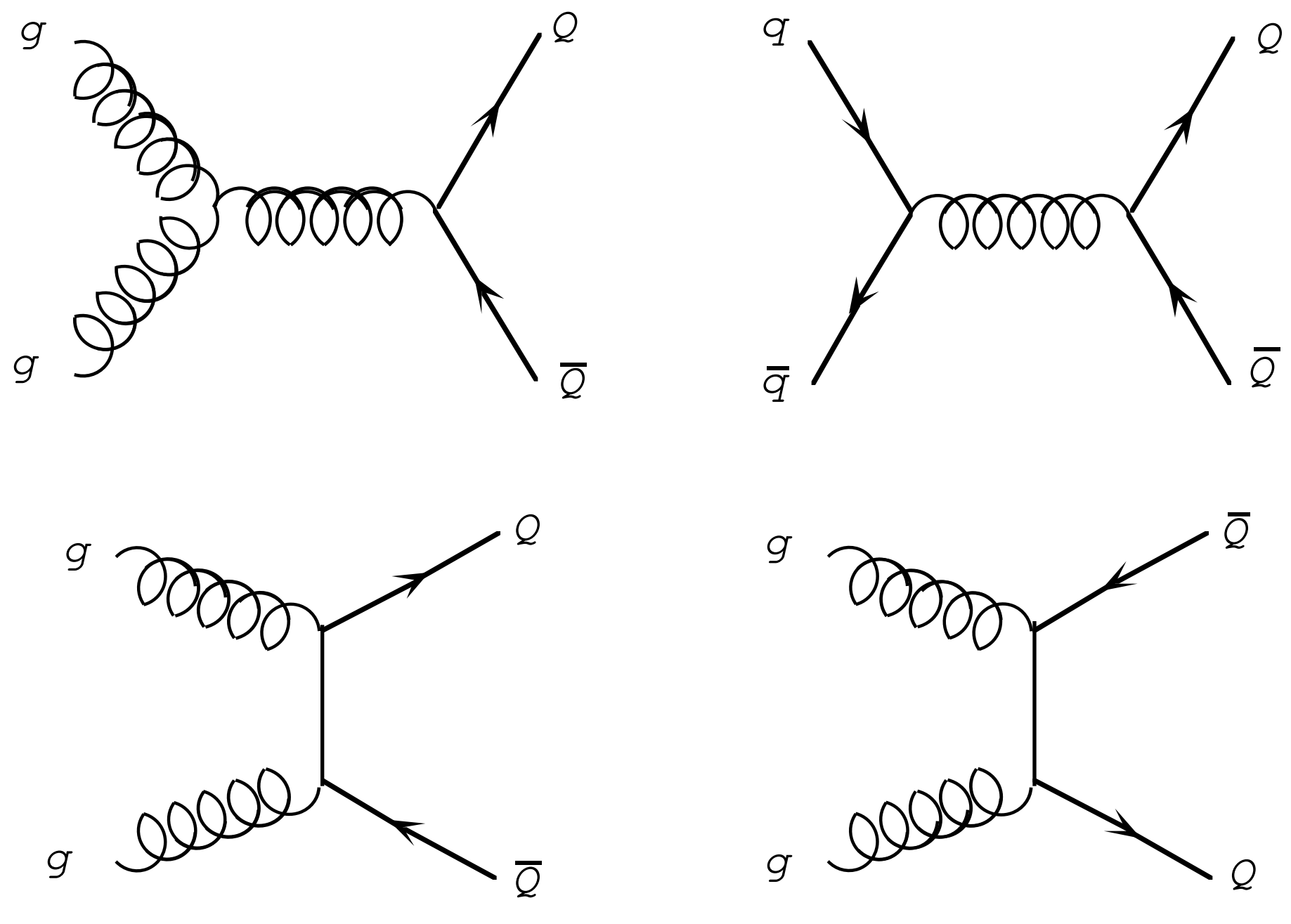

Figure 1.1. Lowest-order QCD subprocess producing $Q\overline{Q}$ heavy quark pairs.

$$\sigma\left(S,m,Q^2\right)=\sum_{i,j}\int_{4m^2/S}^{1}dx_1\int_{4m^2/x_1S}^{1}dx_2\; f_i\left(x_1,Q^2\right)\, f_j\left(x_2,Q^2\right)\hat{\sigma}_{i,j}\left(\hat{s},m,Q^2\right), \qquad (1.1)$$

where $S$ is the square of the total center-of-mass energy in the hadron-hadron system, $m$ is the mass of the heavy quark, and $Q^2$ is the mass factorization scale. The $i$ and $j$ indices run over all partons (quarks, antiquarks, and gluons) in the hadrons $h_1$ and $h_2$, respectively. The variable $x_1$ is the fractional four-momentum vector of the parton in hadron $h_1$, and $x_2$ is that of the parton in hadron $h_2$. Thus, in the parton-parton system the square of the center-of-mass energy is $\hat{s} \approx x_1 x_2 S$. $f_i\left(x_1,Q^2\right)\, dx_1$ is equal to the probability that parton $i$ in hadron $h_1$ carries a fraction $x_1$ of $S$, analogously, $f_j\left(x_2,Q^2\right)\, dx_2$ for parton $j$ in hadron $h_2$. These parton densities are empirically determined from electron and neutrino deep-inelastic scattering (see Ref. [17]), and the parton-parton scattering cross-section, $\hat{\sigma}_{i,j}\left(\hat{s},m,Q^2\right)$, can be calculated in perturbative QCD [18].

The partonic cross-sections in a next-to-leading-order (NLO) or $O\left(\alpha_s^3\right)$ power-series expansion are usually written in terms of so-called scaling ratios as

$$\hat{\sigma}_{i,j}\left(\hat{s},m^2,Q^2\right)=\frac{\alpha_s^2}{m^2}\left[f_{i,j}^{(0)}(\eta)+4\pi\alpha_s\left\{f_{i,j}^{(1)}(\eta)+\bar{f}_{i,j}^{(1)}\ln\left(\frac{Q^2}{m^2}\right)\right\}\right], \qquad (1.2)$$

where $\eta=\dfrac{\hat{s}}{4m^2}-1$ or $\eta=\dfrac{4m^2}{\hat{s}}$. The dimensionless functions $f_{i,j}^{(0)}(\eta)$ and $f_{i,j}^{(1)}(\eta)$ represent the Born approximation and the $O(\alpha_s)$ corrections, respectively, and $\bar{f}_{i,j}^{(1)}(\eta)$ appears when the mass factorization scale, $Q^2$, deviates from the square of the heavy flavor quark mass, $m^2$. NLO calculations of the partonic cross-sections have been computed by Nason, Dawson, and Ellis (NDE) [19]. The $b\bar{b}$ total cross-section for hadron-hadron collisions at current fixed-target energies has been calculated to NLO using NDE results by Berger [20], and also by Mangano, Nason, and Rodolfi (MNR) [21]. Experimentally, the $b\bar{b}$ total cross-section has been measured at fixed-target energies by experiments WA78 [22] and NA10 [23] at CERN and E653 [24] and E672/E706 [11] at Fermilab; Figure 1.2 shows the results for $\pi^{-}$N collisions and also the theoretical predictions of Berger and MNR for several different mass factorization scales, $Q^2$, and heavy-flavor quark masses, $m_b$. It is important to mention here that in hadron-hadron collisions at collider energies such as that in CDF and D0 ($\sqrt{s}$ = 1800 GeV) the $b\bar{b}$ total cross-section is three orders of magnitude larger than at our fixed-targert energies [25].

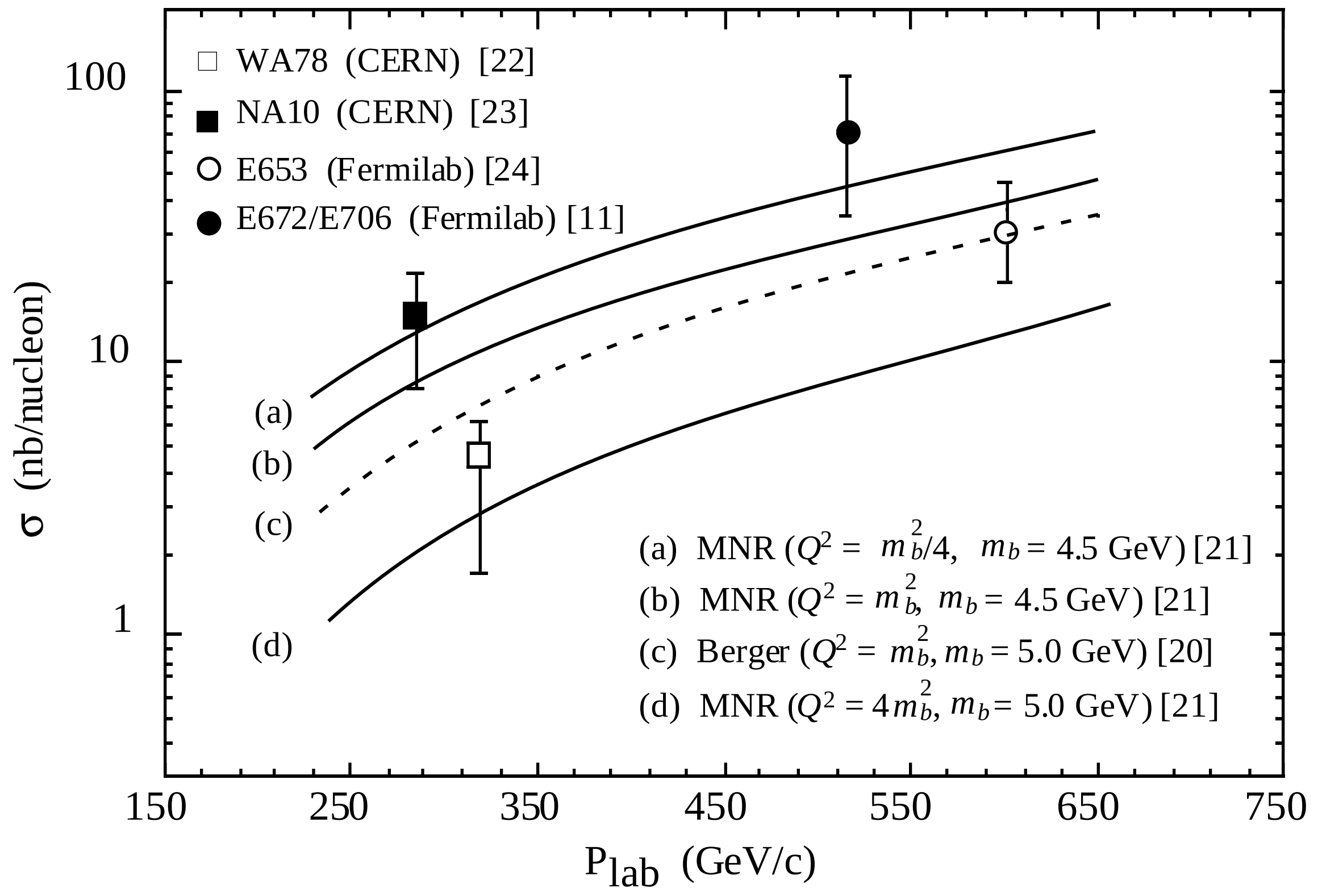

Figure 1.2. The $b\bar{b}$ total cross-section for $\pi^-$N collisions as measured by references. [22], [23], [24], and [11], and the theoretical prediction by Berger [20] and MNR [21]. The uncertainties shown for the measured values are the statistical and systematic contributions added in quadrature.

### 1.1.2 Fragmentation of b quarks

After production of $Q\overline{Q}$ pairs, *fragmentation*, or *hadronization*, occurs subsequently. During this process the color forces organize the $Q\overline{Q}$ pairs with other quarks created in this process into colorless hadrons. Fragmentation is governed by soft, non-perturbative processes that cannot be calculated from first principles and can only be modeled. In the LUND string-breaking model [26] for example, the color energy stored in the color field increases as the individual $Q\overline{Q}$ quarks separate; then, at some point, the color field has enough energy to produce a $q\overline{q}$ (light quark pair) or a $qq\overline{q}\overline{q}$ (light diquark pair) and the string breaks. This effect is shown in Figure 1.3. In this model the ratio of the probability for $qq\overline{q}\overline{q}$ production to that for $q\overline{q}$ is 0.09 : 1.0, where the probability for $q\overline{q}$ production approximately 0.3. [27]. If a *b* (heavy) quark combines with a single light quark, the result is a beauty meson; if it combines with a light quark pair, the result is beauty baryon. The probability $F(\Lambda_b)$ of a *b* quark fragmenting into a $\Lambda_b$ or into a beauty baryon that subsequently decays into a $\Lambda_b$, has been measured to be 7.2 ± 1.2 % [28].

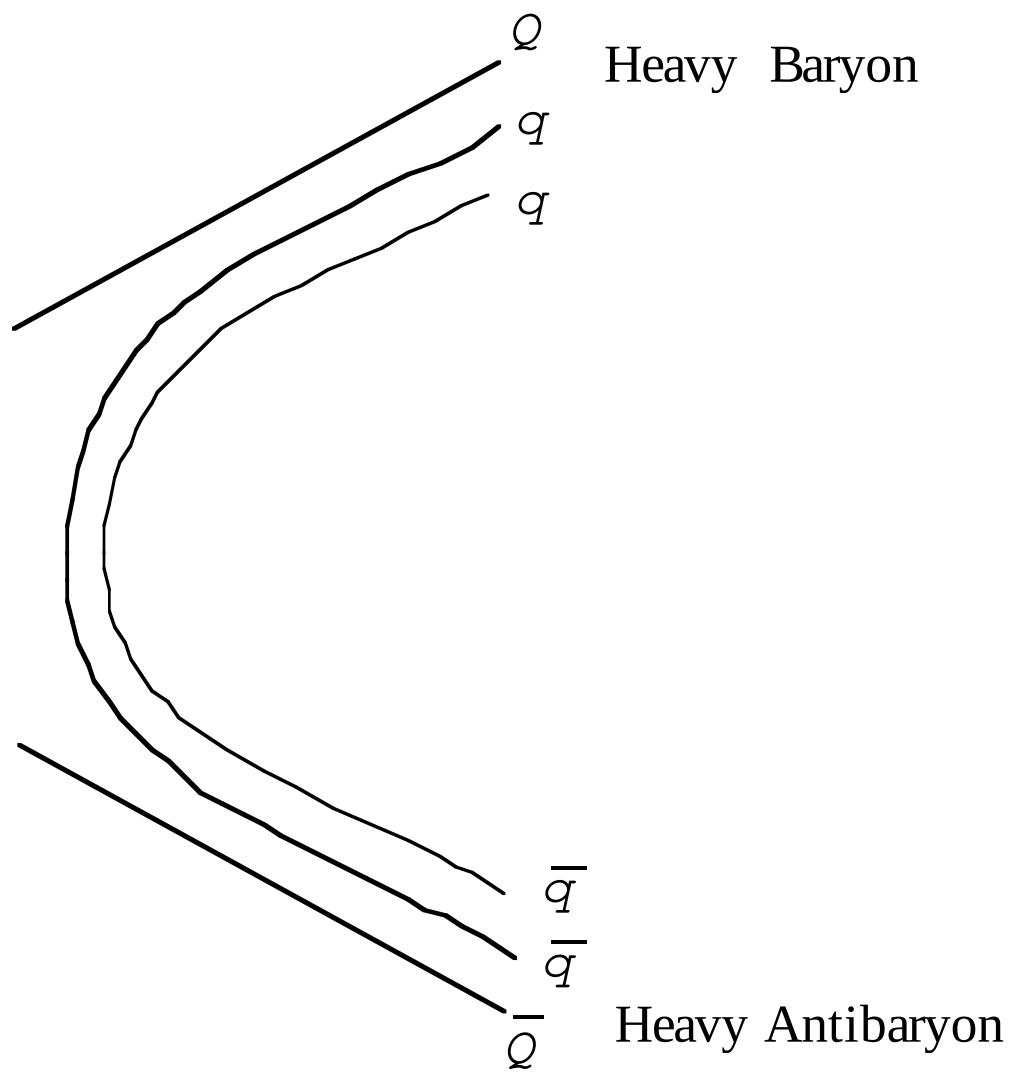

Figure 1.3. Schematic picture of fragmentation in which breaking of the lines of the color force between the separating heavy quarks of a $Q\overline{Q}$ pair produce a light $qq\overline{q}\overline{q}$ diquark pair which combines with the heavy quarks to form a heavy baryon pair .

### 1.1.3 b-hadr on decay

Several kinds of Feynman diagrams contribute to the non-leptonic decays of b hadrons; examples are shown in Figure 1.4. The processes shown are called spectator (external or internal), exchange, annihilation, and penguin (electromagnetic or hadronic) diagrams. In spectator diagrams the *b* quark decays in to a *c* quark (shown), or *u* quark (not shown) and a $W^-$ boson which, in turn, decays primarily into $\bar{u}d$ or $\bar{c}s$ (favored in the Cabbibo-Kobayashi-Maskawa (CKM) scheme), or into $e^-\bar{\nu}_e$, $\mu^-\bar{\nu}_\mu$, or $\tau^-\bar{\nu}_\tau$. In these decays, the lighter quarks in the b hadron act only as spectators and do not participate in the decay. It is important to mention the difference between “external” and “internal” spectator diagrams. In the external diagram, the colorless $W^-$ is allowed to decay into three possible $\bar{u}d$ or $\bar{c}s$ color singlet final states. In contrast, in the internal diagram, in order to form colorless final states only one of the three possible virtual $W^-$ decays is allowed. Therefore, the internal diagram is suppressed by a factor of $1/N_c$ with respect to the exterior diagram, where $N_c$ is the number of colors.

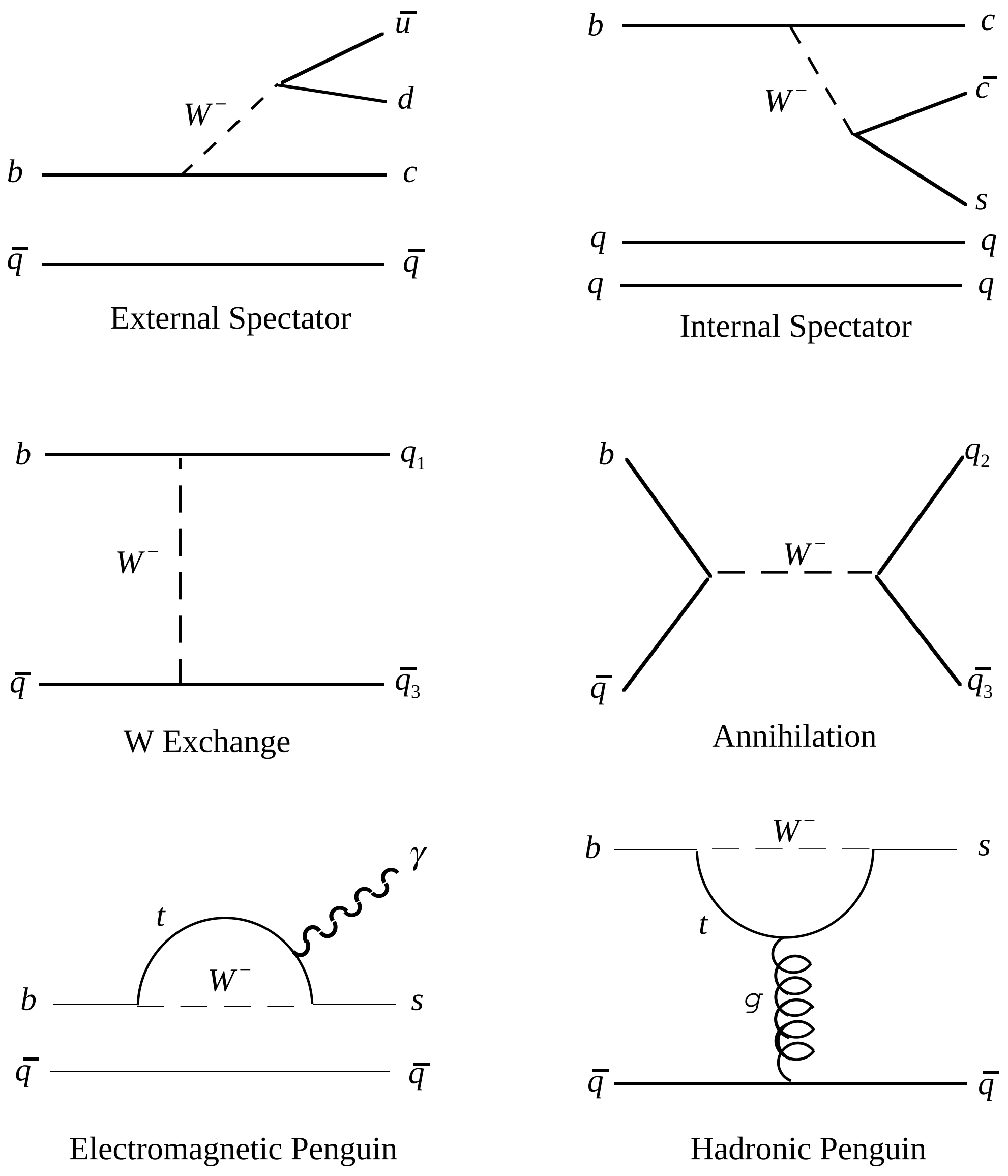

Figure 1.4. Feynman diagrams for the non-leptonic decay of a b hadron.

The total b-quark decay width from the spectator diagrams, $\Gamma_b^{tot\,SP}$, has been computed to be [29]

$$\Gamma_b^{tot\,SP} = \frac{G^2 m_b^5}{192\pi^3}\left[2.72|V_{bc}|^2 + 6.92|V_{bu}|^2\right], \qquad (1.3)$$

where $G$ is the Fermi coupling constant, $m_b$ the mass of the *b* quark, $V_{bc}$ and $V_{bu}$ are the CKM matrix elements that represent the $b \to c$ and $b \to u$ quark transitions, respectively.

In the W-exchange and annihilation decays, both initial-state quarks are involved in the weak vertex; the classic example is $\pi^- \to \mu^- \bar{\nu}_\mu$ (or $\pi^- \to e^- \bar{\nu}_e$). For b-hadrons these types of decays are helicity-suppressed. To zeroth order in QCD, the total hadronic width for W exchange, $\Gamma_b^{W\,ex}$, is given in Ref. [29] to be

$$\Gamma_b^{W\,ex}(b\bar{q} \to q_1\bar{q}_3) = \frac{1}{3}\frac{G^2}{8\pi}|V_{bq_1}|^2|V_{q\bar{q}_3}|^2 f_{b\bar{q}}^2\, m_b\left(m_1^2 + m_3^2\right)M(m_1, m_3, m_b), \qquad (1.4)$$

and that for annihilation is

$$\Gamma_b^{annih}(b\bar{q} \to q_2\bar{q}_3) = 3\frac{G^2}{8\pi}|V_{bq}|^2|V_{q_2\bar{q}_3}|^2 f_{b\bar{q}}^2\, m_b\left(m_2^2 + m_3^2\right)M(m_2, m_3, m_b), \qquad (1.5)$$

where the $V_{qq'}$ are CKM matrix elements,

$$M(m_1, m_3, m_b) = \left[1 - \frac{\left(m_1^2 - m_3^2\right)^2}{m_b^2\left(m_1^2 + m_3^2\right)}\right]\left\{\left[1 - \left(\frac{m_1 + m_3}{m_b}\right)^2\right]\left[1 - \left(\frac{m_1 - m_3}{m_b}\right)^2\right]\right\}^{\frac{1}{2}},$$

and $f_{b\bar{q}}$, called the decay constant, has dimensions of mass and is related to the amplitude of the $b\bar{q}$ wave function at the origin by $f_{b\bar{q}}^2 = 12|\psi(0)|^2 / M_{b\bar{q}}$ (where $M_{b\bar{q}}$ is the mass of the $b\bar{q}$ system. The value of $|\psi(0)|^2$ can be calculated in a similar fashion as for the hydrogen atom. It involves the reduced mass of the system, $m_b m_q / (m_b + m_q) \approx m_q$, and the coupling constant $\alpha_s$; that is, $|\psi(0)|^2 \propto \alpha_s^3 m_q^3$. Thus, by taking $M_{b\bar{q}} \approx m_b$ there follows

$$\frac{\Gamma_b^{W\,ex+annih}}{\Gamma_b^{total\ SP}} \propto \alpha_s^3 \left( \frac{m_q}{m_b} \right)^5 . \qquad (1.6)$$

Since widths are proportional to rates, this indicates that the ratio of the (W exchange + Annihilation) decay rate to the total spectator decay rate is, in general small, since $m_b >> m_q$.

The decay width for the penguin diagram is given by [29]

$$\Gamma_b^{Pen}(b\bar{q} \to s\bar{q}) = \frac{1}{3}(c_5 + 3c_6)^2 \frac{G^2}{8\pi} |V_{tb}|^2 |V_{ts}|^2 f_{b\bar{q}}^2 m_b^3 , \qquad (1.7)$$

where $c_5$ and $c_6$ are constants with values of 0.02 and -0.04, respectively, defined in Ref. [29]. The penguin contribution to b-hadron decay is negligible (a few per cent) with respect to the spectator contribution.

In the spectator-type decay where $b \to c\bar{c}s$, the $c\bar{c}$ can bind to form a $\eta_c$, J/ψ, $\chi_1$, $\psi'$, or a higher-mass charmonium state. The relative ratios for the production of some of these states is predicted in Ref. [30] to be 0.57 : 1.0 : 0.27 : 0.31 for $\eta_c$:J/ψ:$\chi_1$:$\psi'$.

The $\Lambda_b \to J/\psi\ \Lambda^0$ decay has no nonspectator contributions; thus, the theoretical computation of the branching ratio for this decay is relatively clean. The theoretical prediction is that the Br($\Lambda_b \to J/\psi\ \Lambda^0$) is less than $10^{-3}$ [31].

## 1.2 Evidence for the existence of the $\Lambda_b$

While b-flavored baryons have long been predicted by the quark model, only very recently (1992) has their existence been confirmed. Evidence for $\Lambda_b$ baryons was reported by LEP experiments ALEPH, DELPHI, and OPAL [10] through observation of the semi-leptonic decay channels $\Lambda_b \to \Lambda^0 l^- \bar{\nu}_l X$ and $\Lambda_b \to \Lambda_c^+ l^- \bar{\nu}_l X$ in $Z^0 \to b\bar{b}$ events. The analysis was based on correlations between the baryon and lepton produced in the decays, since b baryons are expected to produce $\Lambda^0 l^-$, $\bar{\Lambda}^0 l^+$ or $\Lambda_c^+ l^-$, $\Lambda_c^- l^+$ pairs, and not $\Lambda^0 l^+$, $\bar{\Lambda}^0 l^-$ or $\Lambda_c^- l^-$, $\Lambda_c^+ l^+$ pairs (see Figure 1.5). The experiments found an excess of events of the correct baryon-lepton combination pairs, and interpreted this excess as $\Lambda_b$ decays. The results, in the form of the product of production fraction and branching ratio are, summarized in Table V.

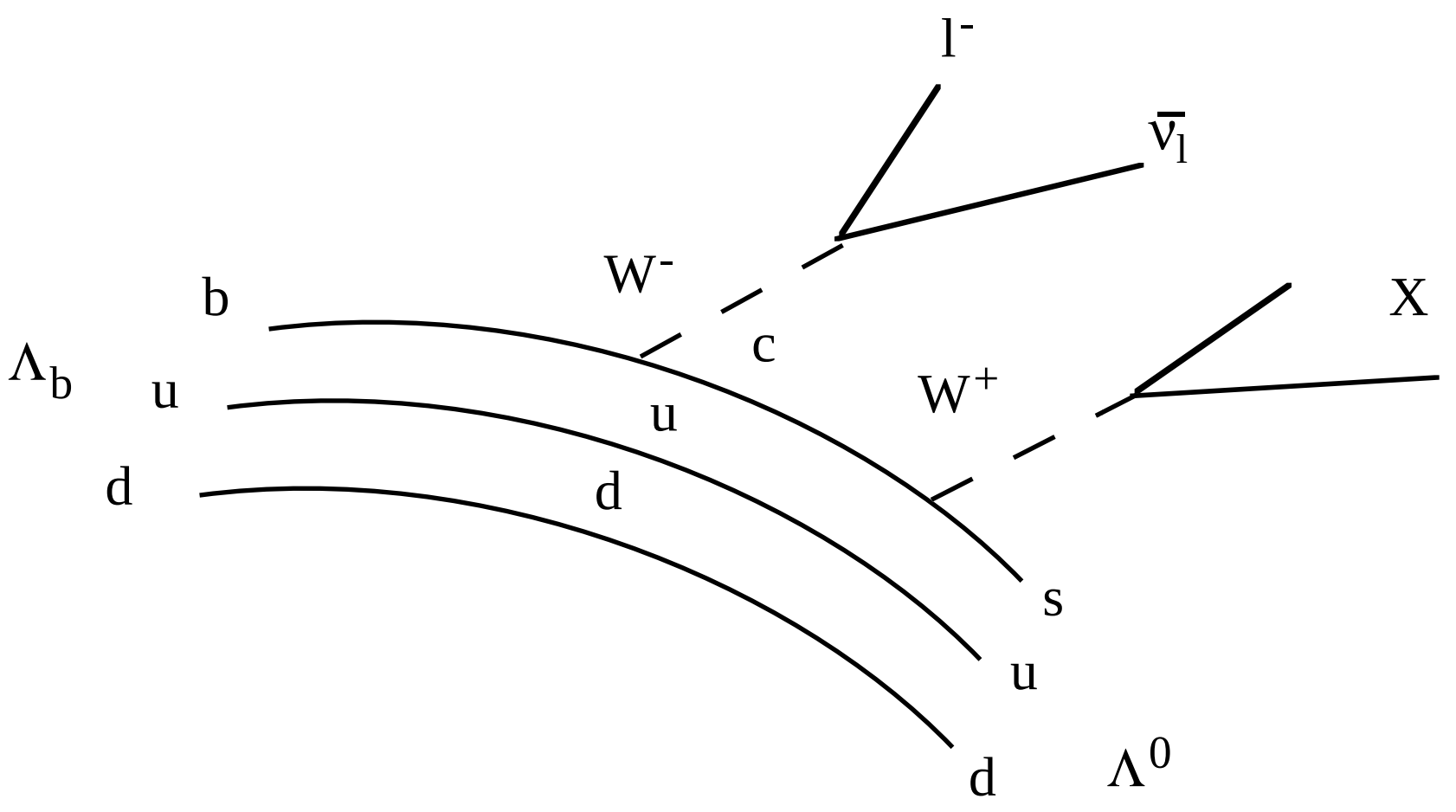

Figure 1.5. The $\Lambda_b \to \Lambda^0 l^- \bar{\nu}_l X$ decay channel quark diagram.

Table V. Results on $\Lambda^0$-lepton and $\Lambda_c^+$-lepton correlations in $Z^0$ decays at LEP. The first uncertainty is statistical and the second is systematic.

| | $F(\Lambda_b) * Br(\Lambda_b \to \Lambda^0 l^- \nu_l X)$ $(10^{-3})$ | $F(\Lambda_b) * Br(\Lambda_b \to \Lambda_c^+ l^- \nu_l X)$ $(10^{-2})$ |
|---|---|---|
| ALEPH | $6.1 \pm 0.6 \pm 1.0$ | $1.51 \pm 0.29 \pm 0.23$ |
| DELPHI | $3.0 \pm 0.6 \pm 0.4$ | $1.18 \pm 0.26 \ ^{+0.31}_{-0.21}$ |
| OPAL | $2.91 \pm 0.23 \pm 0.25$ | $0.83 \pm 0.28$ |

## 1.3 Previous sear ches for the $\Lambda_b \to J/\psi\ \Lambda^0$ decay channel

The quark diagram for the $\Lambda_b \to J/\psi\ \Lambda^0$ decay channel is shown in Figure 1.6. This diagram is an internal spectator diagram. Thus, it is colored-suppressed, since the $c\bar{c}$ pair form a color-singlet state in order to be bound. Only CERN experiment UA1 has observed $\Lambda_b$ s in the $J/\psi\ \Lambda^0$ decay mode. They found a signal of 16 ± 5 events above a background of 9 ± 1 events [7], and measured the mass of the $\Lambda_b$ to be (5640 ± 50 (*stat*) ± 30 (*sys*)) MeV/$c^2$ and F($\Lambda_b$) * Br($\Lambda_b \to J/\psi\,\Lambda^0$) to be (1.8 ± 0.6 (*stat*) ± 0.9 (*sys*))x $10^{-3}$. ALEPH, OPAL, and CDF have failed to confirm UA1's signal (see Refs. [8] and [9]). The results found in their search are summarized in Table Table VI.

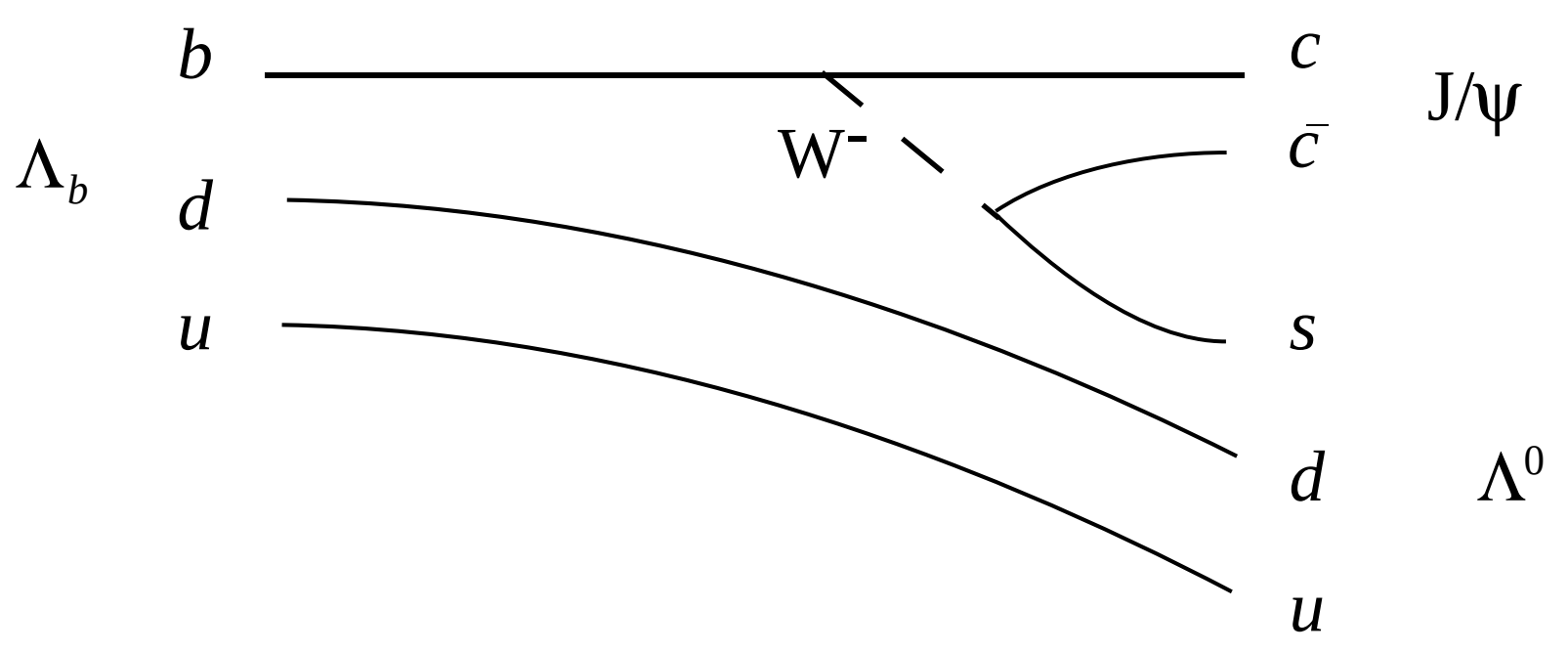

Figure 1.6. The $\Lambda_b \to J/\psi\,\Lambda^0$ decay channel quark diagram.

Table VI. Results from other searches for the $\Lambda_b \to J/\psi\,\Lambda^0$ decay channel.

| Experiment | $F(\Lambda_b) * Br(\Lambda_b \to J/\psi\,\Lambda^0)$ |
|---|---|
| UA1 | (1.8 ± 0.6(*stat*) ± 0.9 (*sys*)) x $10^{-3}$ |
| CDF | < 0.5 x $10^{-3}$ |
| ALEPH | < 0.4 x $10^{-3}$ |
| OPAL | < 3.4 x $10^{-4}$ |

## 1.4 Analysis objective

The object of this thesis is to search for the $\Lambda_b$ beauty baryons in the $\Lambda_b \rightarrow J/\psi\ \Lambda^0$ channel (and the charge conjugate reaction) using the 1990 data from fix-target experiment E672 at Fermilab. The J/$\psi$ s are identified by their decay into $\mu^+ \mu^-$ and the $\Lambda^0$ through the decay channel $\Lambda^0 \rightarrow p\ \pi^-$. By relating the number of $\Lambda_b \rightarrow J/\psi\ \Lambda^0$ events to the E672 measurement of the b-quark cross-section obtained from the same data, the production rate $F(\Lambda_b) * Br(\Lambda_b \rightarrow J/\psi\ \Lambda^0)$ will be extracted. (Throughout this thesis, whenever a state is mentioned its charge conjugate state is also implicitly implied.)

# CHAPTER 2

# THE MWEST SPECTROMETER

The MWEST spectrometer, was located at the meson-west area at Fermilab. It was used simultaneously by experiments E672 and E706. The experiments accumulated data in 1990 and 1991. The output signals from all detector systems were available to both experiments. E672 concentrated on dimuon events, whereas E706 was interested in events that contained direct photons [32]. Experiment E672 was designed to study the hadroproduction of high-mass muon pairs which resulted from heavy-quark production, particularly charmonium [33]. This chapter begins with an overview of the complete spectrometer and then presents a detailed description of the individual components of the apparatus that were used to obtained the data for the analysis in this thesis.

## 2.1 Overview of the detector

The physical layout of the combined E672/E706 apparatus is shown in Figure 2.1. A right-handed orthogonal coordinate system was associated with the spectrometer, with its origin near the target and with the z-axis aligned with the beam direction. The x and y axes were in the horizontal and vertical directions, respectively. The apparatus began with a hadron shield used to absorb beam halo and a set of scintillation-counter veto walls used to veto events with halo particles at the pretrigger level. Downstream was a Cu and Be target, a tracking system composed of a set of 16 silicon-strip detectors (SSDs), a dipole magnet with average $p_T$ kick of 0.4457 GeV/c, 16 proportional-wire-chambers (PWCs), and 4 straw drift-tube detector planes. Following this was a liquid-argon electromagnetic (EMLAC) and hadronic (HADLAC) calorimeter, and a forward calorimeter for energy measurement of forward-going electrons, photons, and hadrons. Finally, about 20 meters downstream of the target was the muon spectrometer consisting of a concrete hadron shield (beam dump) to prevent hadrons produced in the underlying event into the muon system, a toroid magnet with average $p_T$ kick of 1.3 GeV/c, 12 muon PWCs, and two scintillator-hodoscope planes (muon hodoscopes H1 and H2) with associated electronics for generating a trigger signal upon the detection of two muons. Each of the elements pertinent to the analysis in this thesis are discussed in more detail.

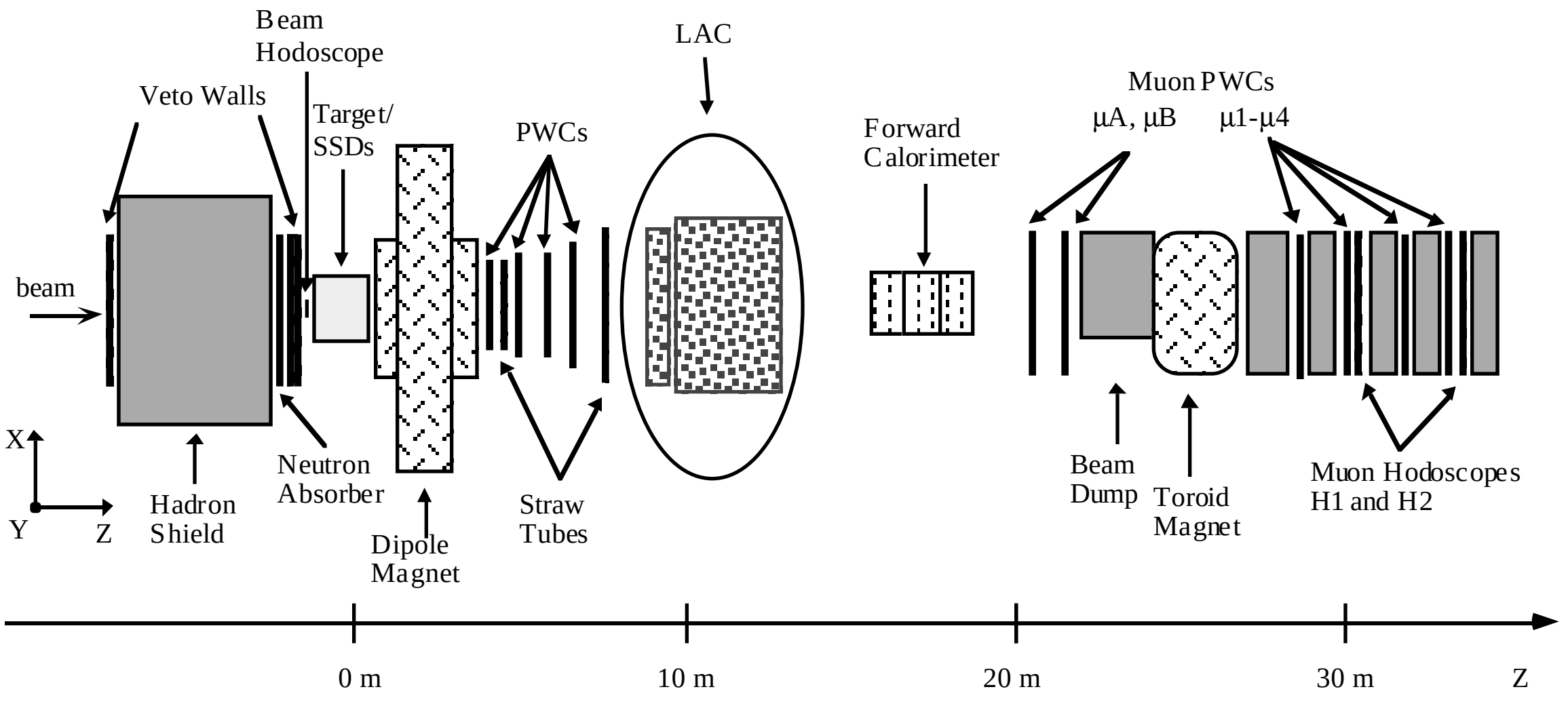

Figure 2.1. The physical layout of the MWEST spectrometer.

## 2.2 Beam

The Fermilab Tevatron operated on a 57.2 second cycle. During which, 23 seconds were used for the spill of the beam to the experimental areas. The Tevatron provided an 800 GeV/c proton beam to the switchyard, the area where the beam was split and directed to the three major beam areas: meson, proton, and neutrino. The beam had a 19.7 ns bucket structure from the characteristic radio frequency of the Tevatron.

### 2.2.1 Beamline

The beam used in the experiment was produced in the following way: The portion of the beam sent to the meson west area, which had a typical intensity of about 2 x $10^{12}$ protons per spill, collided with a beryllium target of dimensions, 46.5 cm in length and 2.22 cm in diameter, corresponding to 1.14 interaction lengths. This produced a secondary $\pi^-$ beam of 515 ± 3 GeV/c average momentum [34], and intensity at its maximum of 2x$10^8$ $\pi^-$ per spill [35]. The same bucket structure applied to the secondary beam, where the probability of a bucket being occupied by a single particle was about 10%, and the probability of being occupied by more than one particle was approximately 2%.

Both pion and proton beams were delivered to the meson-west area through the Fermilab meson-west beamline, which was built specifically for the experiment. This above ground beamline was designed to transport positively and negatively charged particles with a momentum of up to 1 TeV/c.

### 2.2.2 Cerenkov counter

The secondary beam used by the experiment in 1990 was not a pure pion beam, 97.0% of the beam particles were $\pi^{-}$s, 2.9% $K^{-}$s, and 0.1% antiprotons [35]. A Cerenkov detector was used to tag the beam particles. The detector had dimensions of 42.1 m long and radius of 24.4 cm, and it was positioned in the direction of the beam. Helium gas, at pressure between 4 - 8 psi, was used as the radiator. For a more detailed explanation of the Cerenkov counters see reference [35].

## 2.3 Hadron shield and veto walls

A large stack of iron, of dimensions 4.7 m long, 4.3 m wide, and 3.7 m high, with a hole in its center (for the beam particles to pass through), was placed at the end of the beamline, just inside the experimental hall. This served as a hadron shield to absorb the beam halo, those particles traveling in the beam direction, but not on the beam axis. Also "spoiler" magnets in the upstream beamline were employed to sweep halo particles away from the beam.

Halo muons, produced primarily from $\pi^{-}$ decays from the beam, could however, pass through the iron shield very easily, and subsequently throughout the entire spectrometer. To tag the muons so that they would not start the dimuon trigger (for example, one muon from the target and a halo muon), a series of three "veto walls" scintillation counters were used. One was placed just upstream of the hadron shield (VW3) and the other two just down stream (VW1 and VW2), see Figure 2.1. The veto wall VW3 consisted of 18 rectangular scintillator counters, and it covered an area of about 3 m x 3 m.

Veto walls VW1 and VW2 each consisted of matrix of thirty two 50 cm x 50 cm scintillator counters, also covering an area of approximately 3 m x 3 m. These two walls had a 10 cm offset in the x-y plane to cover the gaps from one another. The three walls had a hole in the central region to allow the beam to pass without generating a signal. A coincidence (VW1+VW2) * VW3 between the walls was established to veto events containing halo muons, and this was done at the pretrigger level.

## 2.4 Target

The experiment used a target composed of two pieces of Cu and two pieces of Be, as shown in Figure 2.2. The two Cu targets were 0.8 mm thick, and were separated by a 0.26 cm air-gap. Downstream of this, 0.55 cm, was the first piece of Be, which had a thickness of 3.71 cm, and was followed by a 1.02 cm air-gap. Finally, there was the second piece of Be, which was 1.12 cm thick.

## 2.5 Upstream tracking system

This tracking system, located near the target, was used to reconstruct the tracks (trajectories), and the vertices made by the tracks of the outgoing charged particles accepted by the spectrometer. It had four major parts: the SSDs, the upstream PWCs, the dipole magnet, and the STRAWS. The STRAWS will not be discussed here, since they were not used for this analysis.

### 2.5.1 Silicon-strip detector system

The arrangement of the SSD system is shown in Figure 2.2. It consisted of 16 separate planes, 6 of them upstream of the target and 10 planes downstream. The SSD planes measured the position of a charged particle by putting out a signal induced by the particle on the strips etched onto the silicon wafers. The 16 planes were setup in pairs, consisting of one plane with its silicon-strips oriented in the x-direction (horizontal) and the other in the y-direction (vertical). The 3 x-y planes upstream of the target, had an area of 3 cm x 3 cm, and a pitch of 50 μm. The 5 pairs of planes downstream of the target all had a 5 cm x 5 cm area, and except for the plane just after the target, which had a pitch of 25 μm, they all had a 50 μm pitch.

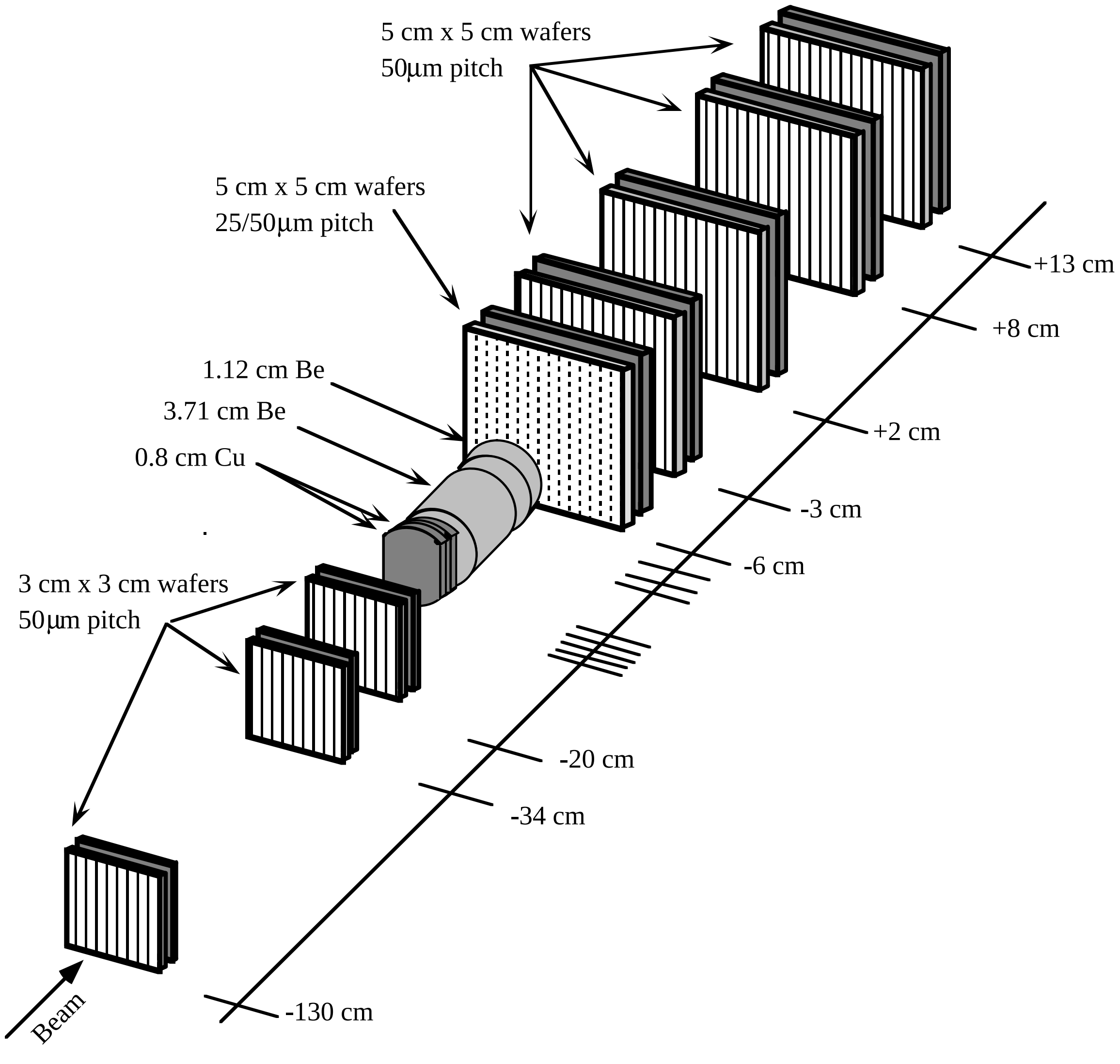

Figure 2.2. The physical layout of the target and SSD system.

The vertex position resolution of the SSD system was of 10 μm in the x- and y-direction, and 350 mm in the z-direction. The angular resolution was 0.1 milliradians. This tracking device played an important role in the reconstruction of secondary vertices, those produced from decays in flight of particles emerging from the primary interaction.

### 2.5.2 Dipole magnet

A conventional liquid helium dipole magnet was used to measure the momentum of the charged particles. This was done by measuring the deflection of their trajectory as they passed through the magnet. The dipole magnet was operated at a current of 1050 A, giving rise to an approximately uniform magnetic field of magnitude 0.6115 Tesla. The magnetic field was oriented in the negative y-direction, and it extended through 2.416 m (length of the magnetic field). A mirror plate was placed in the front (upstream side) and back of the magnet to reduce the fringe field. Each plate had a rectangular opening to allow particles to pass through, the one in front was 35.56 cm wide x 25.4 cm high, and the one in the back 127.0 cm x 91.44 cm. A polyethylene bag filled with helium gas was installed in the central region of the magnet to reduce the multiple scattering of the practices when traversing through the magnet.

Since the direction of the magnetic field was aligned with the y-direction (vertical), the trajectories of the charged particles were only deflected in the horizontal x-view (x-z plane). The effective change, in the x-component of the particle's momentum was by the so called " $p_{T\,kick}$" of the magnet,

$$p_{T\,kick} = 0.3 B_0 L, \tag{2.1}$$

where $B_0$, the magnitude of the field, is in Tesla, and L, the length of the field, is in meters. This was equal to 0.4457 GeV/c in the x-direction.

### 2.5.3 Upstream proportional wire chambers

Downstream of the dipole were a set of 16 proportional wire chambers (PWCs). They were grouped into four modules, named: PWC1, PWC2, PWC3, and PWC4. Each module contained a plane with its anode wires aligned with the x- (horizontal), y- (vertical), u- (rotated $+37^0$ from the vertical), and v- (rotated $-53^0$ from the vertical) directions. The physical dimensions of the four modules were different in order to obtain a constant angular acceptance. PWC1 was 1.63 m wide x 1.22 m high, PWC2 and PWC3 were 2.03 $m^2$, and PWC4 was 2.44 $m^2$. The anode signal planes were made of gold-plated 0.8 mm in diameter wires, spaced at 2.54 mm. For each anode plane there were two cathode planes, one on each side, separated by a distance of 5.74 mm. These cathode planes were made of graphite-coated 1.0 mm thick mylar sheets. The chambers were filled with a gas mixture of 18% isobutane, 2.2% isopropyl alcohol, 0.1% freon, and 79.7% argon. The PWC system is described in grater detail in [36]

## 2.6 Muon system

The muon system was located at the down stream end of the spectrometer about 20 meters from the target. The physical layout of the muon detector is shown in Figure 2.3. It consisted of: 8 upstream muon PWCs, a beam dump (made of steel and tungsten), a conventional toroid magnet, four downstream muon PWCs, two 16-segment scintillating counter hodoscopes, and concrete walls to shield the muon PWCs.

### 2.6.1 Upstream muon proportional wire chambers

The first two muon PWCs named μA and μB were installed to reduce the number of triggers caused by the halo muons and to improve the linking tracks between the upstream tracking system and the muon system. Both μA and μB consisted of four anode signals planes: x, y, u, and v. The x and y planes were oriented in the horizontal and vertical directions, respectively. The u planes were rotated $10^0$ from the vertical, and the v planes were rotated $-10^0$ from the vertical. The anode planes were made of 1.0 millimeter diameter gold-plated wires, separated 3.05 mm. The dimensions of the active areas and the numbers of wires for each anode plane are listed in Table VII.

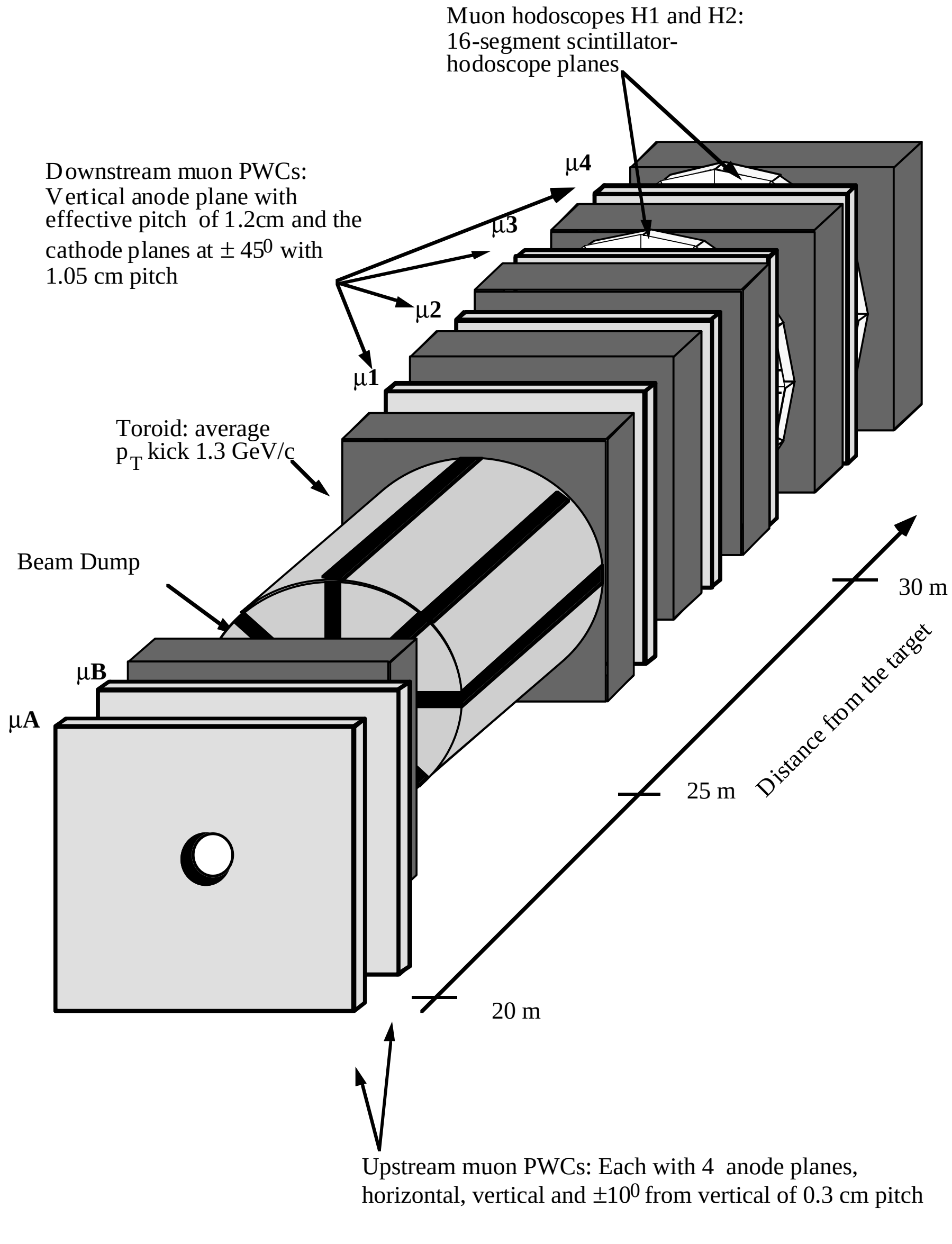

Figure 2.3. The physical layout of the muon spectrometer

The cathode planes consisted of 1650 3.5 mil diameter Cu/Be wires, separated 1.02 mm. The distance between adjacent cathode and anode planes was 0.95 cm. Two ground planes were also included in each PWC module, one in front of the first cathode plane and the other after the last cathode plane. The ground planes were made of 1725 3.5 mil diameter Cu/Be wires, separated 1.02 mm [37]. The gas mixture used in the PWCs was 76 % argon, 15 % isobutane, 8.9 % methylal, and 0.1 % freon. For more information on these PWCs see reference [37].

Table VII. Upstream muon PWC specifications.

| PWC plane | Orientation from the vertical | Number of wires | Active area ($m^2$) |
|---|---|---|---|
| μA x | $90^0$ | 256 | 1.7 x 0.8 |
| μA y | $0^0$ | 560 | 1.7 x 1.7 |
| μA u | $10^0$ | 352 | 1.7 x 1.1 |
| μA v | $-10^0$ | 352 | 1.7 x 1.1 |
| μB x | $90^0$ | 544 | 1.7 x 1.7 |
| μB y | $0^0$ | 544 | 1.7 x 1.7 |
| μB u | $10^0$ | 448 | 1.7 x 1.3 |
| μB v | $-10^0$ | 448 | 1.7 x 1.3 |

### 2.6.2 Toroid

An iron polarized toroid magnet was used to provide a second measure of the momentum of the muons. The toroid was 2.44 m long, had an outer radius of 1.35 m, an inner radius of 16.8 cm at the upstream end, and an inner radius of 19.7 cm at the downstream end. The magnetic field created by the toroid (operated normally at 1700 amperes) was in the azimuthal direction. It had a magnitude of 2.24 T at the inner radius and 1.74 at the outer radius. The field had no measurable dependence in the azimuthal angle. Muons received a momentum kick in the radial direction which decreases linearly from 1.64 GeV/c at the inner radius, to 1.27 GeV/c at the outer radius. The measurement of the momentum of the muons was used at the trigger level and was later compared to momentum measurement of the upstream tracking system.

### 2.6.3 Muon pr etrigger hodoscopes

Two scintillating hodoscopes were used in the muon spectrometer, arranged as shown in Figure 2.3. The muon hodoscopes both had a radius of 1.5 m to match the active area of the downstream muon PWCs. They consisted of 16 triangular scintillator segments ordered in a circular petal pattern, as shown in Figure 2.4. Each of the segments was made of PS-10 plastic scintillator encased in aluminum sheets. The scintillating light produced upon the passage of a muon through a scintillator segment was collected by a BBQ doped waveshifting bar and re-emitted at a wavelength of 410 nm. This light was collected by RCA 8575 photomultiplier tubes operating at 2100 volts. At this voltage the hodoscopes were at least 90 % efficient. The daisy counters were used at the pretrigger level to tag the muons produced in the underlying event .

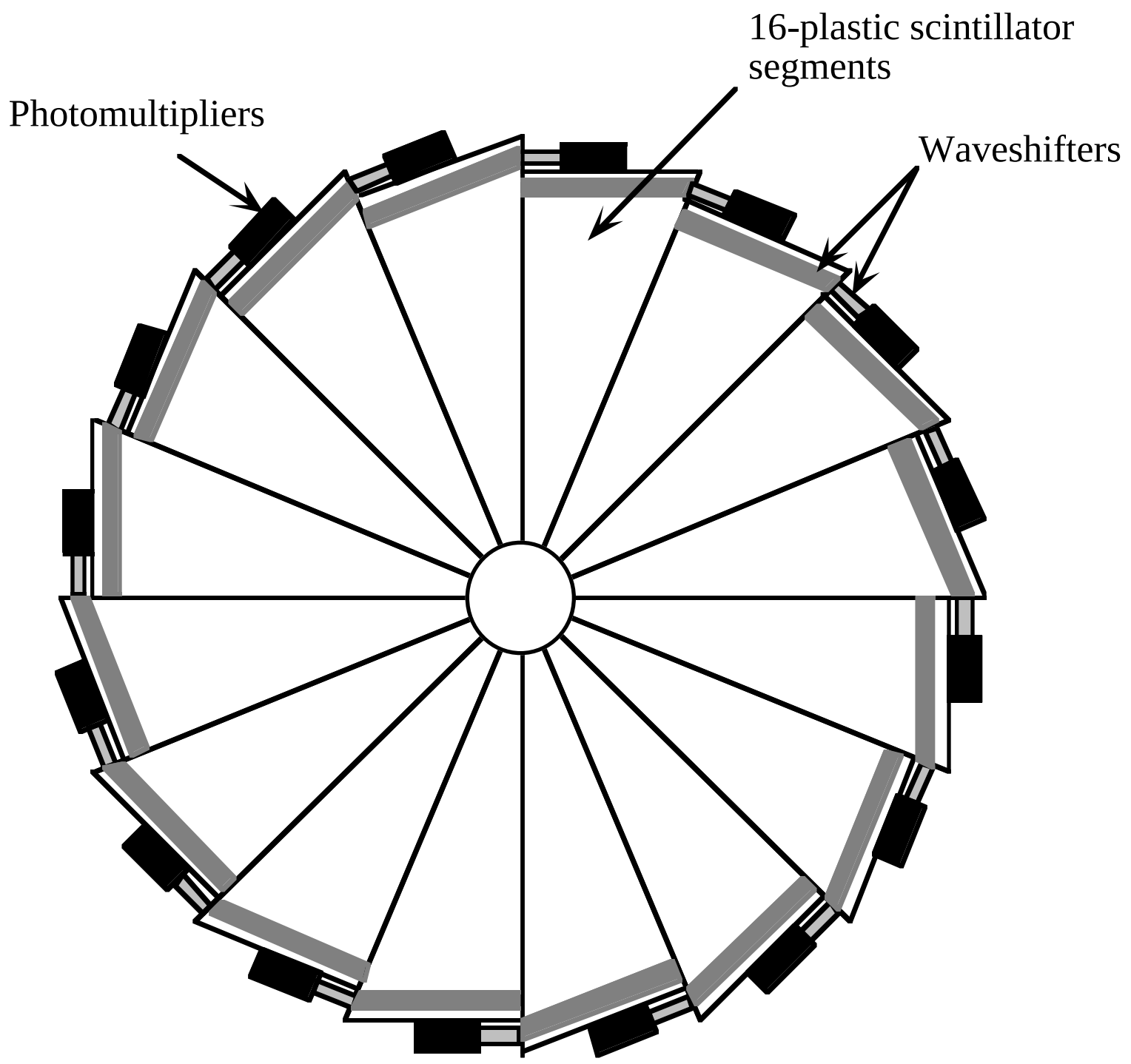

Figure 2.4. The schematic view of the muon pretrigger hodoscopes

### 2.6.4 Downstream muon proportional wire chambers

The main components of the muon system were the four downstream PWCs, named: μ1, μ2, μ3, and μ4. These chambers detected the trajectories of the muons as they passed throughout the muon system. Their arrangement is shown in Figure 2.3. The physical dimensions of PWC μ1 are shown in Figure 2.5 (a), and the specifications of all the downstream muon PWCs are listed in Table VIII. Each chamber contained two cathode and one anode signal planes (see Figure 2.5(b)). The cathode planes were made of copper clad G-10 epoxy-fiberglass sheets of 1.6 mm thickness, etched to give a pattern of parallel strips 9.5 mm wide and separated 1 mm. The strips were oriented in the u-

($+45^0$ from the vertical) and v- ($-45^0$ from the vertical) directions (see Figure 2.5(c)). The anode planes were oriented in the y-direction (vertical) and made of 25 μm diameter gold-plated tungsten wires, separated by 6 mm, except for μ2 which had 4 mm pitch. The anode wires of μ1, μ3, and μ4 were connected in groups of three adjacent wires to each amplifier card, and in pairs for μ2, so the effective pitch of each PWC was 1.2 cm. The detail explanation of the muon PWC readout, and in general the data acquisition system (DA), is explained in the next chapter.

Table VIII. Downstream muon PWC specifications.

| | | |
|---|---|---|
| Inner radius | μ1 | 15 cm |
| | μ2 | 15 cm |
| | μ3 | 18 cm |
| | μ4 | 20 cm |
| Outer radius | all | 1.35 m |
| Effective pitch | u,v | 10.5 mm |
| | y | 12.0 mm |
| Anode-Cathode separation | all | 9.5 mm |

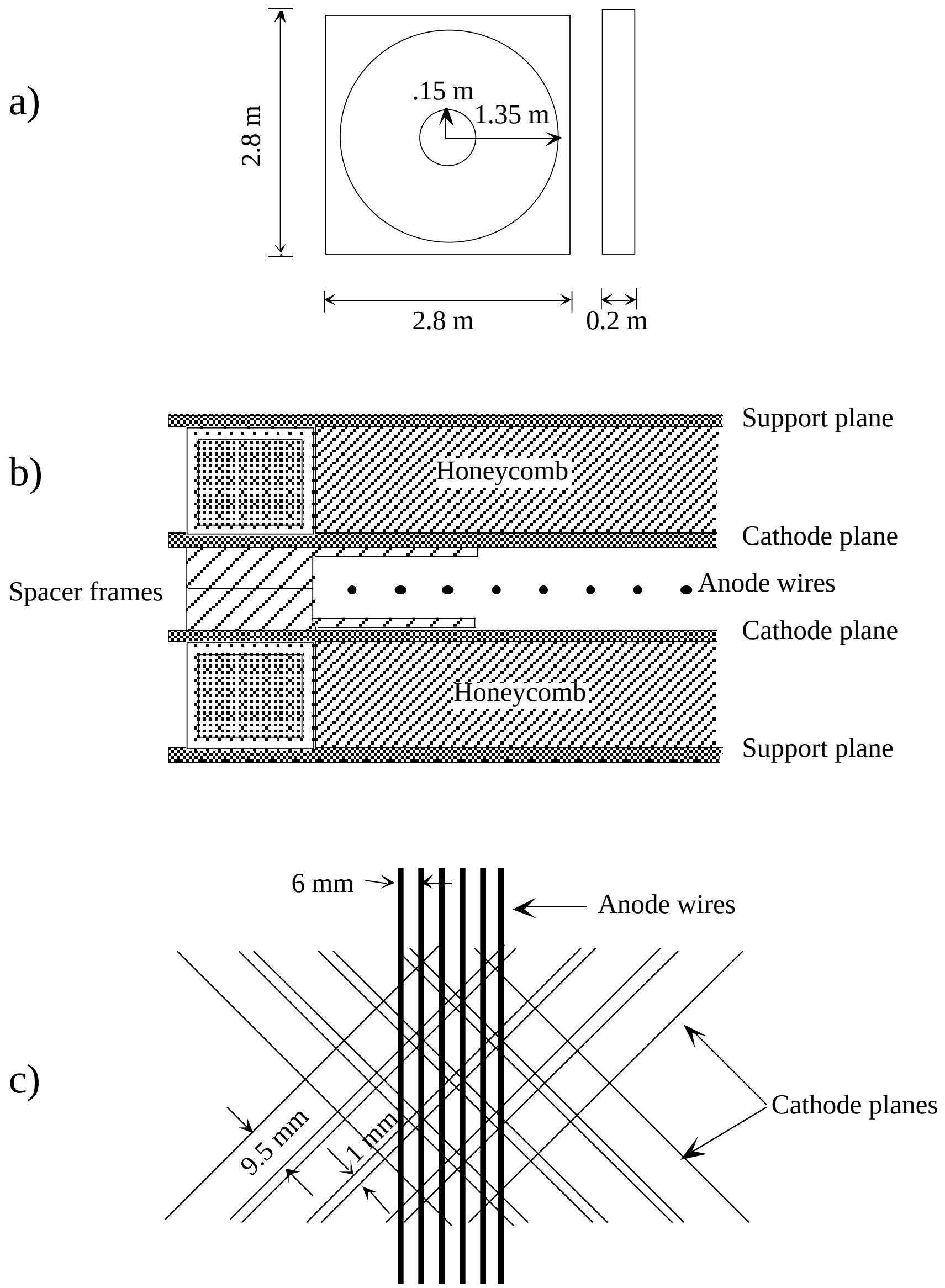

Figure 2.5. Downstream muon proportional wire chamber: (a) physical dimensions; (b) cross section of the chamber; (c) cutaway view of the chamber.

# CHAPTER 3

# DATA ACQUISITION SYSTEM

The data acquisition system of the MWEST spectrometer recorded the events of interest for both experiments, E672 and E706. E672 triggered on events that contained at least one high-mass dimuon, and E706 had several LAC based triggers [38]. This chapter describes the data acquisition system of the whole detector, however, only the muons system readout is described in detail.

## 3.1 Overview of the data acquisition system

The data acquisition system (DA) was divided into four parts, three CAMAC based PDP-11 systems and one FASTBUS system. A schematic diagram of the DA is shown in Figure 3.1. The muon system was readout using a PDP-11/34, named MU. It communicated with the muon system CAMAC interface through a Jorway 411, which read the CAMAC crates on a serial CAMAC highway. The upstream PWCs, SSDs, and CERENKOV counters were readout through a PDP-11/34, named NEU, and the forward calorimeter (FCAL) through a PDP-11/34, named ROCH. The FASTBUS system was used to readout the LAC and straw drift-tubes planes (STRAWS). The PDP-11s and FASTBUS were connected to a device developed by Fermilab, called a Bison Box [39], which controlled the beginning and ending of the spill interrupts, as well as the event interrupts (triggers). The PDP-11s were also connected to a DEC Micro VAX II computer, using communication device (CD) links [40]. The Micro VAX was running the VAXONLINE software system [41], which concatenated the readout data of the PDP-11s. This data, along with the FASTBUS data, constituted a complete event. THE VAXOLINE also controlled the beginning and ending of the data taking process, kept a dynamical event pool on the Micro VAX, for on-line data monitoring, and wrote the collected data to 8 mm exabyte tapes

## 3.2 Muon system readout

The electronic readout of the muon system consisted of several CAMAC crates modules which included a LeCroy PCOS III DataBus Interface (LeCroy 4299). This collected data from the muon chambers, ADCs, TDCs, scalars, the dimuon trigger processor (DTP), and sent it to the PDP-11 MU (see Figure 3.1).

### 3.2.1 Muon pretrigger hodoscope readout

The output signals from the 16 phototubes of each hodoscope were sent through a RG58-C coaxial cable to two 16-channel CAMAC multiplicity logic units [42]. The two modules discriminated the received signals and set a latch for each channel that exceeded a certain preset voltage threshold during an external set gate. The multiplicity of the tracks in the muon hodoscopes was then calculated from the multiplicity of the latch bits. The CAMAC module could be preset to a certain multiplicity threshold, so if the calculated multiplicity was grater than or equal to the set threshold, a NIM level was output. During normal data taking the multiplicity threshold was set to two for both muon hodoscopes.

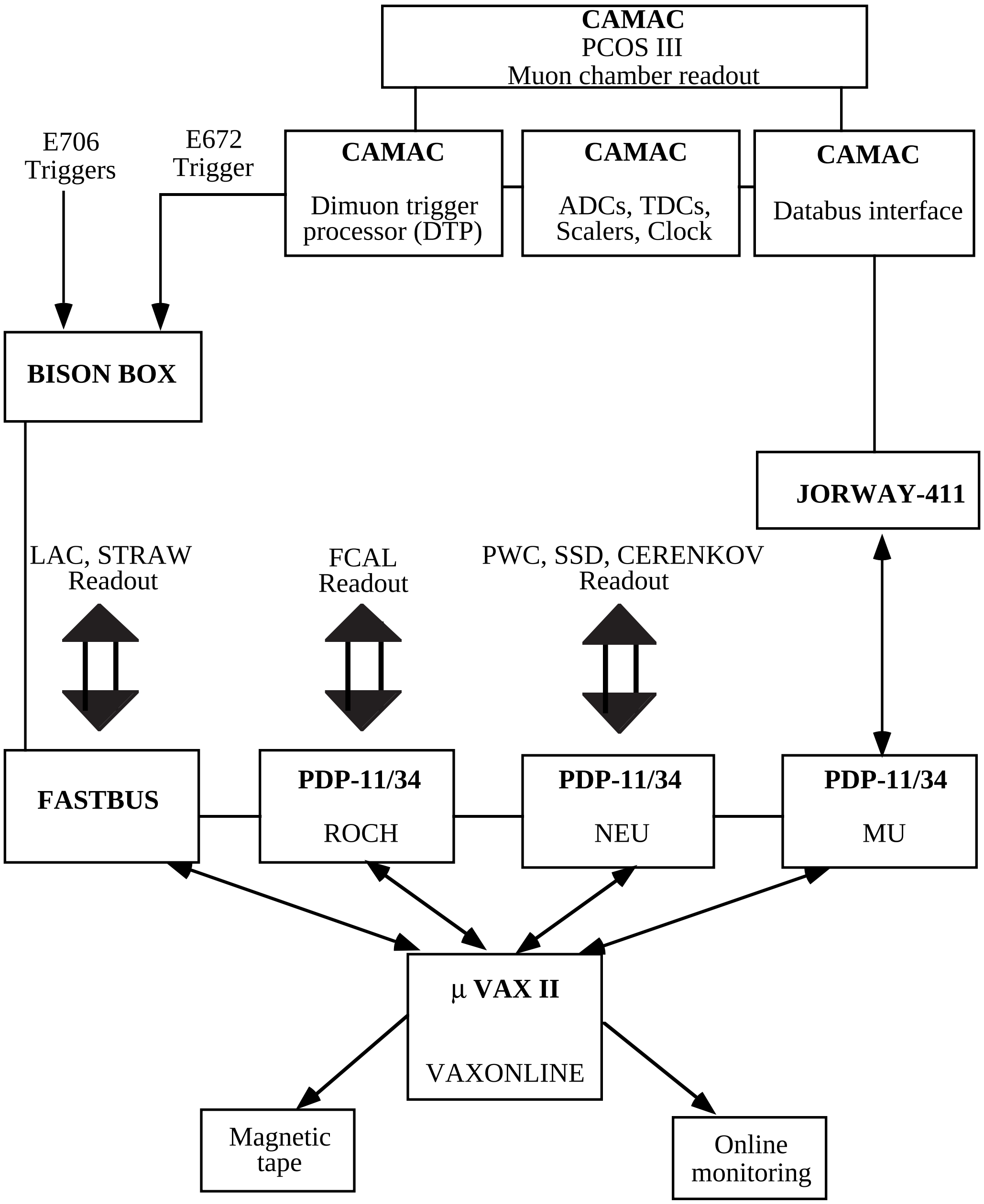

Figure 3.1. Diagram of the MWEST detector data acquisition system.

### 3.2.2 Upstream muon proportional wire chamber readout

All the muon chambers were readout using a LeCroy PCOS III system (Proportional Chamber Operating System) [43]. The readout signals were sent to LeCroy 2731 latch modules, which resided in CAMAC crates. Each crate could hold 23 latch modules, and one controller (LeCroy 2738). The controller module, which queried each latch module, stored the addresses of the active channels in an internal buffer, then sent them to the data acquisition system. The controller also had an ECL port which transferred data ( at a rate of 10 times that of CAMAC) to the trigger logic circuit.

### 3.2.3 Dimuon trigger processor

The dimuon trigger processor (DTP) was used to do an on-line calculation of the dimuon invariant-mass. It consisted of 7 CAMAC double-width modules. These received the data output of the muon PWCs PCOS III. The data was sent by the LeCroy 2738 crate controller through the fast ECL port. A flow chart diagram of the DTP is shown in Figure 3.2. The wire-hit data information from the chambers, $\mu 1$ and $\mu 4$, was sent to two modules named POINT. These modules computed in parallel (using the wire hits) the space points of the trajectories of the muon tracks. The location of these spatial points was checked against a look-up table stored in Programmable Read Only Memory (PROM) to assure that the points were in the feducial volume of the muon spectrometer. The list of valid points was stored in a Random Access Memory (RAM) and sent to two other modules named TRACK.

The TRACK modules also received the wire hit information from the $\mu 2$ and $\mu 3$ muon PWCs. Using this additional information and the valid space points, they calculated in parallel the possible associated tracks in the x (x-z plane) and y (y-z plane) views. This list of tracks was sent to two other modules named MOMENTUM and TARGET. The TARGET module, which also received the hit information from the upstream muon chambers $\mu$A, and $\mu$B, checked to see if the tracks were consistent with the hits found in the $\mu$A and $\mu$B. This was done in the following way : An imaginary straight line was constructed from the center of the target to the point in space were the track in question met the center of the toroid; the imaginary line was compared with the wire-hits of $\mu$A and $\mu$B, and if the hits on at least 3 out of 4 planes in each chamber were consistent with the imaginary line, the associated track was said to be a valid track. Again, in parallel with the TRACK module, the MOMENTUM module calculated the momenta of the tracks from their bend in the toroid assuming they originated from the target. The last step of the DTP was to compute the invariant-mass of every pair-combination of valid tracks. This was done (assuming the tracks were from muons) by a module named MASS, which received the list of valid tracks and their associated momenta . If at least one pair-combination of valid tracks gave an invariant-mass greater than a preset threshold, a success signal was output. In the 1990 run, the mass threshold was usually set between 0.5 GeV/$c^2$ and 1.0 GeV/$c^2$.

## 3.3 Dimuon trigger

The E672 dimuon trigger was activated after the detection of both a beam particle and an interaction, and if the DA was not busy. It was done in two stages: Level 1, or pretrigger; and level 2.

The pretrigger required that the hits in the muon hodoscopes be consistent with a track multiplicity in the muon detector of least two tracks. If this was true and no particles were detected by the vetowalls, level 2 was started. Muons produced in the target required approximately 15 GeV to penetrate the spectrometer material and reach the muon hodoscopes, all particles reaching them were assumed to be muons. The average hit multiplicity was 2.3. The pretigger rate was $2 \times 10^{-4}$ per interaction. The pretigger efficiency for two muons penetrating the system was 0.76 and remained constant over the data taking period.

Level 2 was based on the requirement that at least one of the dimuons had invariant-mass above a certain mass threshold. The invariant-mass calculation was done by the DTP based on the information in the muon chambers. If at least one muon pairs satisfied the DTP conditions the event was recorded to tape. The threshold of 0.7 GeV/$c^2$ resulted in a trigger rate of $2 \times 10^{-5}$ per interaction.. The average DTP processing time was 10 $\mu$s per pretigger, which included 5 $\mu$s to decode the muon chamber data. The combined efficiency of the chambers and the DTP algorithm was 0.77 for dimuon events.

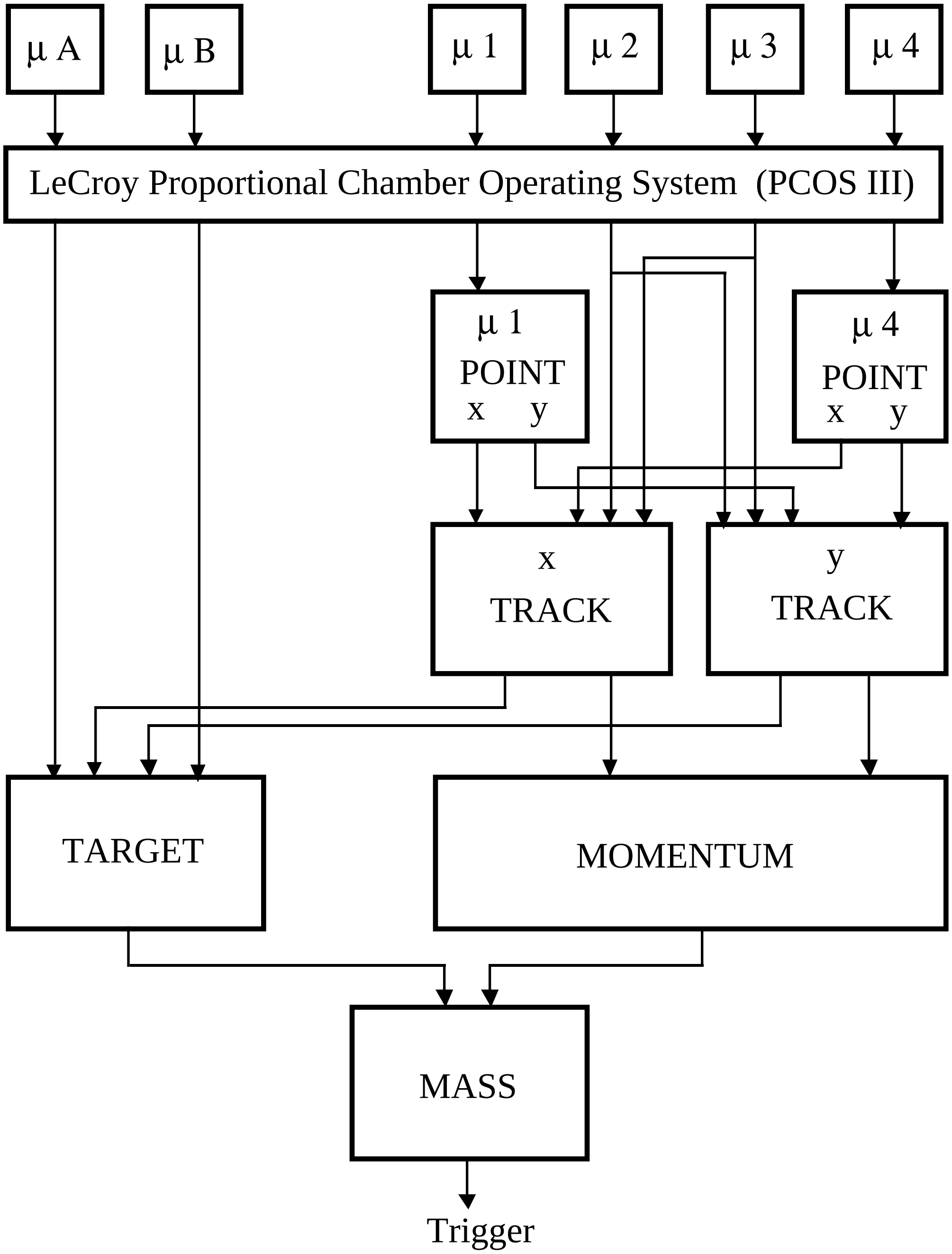

Figure 3.2. Schematic diagram of the dimuon trigger processor.

# CHAPTER 4

# EVENT SELECTION AND RECONSTRUCTION

The raw data were written onto 8 mm exabyte tapes. These contain the complete information from every piece of apparatus in the entire spectrometer for all the different events that where triggered (dimuon and LAC triggers). This chapter describes the data reduction from the raw data to the data sample used in the analysis of this thesis.

## 4.1 Event selection

The first step towards reducing the data was to extract the dimuon triggers from the raw data tapes. This step reduced the amount of the data to about 18 % of the complete data sample, yielding approximately 5 million dimuon triggers (corresponding to a luminosity of 8 $pb^{-1}$). To further reduce the data, only information from the upstream tracking system and the muon system were extracted. The information from the LAC and forward calorimeter was excluded. By excluding this portion of the data, the event size was reduced by a factor of 8. The set of events that were extracted is known as the "MNS" sample.

### 4.1.1 Dimuon preselection

Because of inefficiencies of the veto walls, some halo muons activated the dimuon trigger. Also, muons produced by beam interactions with the forward calorimeter could set off the dimuon trigger. This happened often since the forward calorimeter only had a 3.2 cm diameter hole in the center. If it was not properly aligned, the beam could possibly interact with the calorimeter and produce muons. To avoid events that might have any of the two problems just mentioned, the MNS dimuon sample was run through a preselection program. The program required that at least two muons with a momenta greater than 20 GeV/c each be present in the event. Both muons had to be consistent with originating from the target region. Also, the track segments from the muon system had to be linked with an upstream PWC track segment. These requirements reduced the MNS sample by 86 %. The set of events that survived the requirements of the dimuon preselection program is known as the "DIM" sample. This sample consists of about 750 thousand events. To reduce the data even more, the events were required to have at least one opposite sign dimuon pair with an invariant-mass greater than 2.0 GeV/$c^2$. Approximately 35,000 events survived these requirements. This set of events is called the "PSI" sample. It is the actual data set used in the analysis of this thesis.

## 4.2 Event reconstruction

The event reconstruction program was written by Prof. Jack Martin [44]. The program reconstructed the "tracks" (trajectories) of the charged particles detected by the spectrometer. It first found the track segments detected by the SSD, upstream PWC, and muon PWC systems. Using the track segments, it linked the tracks from the various systems: The SSDs with the upstream PWCs, these are labeled SSD-PWC linked tracks; and the upstream PWCs with the muon PWCs, these are called PWC-MUON linked tracks. It also linked the track segments throughout the complete detector, these type of tracks are called SSD-PWC-MUON linked tracks or fully linked tracks. Finally, with the reconstructed tracks it searched for the location of the vertex in the event. All these steps are described in detail below.

### 4.2.1 Track finding

The track finding algorithm for the muon PWC system began by searching each of the view planes: u, v, and y, for groups of adjacent wire hits. These groups of hits were called hit clusters. The center of each hit cluster (with its corresponding uncertainty of half the width of the cluster) was used as a view hit. Having the list of view hits in each view, a search for single-view tracks was performed. They were found by doing a search of all the possible roads containing only one hit cluster per plane in at least 3 of the 4 planes. A road was defined as a quadrangular portion of the active area of a given plane view of the muon PWC system. It extended from the first PWC ($\mu$1) to the last PWC ($\mu$4). Figure 4.1 shows the first road used in the algorithm, and the subroads into which it is divided. Roads that contained more than the allowed number of hit clusters were further subdivided using a scheme as in Figure 4.1. The algorithm was repeated until the roads satisfied the conditions mentioned above. The roads that did not meet the criteria were disregarded.

As a simple example of the track finding algorithm, consider the hit cluster pattern found in the muon system as shown in Figure 4.2. After the first iteration, it is easy to see that only subroad 3 has no more than one hit per plane in at least 3 out of the 4 planes. In this case, it has hit clusters on all the planes. This road is not subdivided further since it now meets the track finding criteria. All the other roads are disregarded. Of course the hit cluster pattern found in a real event is far more complex than this simple example.

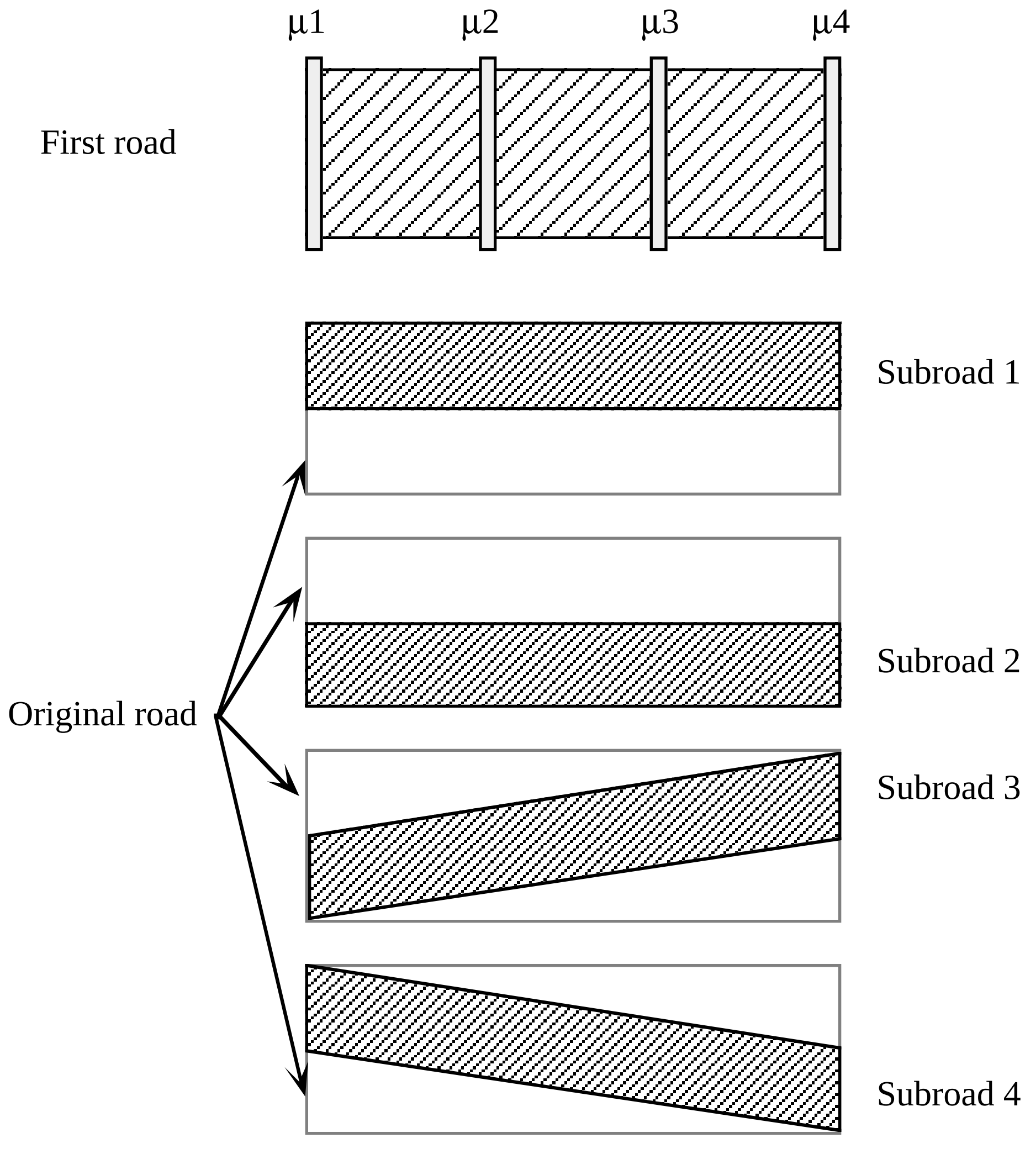

Figure 4.1. Illustration showing the first roads used in the track algorithm and the scheme to subdivide a road into other roads.

Once the final set of roads is found, a straight-line fit is performed to the hit clusters within each road. This fit determines the slope and intercept of the tracks in each view. After this, a three-dimensional space-track finding is performed in a similar fashion as in the single-view track finding. Space-roads are constructed and checked to see if any single-view tracks are consistent with belonging to the space-road. The space-roads that contain more than one view-track are further subdivided. This is done until a set of space-roads which only contain one single-view track per view is found. The hit clusters belonging to each space-road are then fit to a straight line. The fit gives the parameters of the space-track as the slopes and intercepts of the track in the x-z and y-z planes (x and y views, respectively).

The track finding for the upstream PWC and SSD systems are essentially the same as for the muon system. The only difference is that when fitting the hit clusters to find the final parameters of the tracks, single wire hits are used instead of the hit clusters (one wire hit from each cluster). The parameters of the single wire hit track combination, that give the smallest $\chi^2$ from the straight-line fit are kept as the parameters of the track. This is done because the high multiplicity of the wire hits gives rise to spuriously large hit cluster widths. No space-track finding is done for the SSD system, since it only had two plane views.

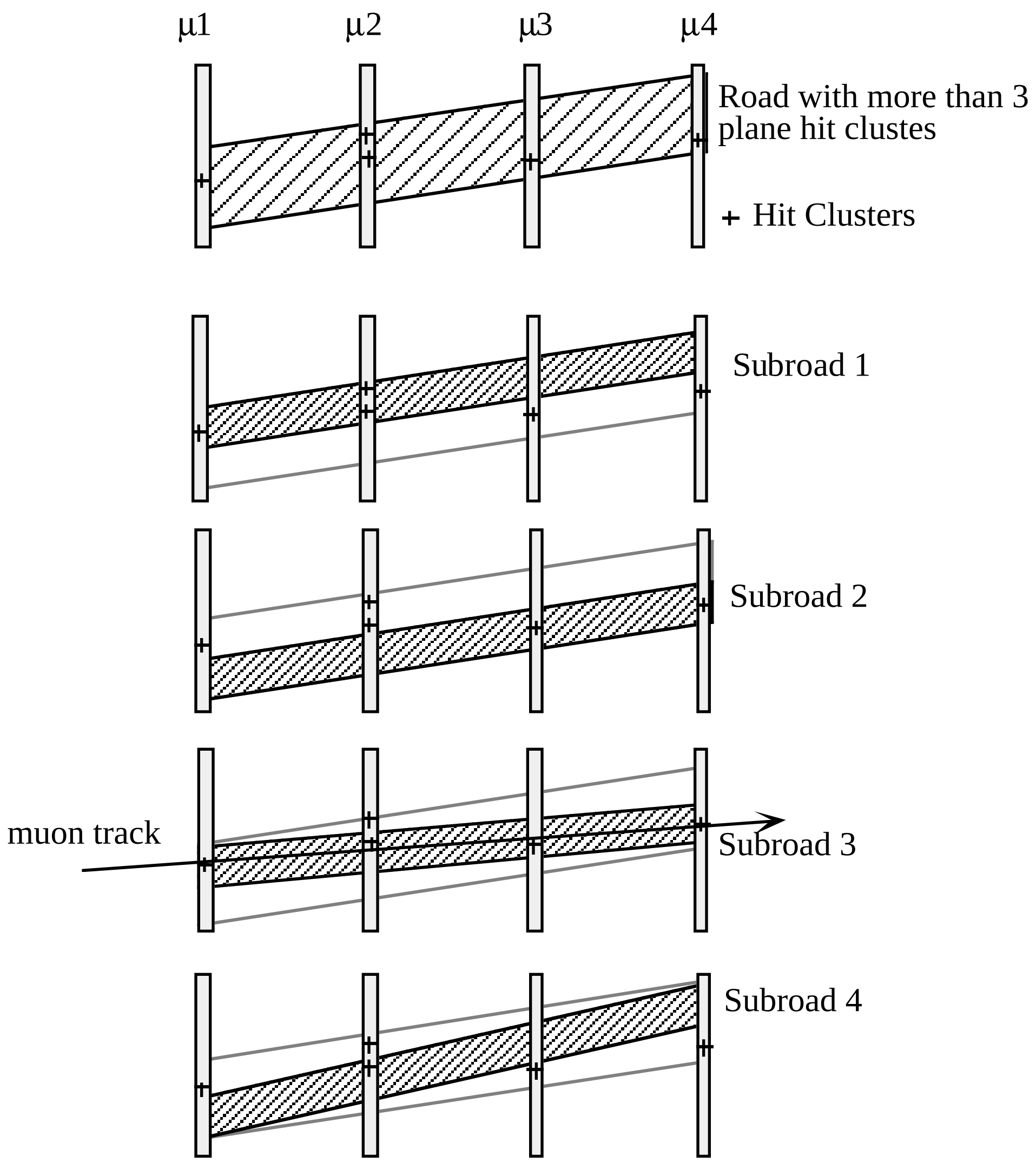

Figure 4.2. Illustration of a simple example of the track finding algorithm, the arrow shows the muon track.

### 4.2.2 Track linking

After reconstructing the track segments in each of the tracking devices, these are linked from one device to another. The track linking fit determines the momenta of the tracks by using the bend in the dipole magnet, or in the toroid, for the case of muon tracks.

The SSD-PWC track linking is performed in the following way: Track segments from the SSD and upstream PWC are extrapolated to the center of the magnet in each view. The segments from the y-view must meet one another within the error of track segments. The segments from the x-view must meet each other within 5 mm. The linking in this case is done in each view, separately, because the SSD system only has single-view tracks and no space-tracks. The reason for the linking requirements to be different in each view is because the dipole magnet only deflected the tracks in the x-view. Thus, the y-view tracks should have approximately the same parameters as seen in the SSD and upstream PWC systems. Figure 4.3 illustrates the SSD-PWC linking. Track segments that meet the linking criteria are fit using the slopes and intercepts of each track segment: a, b, e, f; and c, d, g, h (see Figure 4.3). The y-view fit is done with the constraint that the slope and intercept of the track are the same in both the SSD and upstream PWC systems. In the x-view, the fit outputs the momentum of the linked track (as1/P), the slope, and the intercept of the track as seen by the SSD system in the x-view. Tracks that pass the $\chi^2$ quality cut of the fits are said to be SSD-PWC linked tracks. PWC space-tracks can have up to three SSD-x and three SSD-y link combinations if they pass the $\chi^2$ cut. These ambiguities are resolved at a later stage.

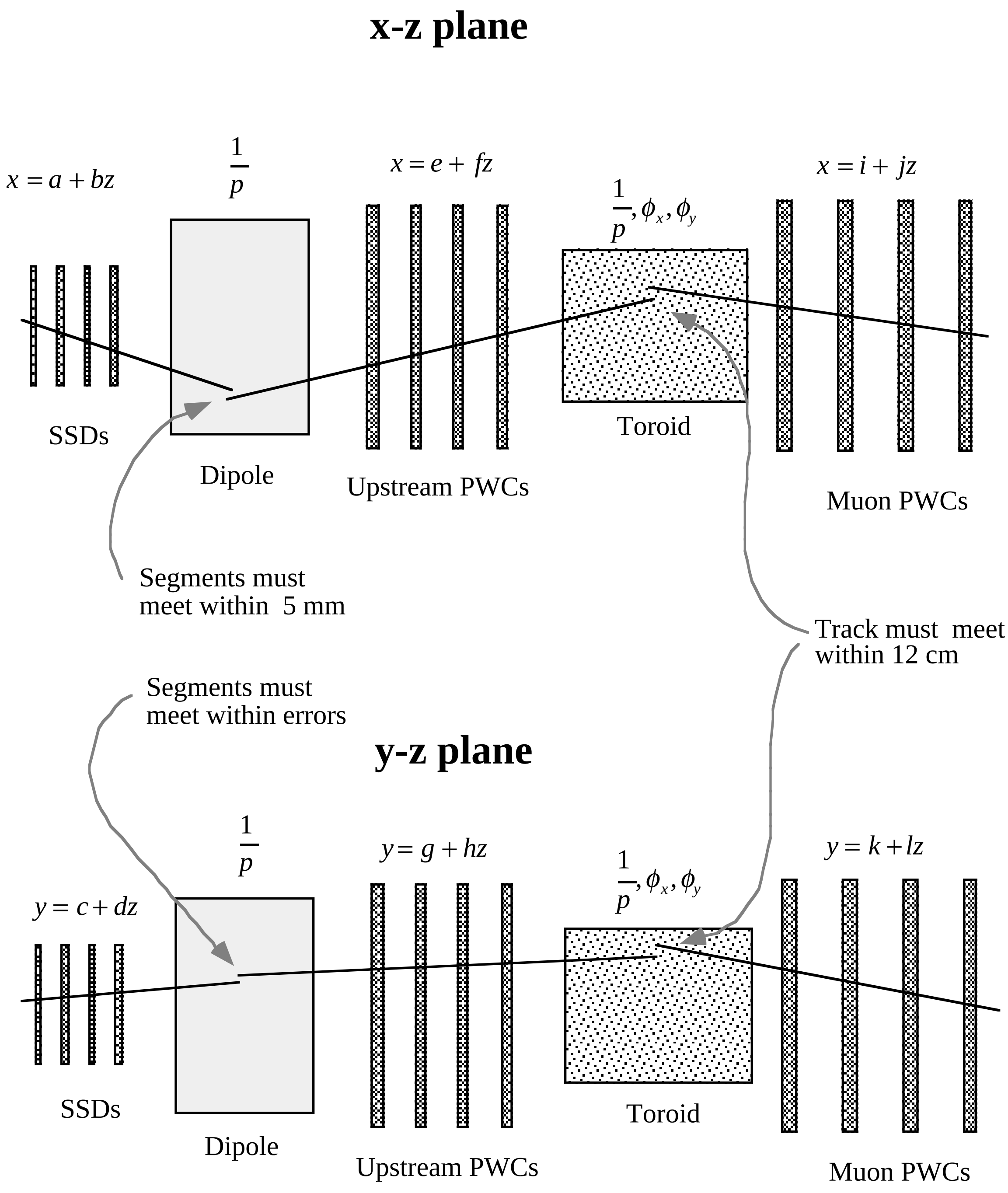

Figure 4.3. SSD-PWC linking scheme.

The PWC-MUON track linking is done in the following way: The track segments from the upstream PWC and muon PWC systems are extrapolated to the center of the toroid. Track segments that meet one another with in 12 cm in both the x and y views are considered as linked track candidates. Figure 4.3 shows the PWC-MUON linking scheme. The linked track candidates are fitted using a standard $\chi^2$ minimization algorithm, using the parameters of the track segments: e, f, g, h, i, j, k, and l. Also, the momentum (in the form 1/P) and two multiple scattering angles ($\phi_x$ and $\phi_y$) are included in the fit. Candidates that pass a $\chi^2$ quality cut are said to be PWC-MUON linked tracks. Again, in the case of ambiguities, up to 3 PWC-MUON linked combinations are kept for each muon track. Matching of these tracks to the SSD-PWC linked tracks and to the vertex position are used to resolve the ambiguities at a later stage. A global fit is done to the muon tracks. These are linked throughout the entire detector. To do this, the SSD-PWC linked tracks are extrapolated to the center of the toroid and matched with the muon tracks in the same manner as the upstream PWC segments, described above. Here, the fit is performed using all the parameters of each segment composing each track: a, b, c, d, e, f, g, h, i, j, k, and l. Also, the momentum of the track in the dipole ( written as 1/P), the momentum in the toroid ( written as 1/P), and two multiple-scattering-angles $\phi_x$ and $\phi_y$ are included (see Figure 4.3). Once again, a $\chi^2$ minimization algorithm is used. Track combinations that pass a $\chi^2$ quality cut are called SSD-PWC-MUON linked tracks, or fully linked muon tracks.

To improve the resolution of the SSD-PWC linked tracks, the SSD segments associated with a vertex were refitted using the vertex position as an additional pseudo-plane hit. A $\chi^2$ fit like the one for the PWC-MUON linked tracks, is now performed on the SSD-PWC tracks, yielding the momenta and the track parameters in each view at the production vertex.

The fractional uncertainty in the momentum measurement of a fully linked (SSD-PWC-MUON linked) muon track is:

$$\frac{\delta p}{p} = 0.0005\,p;$$

as seen in Figure 4.4. The over all efficiency to reconstruct a fully linked muon track is 86%.

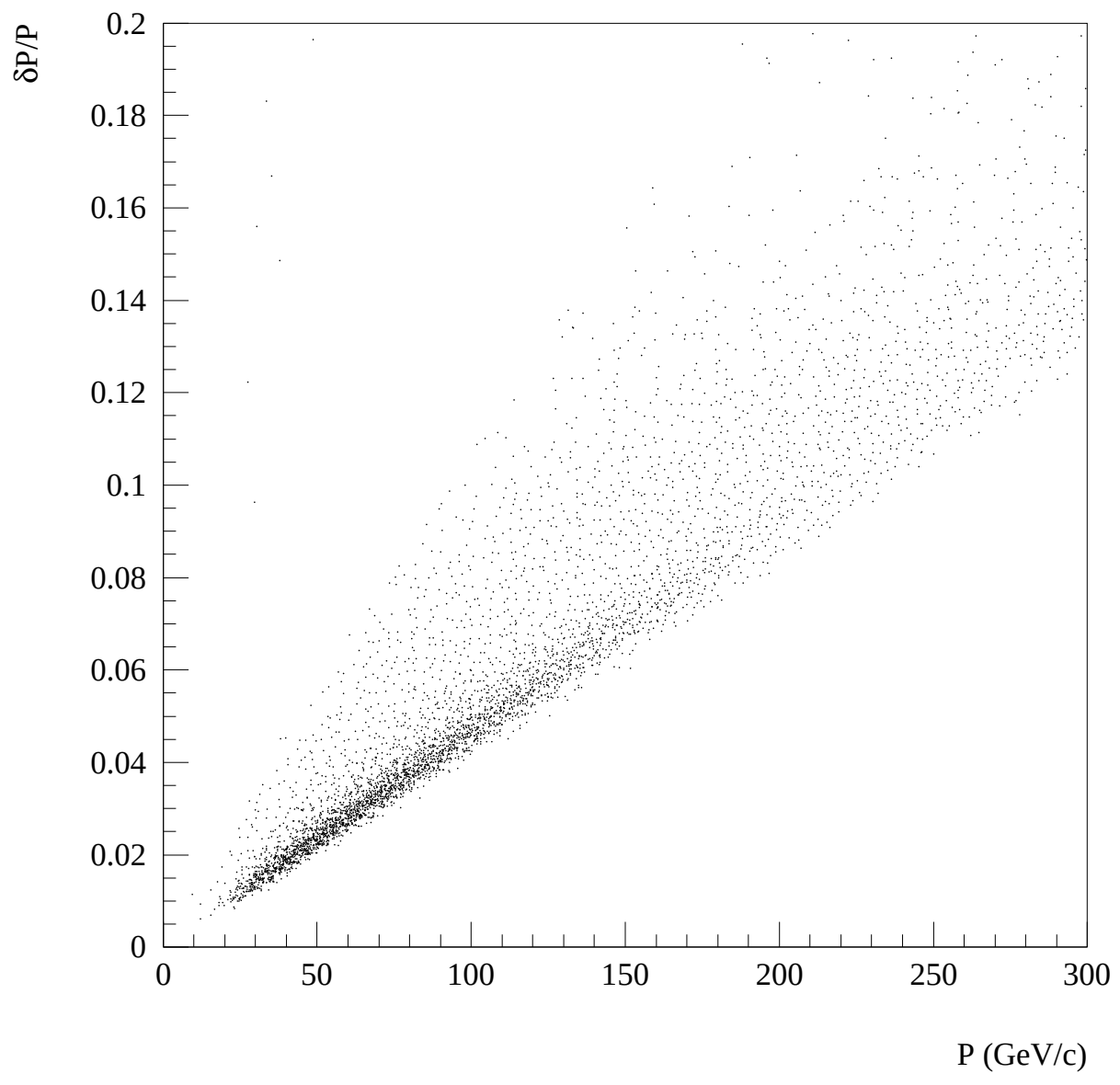

Figure 4.4. The fractional uncertainty of momentum vs. momentum for fully linked muon tracks.

### 4.2.3 Vertex finding

The vertex finding algorithm begins by looking at the SSD x and y views for intersections of two track segments. In each view, intersections that occur approximately at the same position (within errors) are combined. A list of possible multi-segment intersections is made for each view. The two lists are compared, and only intersections that have consistent z positions (within errors) in both views are kept. This results in a three dimensional position (x, y, z) of the vertices and a list of the tracks associated with each one.

A fit minimizing the impact-parameter of each track connected to its vertex is performed to improve the resolution of the measurement of the vertex position. Figure 4.5 shows the distribution of the primary vertices found in the events of the dimuon PSI sample (descried in section 4.1.1) for those dimuon events that are tagged as J/ψs[1]. This distribution clearly shows the target configuration as well as the SSD planes (see sections 2.4 and 2.5.1 for the target and SSD plane configurations). The comparison with the actual physical size of the target indicates a resolution of approximately 3 mm in the longitudinal direction. Figure 4.6 shows the transverse impact-parameter of the SSD-PWC linked tracks to the primary vertex for the events in the PSI sample. This distribution demonstrates a resolution of about 10 μm for the measurement of the vertex position in the transverse plane (x-y plane).

---

[1] See chapter 5 for the definition of J/ψ events.

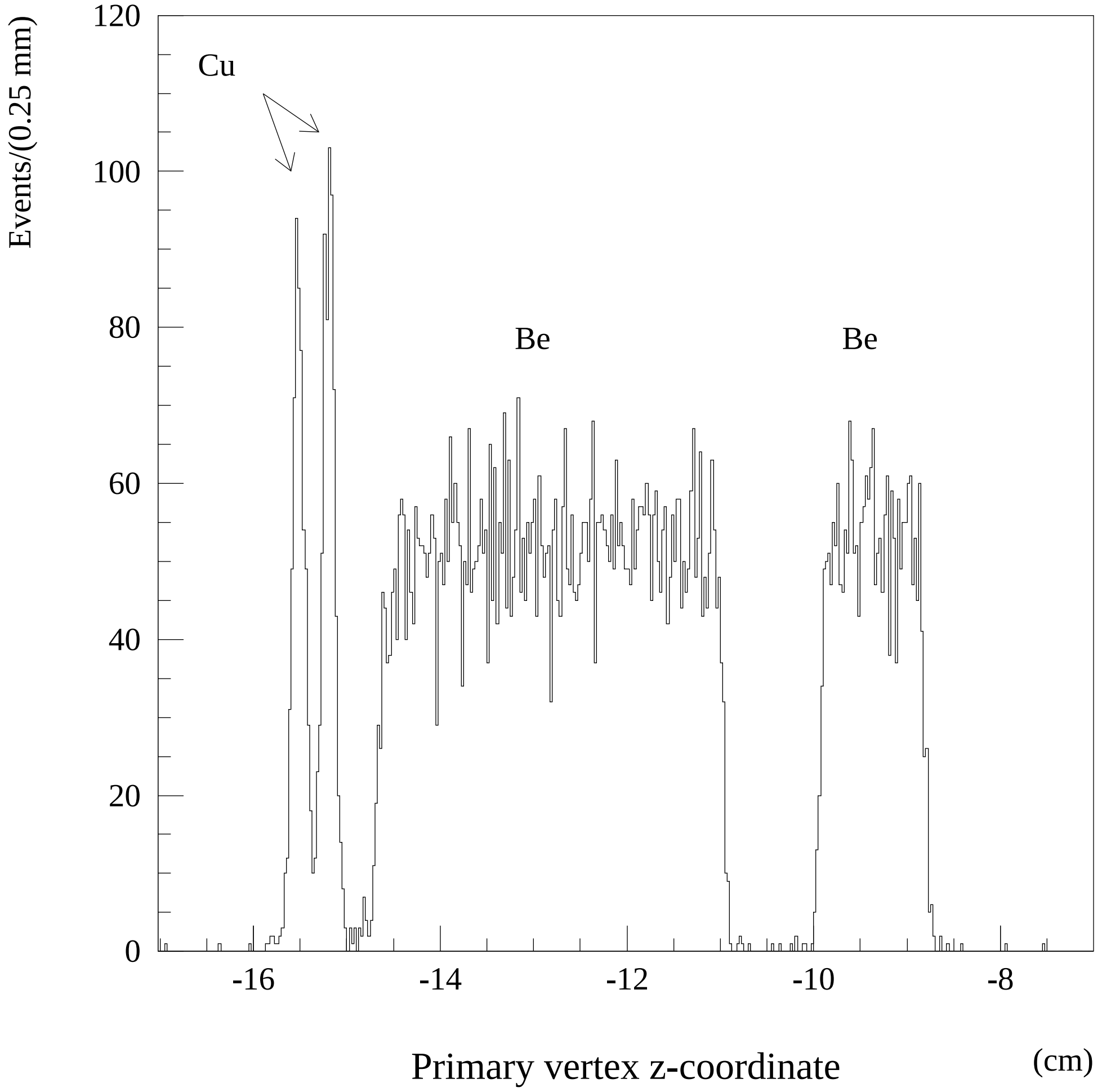

Figure 4.5. Distribution of the primary vertices found in the “PSI” sample for those dimuon events that are tagged as J/ψ s.

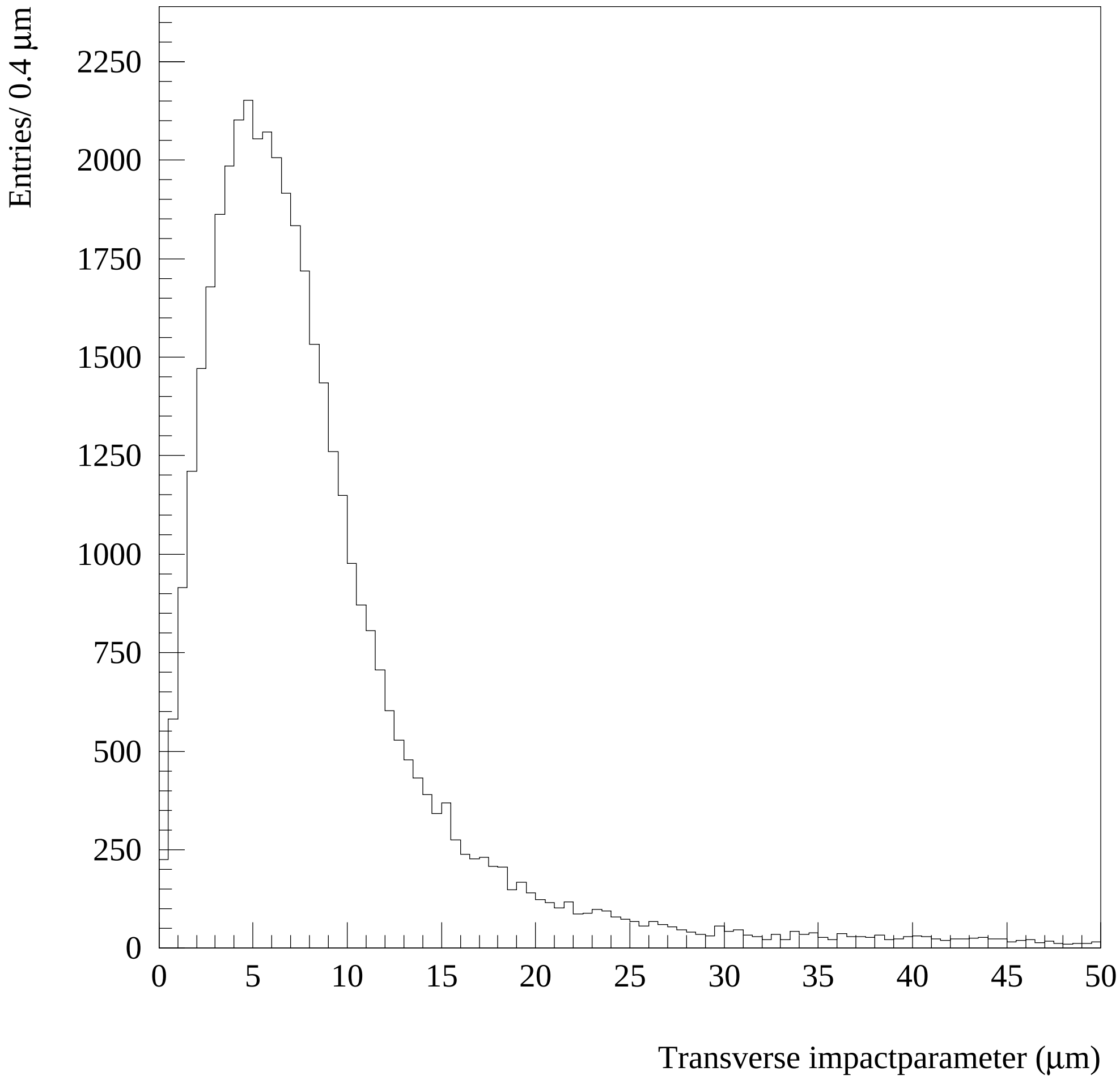

Figure 4.6. Transverse impact-parameter of the SSD-PWC linked tracks with the primary vertex, for events in the “PSI” sample.

## 4.3 Characteristics of reconstructed dimuon events

E672/E706 has studied the hadroproduction of charmoniuom by measuring the $\psi(2S)$, $\chi_1$, $\chi_2$, and directly produced J/$\psi$ fractions contributing to the inclusive J/$\psi$ production (see Ref. [45] and [46]), and compared these values with those predicted by the theoretical charmoniuom models: color evaporation [47] and color singlet [48]. We have also, as mentioned previously, measured the $b\bar{b}$ cross section by using J/$\psi$ s from secondary vertices [11]. This measurement will be used later to extract the production rate for $\Lambda_b$ s, $F(\Lambda_b)*Br(\Lambda_b \rightarrow J/\psi\ \Lambda^0)$. Another E672/E706 study, in which the author has been directly involved is the measurement of the atomic number (A) dependence of the production cross sections for vector mesons $\rho/\omega$, $\phi$, and J/$\psi$ [49]. The results of this study are not included in detail here in order to make this thesis coherent. The following is a brief summary of the results of these three studies.

The invariant mass spectrum of our 1990 opposite sign muon pairs is shown in Figure 4.7. This is our so called "DIM" sample (see 4.1.1 for the for its definition). This sample contains approximately 36 K $\rho/\omega$, 10 K $\phi$, and 12 K J/$\psi$. A closer look at the J/$\psi$ region (see Figure 4.8) shows a $\psi(2S)$ signal which contains 270 ± 35 (*stat*) ± 50 (*sys*), $\psi(2S) \rightarrow \mu^+ \mu^-$ events. The $\psi(2S)$ is observed in both the $\mu^+ \mu^-$ and $J/\psi\ \pi^+ \pi^-$ modes. Figure 4.9 shows the $J/\psi\ \pi^+ \pi^-$ invariant-mass spectrum. A fit to the signal yields 224 ± 44 (*stat*) ± 20 (*sys*), $\psi(2S) \rightarrow J/\psi\ \pi^+ \pi^-$ events. Differential cross sections for J/$\psi$ and $\psi(2S)$ can be found in [45]. The total integrated cross sections were measured to be $Br(J/\psi \rightarrow \mu^+ \mu^-)\ \sigma(\pi^- Be \rightarrow J/\psi + X)/A$ = (9.3 ± 1.1(*sys*)) nb/nucleon for $x_F > 0.1$, and $Br(\psi(2S) \rightarrow J/\psi\ \pi^+ \pi^-)\ \sigma(\pi^- Be \rightarrow \psi(2S) + X)/A$ = (7.4 ± 1.5 (*stat*) ± 1.2 (*sys*)) nb/nucleon for $x_F > 0.1$. The fraction of inclusive J/ψ yield due to ψ(2S) meson decays was found to be 0.083 ± 0.017 (*stat*) ± 0.013 (*sys*) [45].

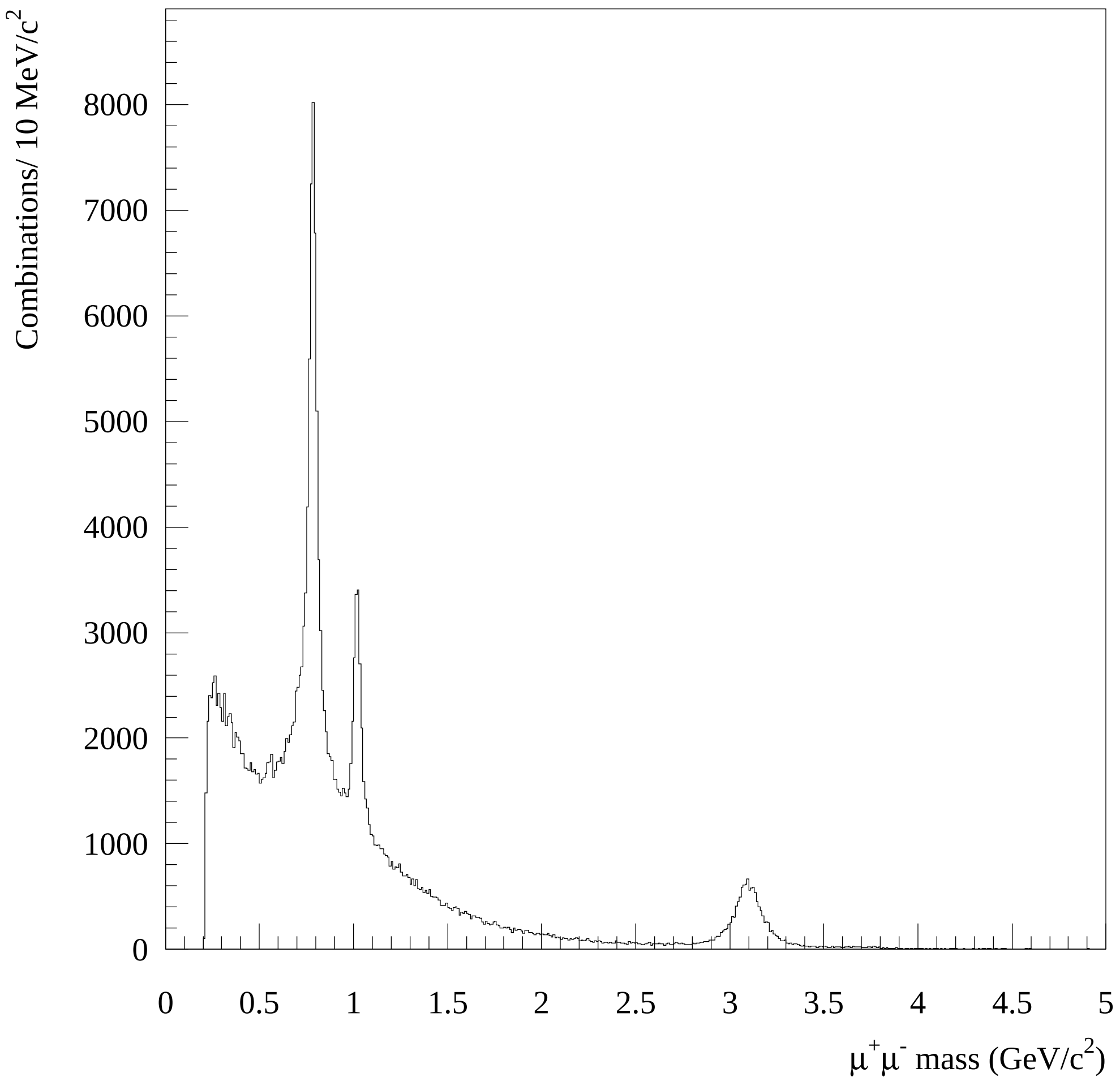

Figure 4.7. Invariant-mass spectrum of the opposite sign muon pairs.

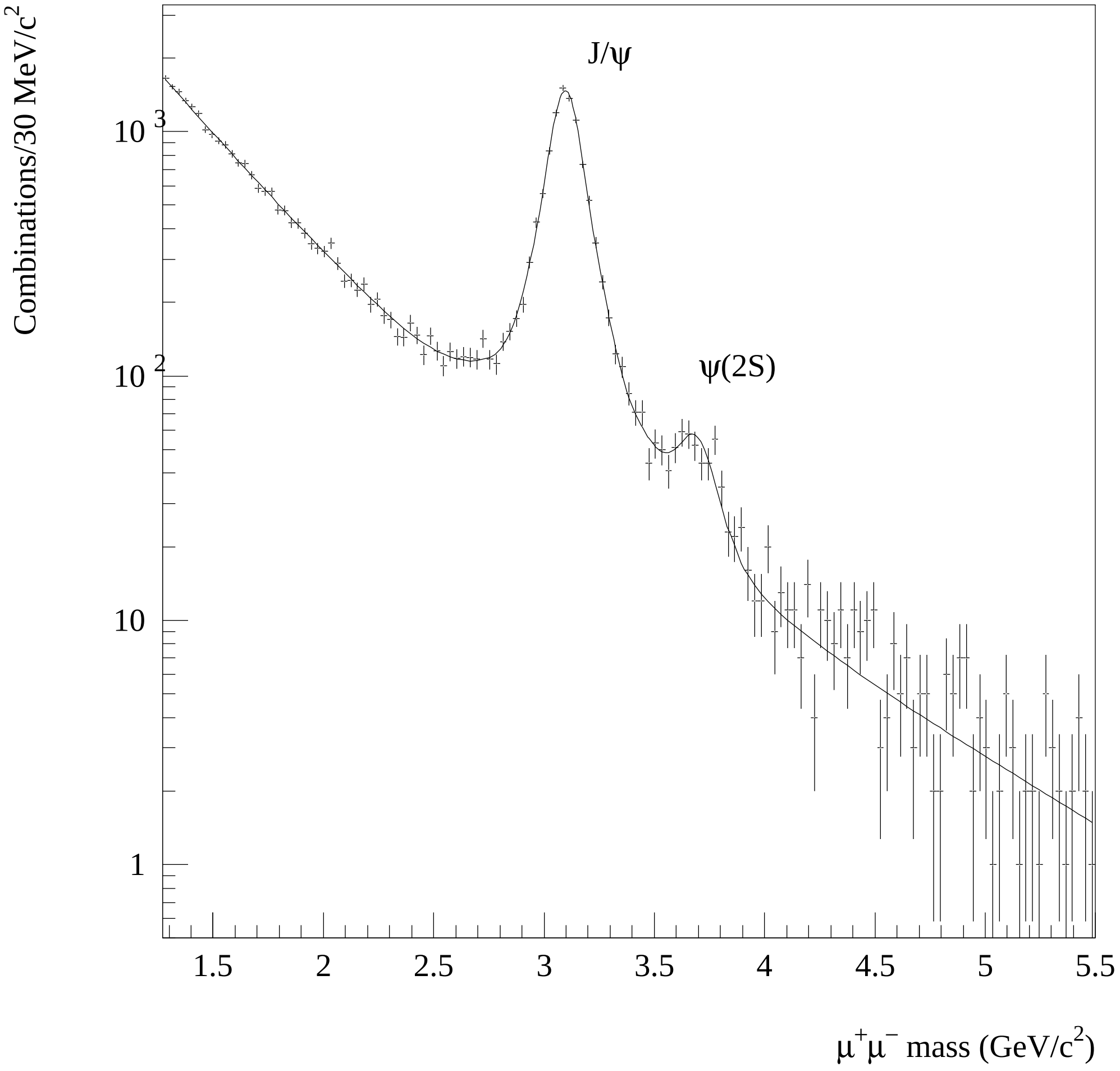

Figure 4.8. Invariant-mass spectrum of the opposite sign muon pairs in the J/ψ region. The solid line is a fit to the data.

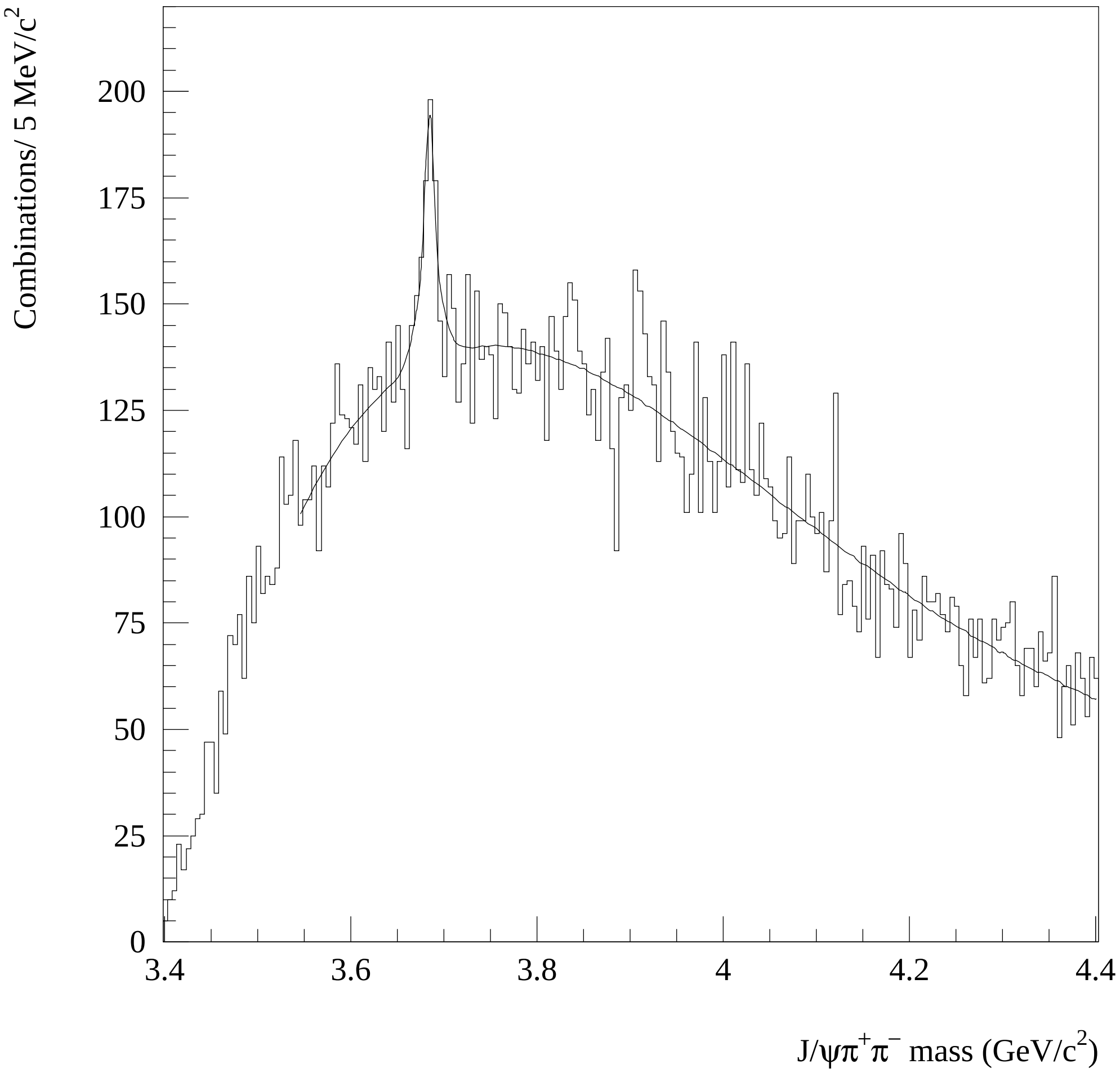

Figure 4.9. Invariant-mass spectrum of $J/\psi\ \pi^+\ \pi^-$ combinations. The solid line is a fit to the data.

$\chi_1$ and $\chi_2$ states are reconstructed via their decay into J/ψ γ. The γ s are detected via their conversions into $e^+ e^-$ pairs. Figure 4.10 shows the J/ψ γ spectrum for converted γ s. Clear peaks are seen for both $\chi_1$ and $\chi_2$. A fit to the signals yields 57 ± 13 (*stat*) ± 16 (*sys*) and 40 ± 10 (*stat*) ± 14 (*sys*) $\chi_1$ and $\chi_2$, respectively. The integrated total cross sections associated with these values are Br( $\chi_1 \rightarrow J/\psi\, \gamma$) $\sigma(\pi^-\ Be \rightarrow \chi_1 + X)/A$ = (50.5 ± 9.0 (*stat*) ± 4.1 (*sys*)) nb/nucleon for $x_F > 0.1$ and Br( $\chi_2 \rightarrow J/\psi\, \gamma$) $\sigma(\pi^-\ Be \rightarrow \chi_2 + X)/A$ = ( 35.5 ± 7.8 (*stat*) ± 7.6 (*sys*)) nb/nucleon for $x_F > 0.1$. Thus the fraction of inclusive J/ψ yield due to $\chi_1$ and $\chi_2$meson decays are 0.24 ± 0.04 (*stat*) ± 0.03 (*sys*) and 0.017 ± 0.04 (*stat*) ± 0.02 (*sys*), respectively.

Combining the 0.083 ± 0.017 (*stat*) ± 0.013 (*sys*) of J/ψ s due to ψ(2S) and assuming 0.02 of the inclusive J/ψ are from $\chi_0$ and b-hadron decays, the fraction of the total inclusive J/ψ produced directly is 0.49 ± 0.05 (*stat*) ± 0.05 (*sys*) [46]. A comparison of our measurements with the theory are shown is Figure 4.11.

.

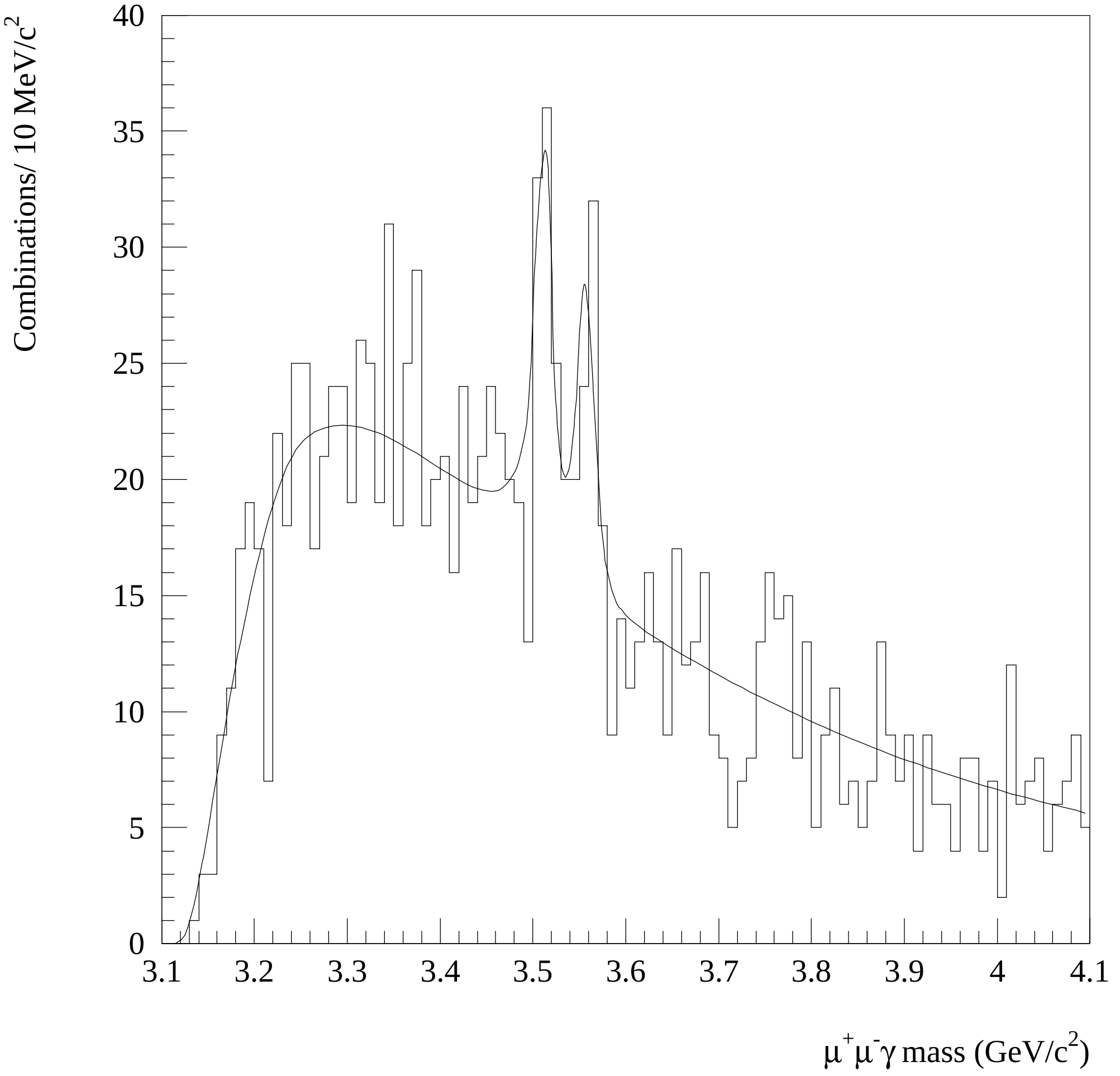

Figure 4.10. Invariant-mass spectrum of J/ψ γ combinations for converted γ s. The solid line is a fit to the data

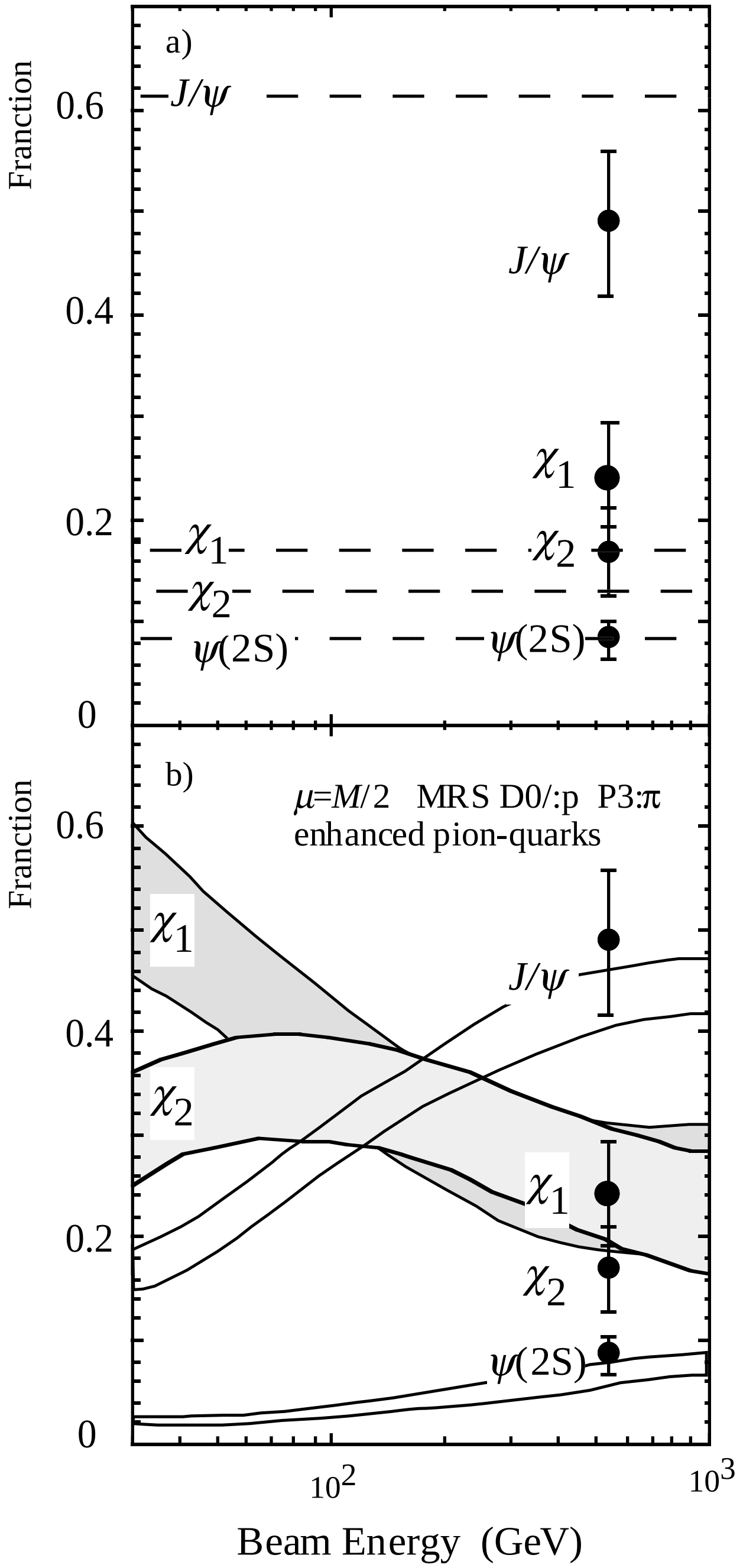

Figure 4.11. (a) A comparison of our measurements of the fraction of inclusive J/ψ yield due to ψ(2S), $\chi_1$, $\chi_2$, and directly produced J/ψ s, to the fractions predicted by the color evaporation model [47]; and (b) to the fractions predicted by the color singlet model [48].

E672 has also measured the $b\bar{b}$ cross section [11]. This analysis was performed by using J/ψ s emerging from secondary vertices in the regions of the target where only air was present. E672 found an excess of 8 events in the air-gap regions of the target, and attributed these to b → J/ψ + X decays. Which correspond to an inclusive cross section of 75 ± 31 (*stat*) ± 26 (*sys*) nb/nulceon for all $x_F$. Some exclusive B hadron events were also reconstructed. Figure 4.12 shows the combined invariant-mass spectrum for $B^{\pm} \rightarrow J/\psi\ K^{\pm}$ and $B^{0} \rightarrow J/\psi\ K^{0*}$ events. There are five events near the nominal mass of the B, 3 of them are from $J/\psi\ K^{\pm}$ and 2 from $J/\psi\ K^{0*}$.

In another study, we look at the atomic number dependence of the total and differential cross sections for the vector mesons ρ/ω, φ and J/ψ and low-mass Drell-Yan pairs in the Feynman-x region $0.1 < x_F < 0.8$. By parameterizing the total cross-section as $\sigma_0 A^{\alpha}$, we find the values for α to be equal 0.74 ± 0.01 (*stat*) ± 0.02 (*sys*), 0.80 ± 0.01 (*stat*) ± 0.02 (*sys*), and 0.92 ± 0.02 (*stat*) ± 0.02 (*sys*), for these vector mesons, respectively. We find the value of α for the Drell-Yan dimuon continuum equal to 1.16 ± 0.08 (*stat*) ± 0.02 (*sys*) for dimuon masses between 4.0 GeV and 7.0 GeV. We found no significant dependence of the α parameter on $x_F$ or the transverse momentum $p_T$ [49].

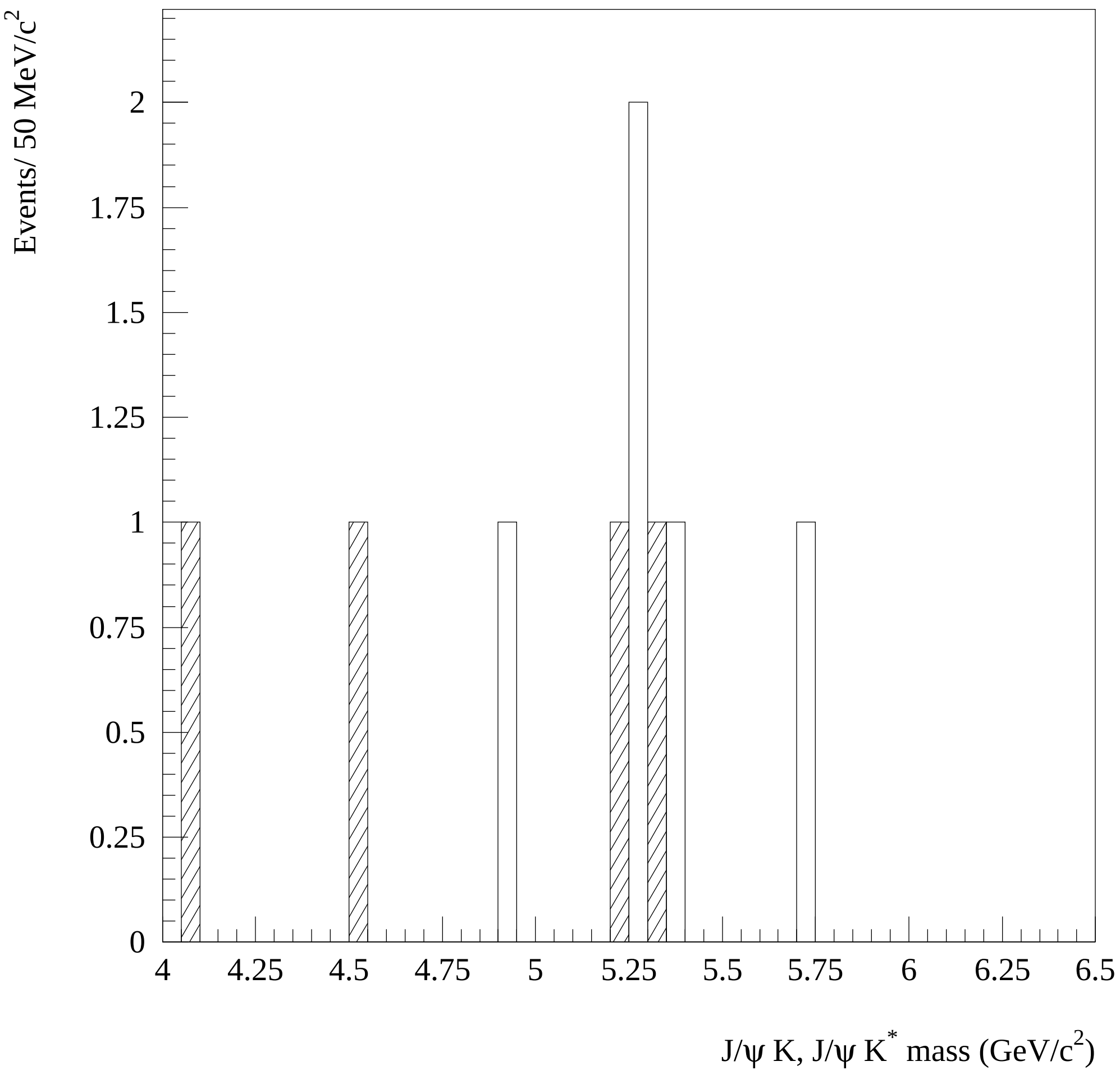

Figure 4.12. Combined invariant-mass spectrum for $B^{\pm} \rightarrow J/\psi\ K^{\pm}$ and $B^{0} \rightarrow J/\psi\ K^{0*}$ (hatched).

# CHAPTER 5

# THE $J/\psi \to \mu^+ \mu^-$ SAMPLE

The $J/\psi \to \mu^+ \mu^-$ event selection begins by including all dimuons from the "PSI" data sample (the procedure to reduce the raw data to the "PSI" sample is explained in section 4.1.1.). Each dimuon is refitted using a standard $\chi^2$ technique, and only dimuons that survive the $\chi^2$ cut are kept, but no more than one dimuon per event. This chapter describes the requirements for the selection of $J/\psi \to \mu^+ \mu^-$ events used in the $\Lambda_b \to J/\psi \Lambda^0$ analysis of this thesis. The details of the dimuon refit are explained in Appendix A.

## 5.1 Selection of the $J/\psi \to \mu^+ \mu^-$ sample

The dimuon invariant-mass distribution for the "PSI" sample is shown in Figure 5.1. A fit to this distribution that included the Monte-Carlo resolution functions for the $J/\psi$ and $\psi(2S)$, and two exponential functions for the background, yields a $J/\psi$ mass of (3.097 $\pm$ 0.001) GeV/$c^2$. This value of the $J/\psi$ mass is in agreement with the world average value from the particle data group [12]. The $J/\psi$ signal region is defined as the mass interval from 2.85 GeV/$c^2$ to 3.35 GeV/$c^2$. This region contains about 13,053 muon pairs, with a background subtracted signal of 11,500 $J/\psi$ events.

The muon tracks in the signal region are then refitted with the constraint that both muons intersect at a common point and that the invariant-mass is equal to 3.097 GeV/$c^2$. This constrained fitting technique improves the momentum resolution as compared to initial track finding by a factor of 2. Also, the vertex position resolution of dimuons improved by 15 %. See appendix A for plots of the momentum and vertex residuals before and after the fit. The quality of the fit was used to reduce combinatorial background. A cut on the $\chi^2$ per degree of freedom of the fit, to be less than 5.0, reduced the $\mu^+ \mu^-$ background in the J/$\psi$ signal region by 50%. A total of 12,340 muon pairs survived the refit criteria.

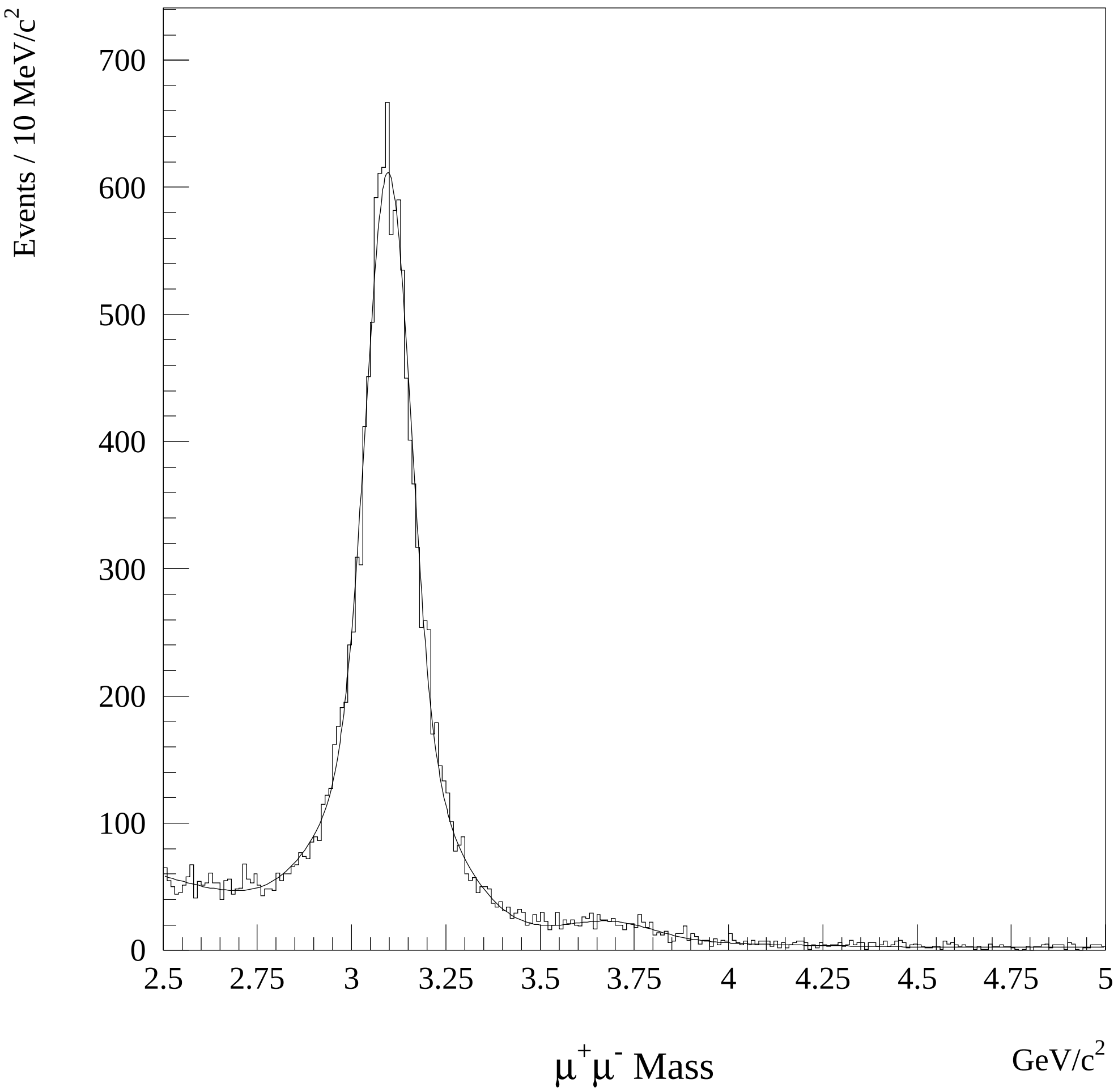

Figure 5.1. The dimuon invariant-mass distribution for the “PSI” sample. The solid line is a fit to the data that includes the Monte-Carlo resolution functions for the J/ψ and ψ(2S), and two exponential functions for the background.

# CHAPTER 6

## THE $\Lambda^0 \rightarrow p\,\pi^-$ (AND CHARGE CONJUGATE) SAMPLE

The great majority of $\Lambda^0$ s produced in this experiment decayed after the target/SSDs. Therefore, the tracks of their daughter decay products (proton and pion) were not detected before their trajectories were deflected by the magnetic field of the dipole magnet. Thus, their reconstruction is more complex. The $\Lambda^0 \rightarrow p\,\pi^-$ decays can be reconstructed if the $\Lambda^0$ s decayed upstream of the PWC system, see Figure 2.1. In particular there are three regions in the spectrometer where this reconstruction can be achieved: near the target/SSDs, upstream of the dipole magnet, and inside the dipole magnet, see Figure 6.1. In the region near the target/SSDs, the proton and pion are detected by the SSD and PWC systems, but, for the other two regions they are only detected by the PWC system. This is important, because the standard way of determining the momentum of a charged particle is by measuring the deflection of its trajectory upon passing through a magnetic field. This chapter describes the procedure and algorithms used to reconstruct $\Lambda^0 \rightarrow p\,\pi^-$ decays. Also, to confirm the validity of the reconstruction algorithms, $K^0_s \rightarrow \pi^+\,\pi^-$ decays (which because of their lifetime have the same topology as $\Lambda^0 \rightarrow p\,\pi^-$ decays) are reconstructed in similar fashion, and they are also used to clean the $\Lambda^0$ (and $\overline{\Lambda}^0$) sample. I remind the reader that the throughout this thesis whenever a particle or a reaction is mentioned the charge conjugate particle or reaction is implicitly implied.

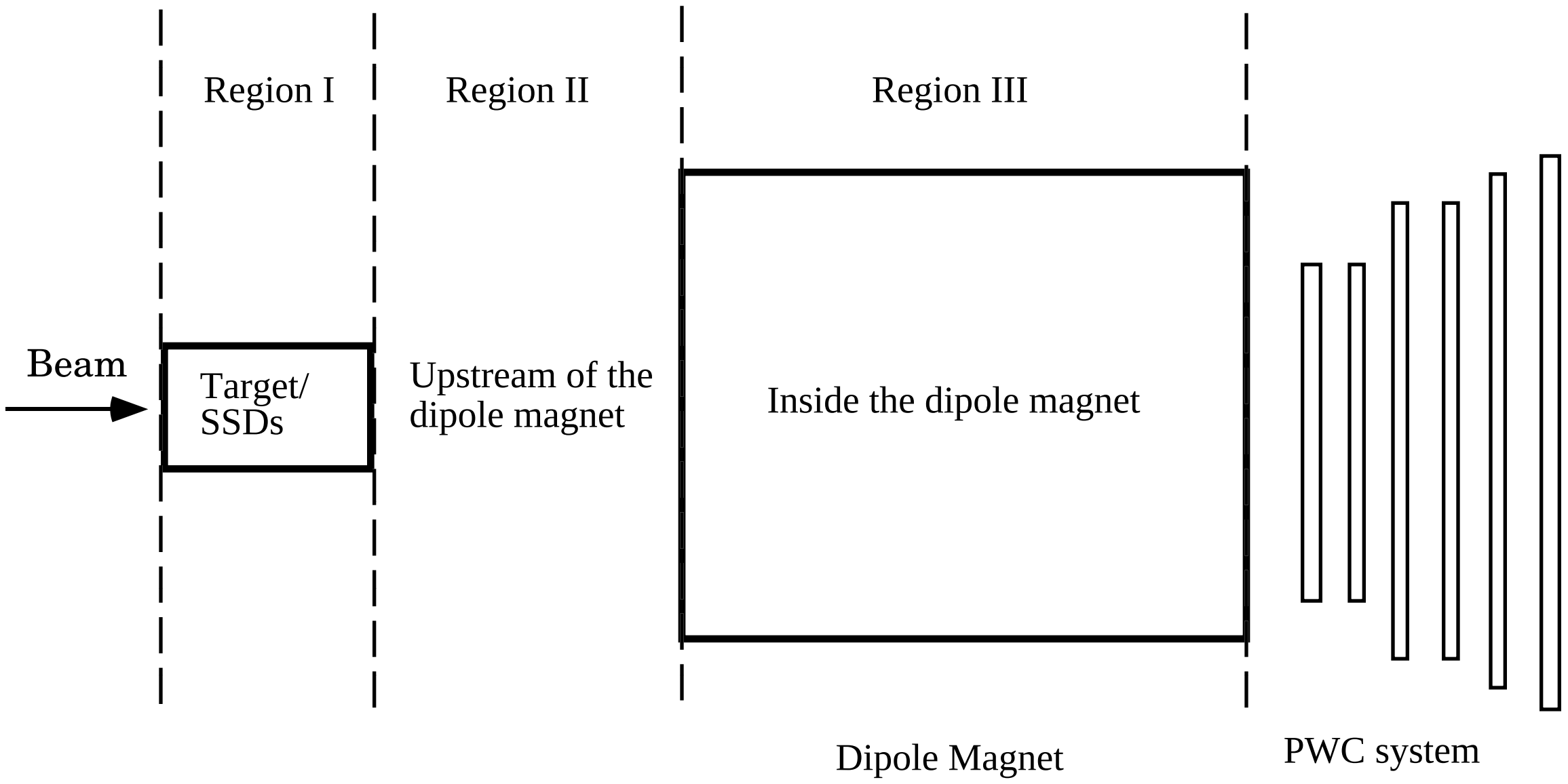

Figure 6.1. Regions in the spectrometer where $\Lambda^0$ s can be reconstructed.

## 6.1 Reconstruction algorithms for the $\Lambda^0$

An important characteristic of $\Lambda^0$ decay in the laboratory frame, is the imbalance in momentum between the decay proton and pion. Because of the large difference in mass between the the proton and pion, the proton from the $\Lambda^0$ decay will have on the average larger momentum than the pion[1]. In the $\Lambda^0 \rightarrow p\, \pi^-$ events accepted by the E672/E706 detector, the proton always had larger momentum than the pion (this was determined by Monte-Carlo). Thus, when performing the two-track combinations to reconstruct the $\Lambda^0$, the track with the largest momentum of the two is always assumed to be the proton.

[1] $M_p$ = 0.938 GeV/$c^2$ , $M_\pi$ = 0.1389 GeV/$c^2$ and in the center-of-mass frame $p^*$ = 0.101 GeV/c, thus, $E^*_p$= 0.943 GeV and $E^*_\pi$= 0.172 GeV. Therefore, after boosting the proton and pion momenta to the laboratory frame, on the average, the proton will have a larger momentum than the pion by a factor of approximately $E_p^* / E_\pi^*$ (= 5.48).

### 6.1.1 Reconstruction near the target/SSDs region

The topology of the $\Lambda^0 \rightarrow p\,\pi^-$ decay in this region is illustrated in Figure 6.2. For these decays, the tracks of the daughter particles are reconstructed using the SSD and PWC systems. Their momenta are measured by linking the SSD tracks to their proper PWC tracks as explained in section 4.2.2. The $\Lambda^0$ s are then reconstructed by taking two SSD-PWC linked track combinations, assigning the proton mass to the track with the largest momentum of the two, and the pion mass to the other one. The proton and pion tracks are required to form a vertex. This means that the distance of closest approach between the two tracks is less than 50 μm. In order to see a clear $\Lambda^0$ signal, this vertex should be at least 1.5 cm downstream of the production vertex of the $\Lambda^0$, which in this case is assumed to be the J/ψ decay vertex. According to Monte-Carlo, an impact parameter of a track to the primary vertex of less than 200 μm means the track is associated with the primary vertex.

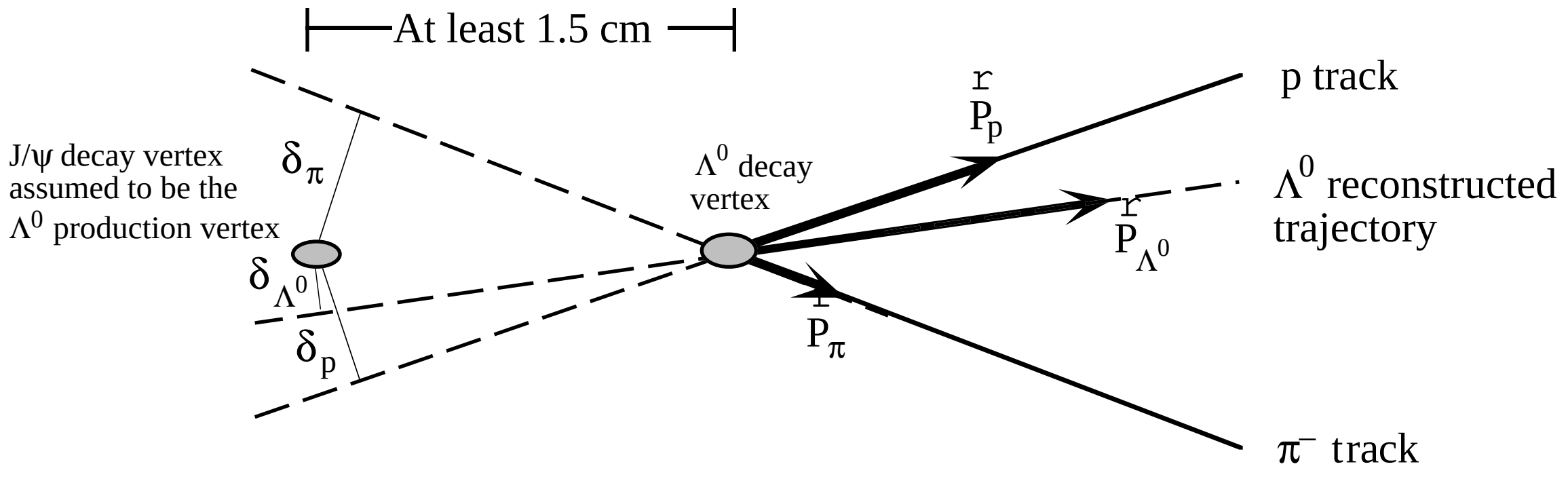

Figure 6.2. The illustration shows the topology of a $\Lambda^0 \rightarrow p\,\pi^-$ decaying near the target/SSD region.

Therefore, the impact parameter of the pion $\delta_\pi$ relative to the $\Lambda^0$ production vertex is required to be greater than 200 μm. Since the proton has a tendency of following the direction of its parent $\Lambda^0$, it is easy for the vertex reconstruction program to assign the proton track to the primary vertex. The cut on the impact parameter of the proton $\delta_p$ relative to the $\Lambda^0$ production vertex is therefore looser than that for the pion and it is required to be greater than 100 μm. Also, according to Monte-Carlo the impact parameter $\delta_{\Lambda^0}$ of the reconstructed $\Lambda^0$ relative to its production vertex should be less than 120 μm. Using all these criteria a clear signal for $\Lambda^0$ (and $\overline{\Lambda}^0$) is seen in Figure 6.3. The $\Lambda^0$(and $\overline{\Lambda}^0$) signal region is defined as the mass interval from 1.112 GeV/$c^2$ to 1.118 GeV/$c^2$, and there are 10 $\Lambda^0$(and $\overline{\Lambda}^0$) candidates. A fit to this distribution that included a gaussian function, and a second order polynomial function for the background, yields 9 ± 3 (*stat*) background-subtracted events, and a $\Lambda^0$(and $\overline{\Lambda}^0$) mass of (1.114 ± 0.0004) GeV/$c^2$, with a FWHM mass resolution of 2.3 MeV/$c^2$. This value of the $\Lambda^0$(and $\overline{\Lambda}^0$) mass is in agreement with the world average value from the particle data group [12].

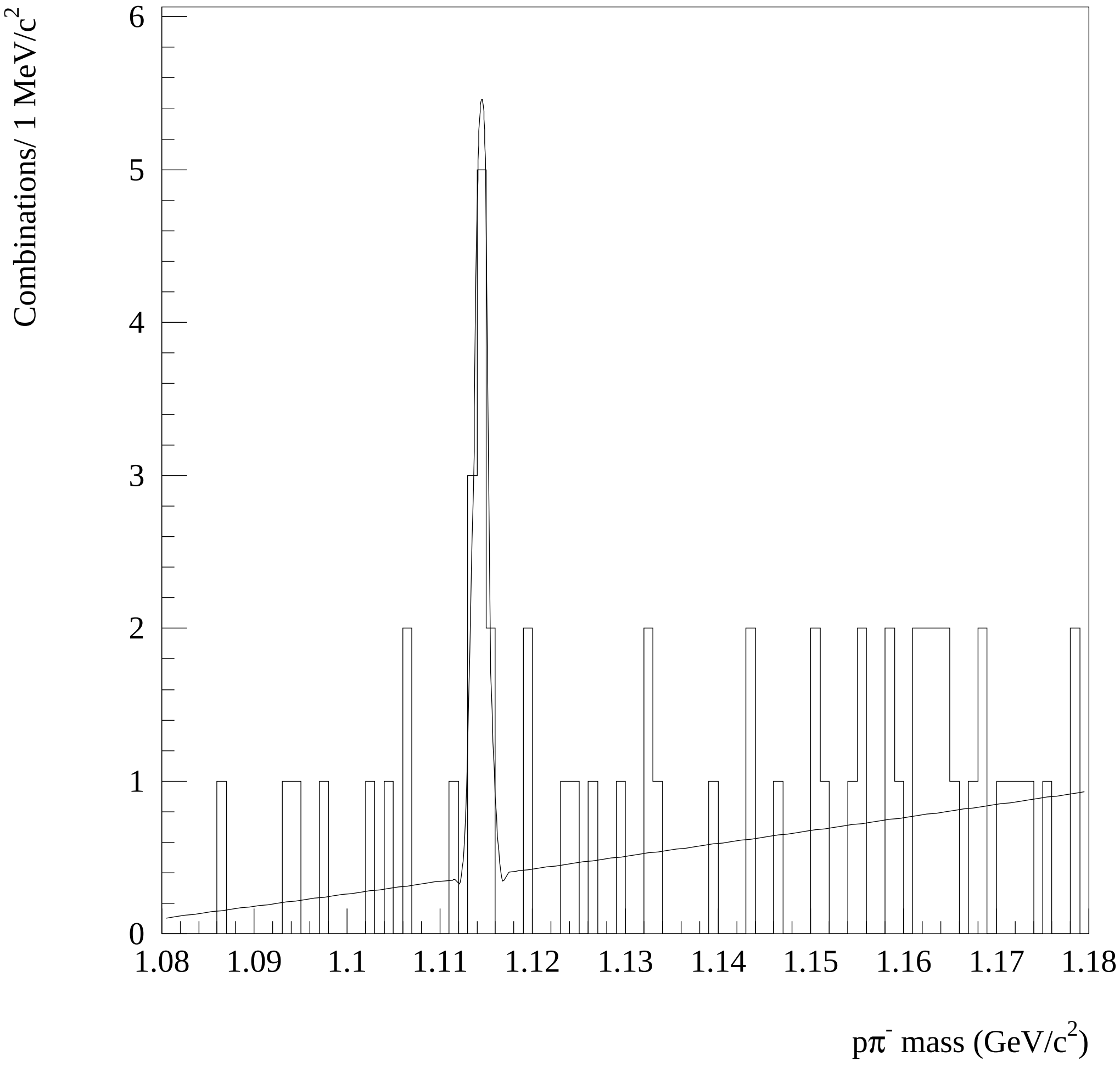

Figure 6.3. Invariant-mass distribution using only SSD-PWC tracks for p $\pi^-$ (and $\bar{p}$ $\pi^+$) track combinations intercepting in the target/SSDs region. The solid line is a fit to the data.

### 6.1.2 Reconstruction in the region upstream of the dipole magnet

Taking advantage of the two-body decay kinematics, and assuming the production vertex of the $\Lambda^0$ s to be the J/ψ decay vertex, the $\Lambda^0$ s that decayed in this region can be reconstructed. The first step is to determine the $\Lambda^0$ decay vertex (x, y, z). The dipole magnet only bent the trajectory of the electricly charged particles in the x-z plane. Therefore, the y and z coordinates are found from the intersection of the proton and pion PWC tracks in the y-z plane, see Figure 6.4. Knowing y and z, and assuming an x coordinate for the $\Lambda^0$ decay vertex, the corresponding momenta of the proton and pion can be computed. This can be determined by extrapolating the PWC tracks in the x-z plane to the center of the dipole magnet, and using the $\Lambda^0$ decay vertex to estimate the deflection of their trajectories throughout the dipole magnet, see Figure 6.4. See also Appendix B for the explicit calculation of the momentum vectors.

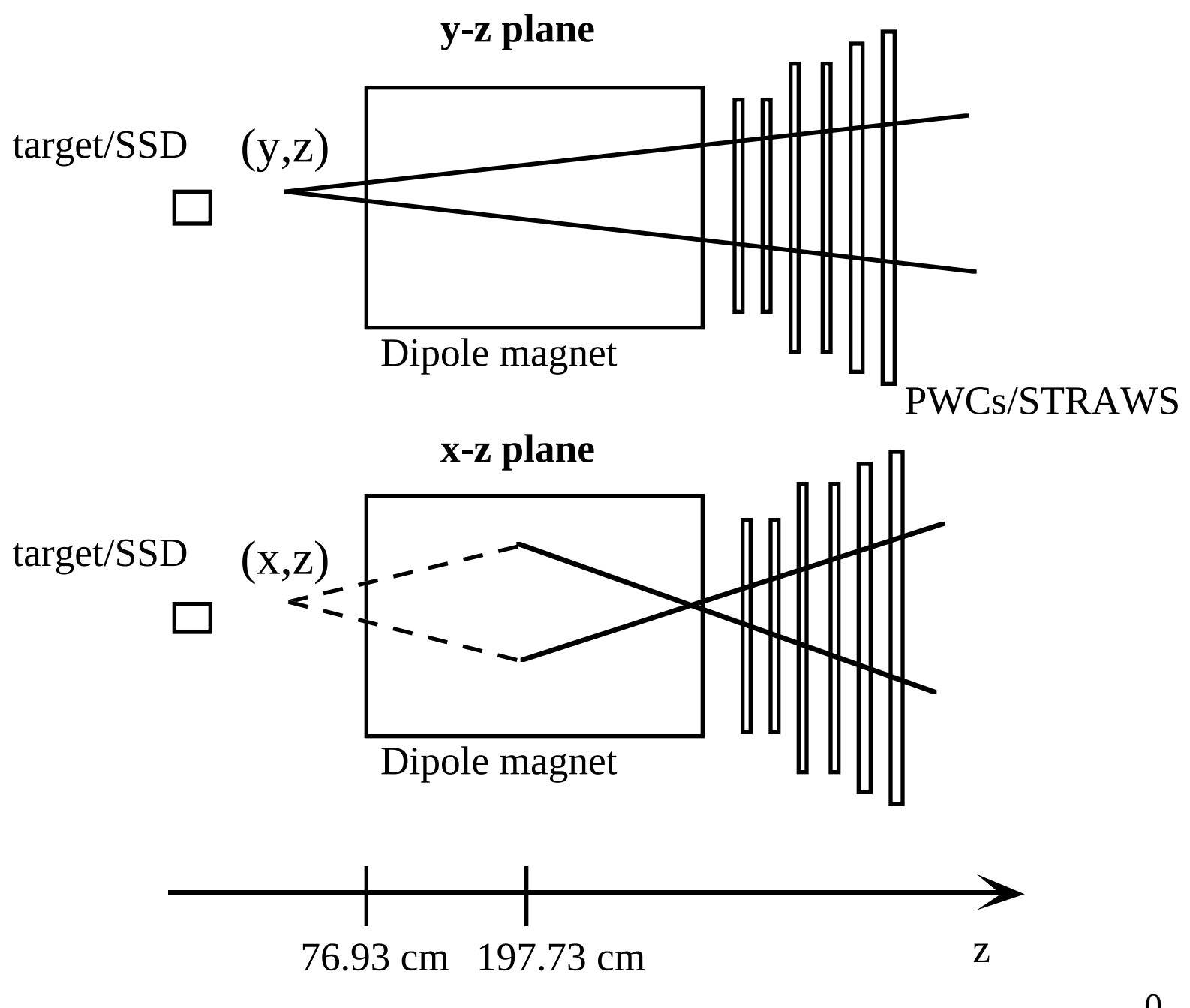

Figure 6.4. Illustration of the proton and pions PWC tracks, for those $\Lambda^0$ s that decayed in the region upstream of the dipole magnet. The solid lines are the reconstructed PWC tracks.

To find the actual x coordinate of the $\Lambda^0$ decay vertex, the assumed position of the x coordinate is scanned across the x-axis for the value such that the vector sum of the transverse momentum of the proton and pion with respect to the direction of flight of the $\Lambda^0$ divided by the sum of their magnitudes (this ratio is called the " relative $p_T$ ", see Figure 6.5) is a minimum. Real $\Lambda^0$ decays should have, in principle, minimum relative $p_T$ equal to zero; thus, cutting on this quantity removes undesired background to $\Lambda^0$ decays originating from accidental intersections of tracks. Figure 6.6 shows the distribution of minimum relative $p_T$ for proton and pion tracks intersecting upstream of the dipole magnet. According to Monte-Carlo studies, 99 % of reconstructed $\Lambda^0$ s have minimum relative $p_T$ less than 0.4. Thus, a cut on minimum relative $p_T$ of less than 0.4 is imposed on the data.

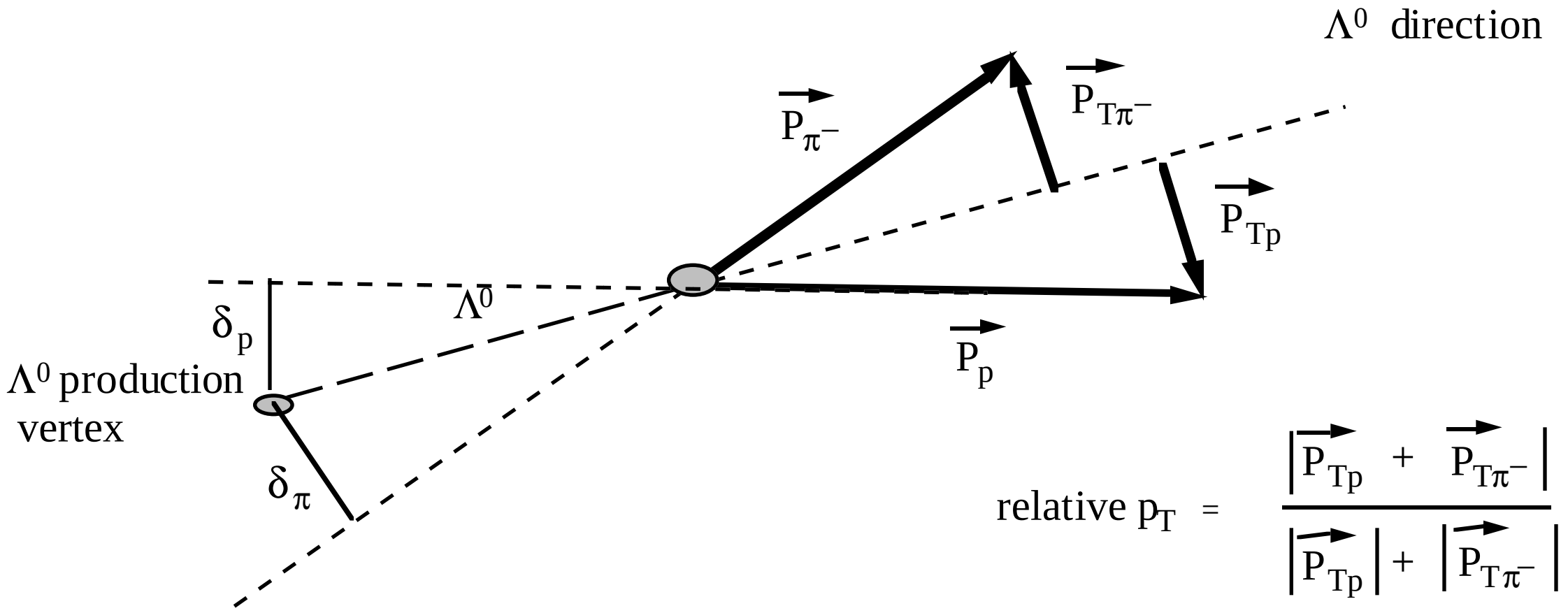

Figure 6.5. The illustration shows the topology of a $\Lambda^0 \rightarrow p\,\pi^-$ decay, and the definition of relative $p_T$.

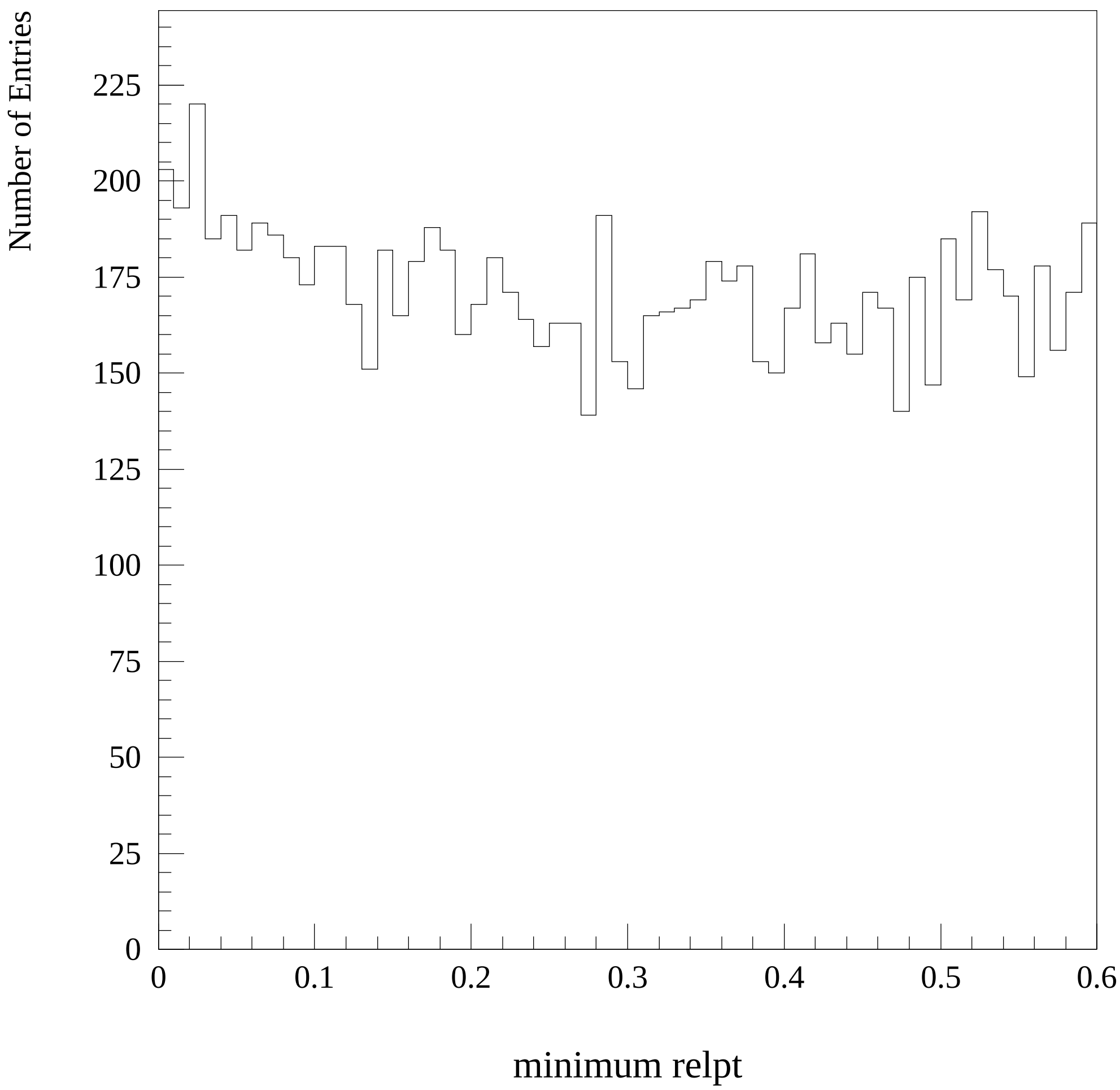

Figure 6.6. Distribution of minimum relative $p_T$ for tracks intersecting upstream of the dipole magnet.

To reject more undesired background to $\Lambda^0$, the proton and pion PWC tracks are required to not be linked to any SSD track. Also, because of the small difference between the $\Lambda^0$ and proton mass, the proton tends to follow the trajectory of the parent $\Lambda^0$. Thus, the impact parameter of the proton $\delta_p$ relative to the $\Lambda^0$ production vertex is required to be less than 1.5 cm (this value was determined from Monte-Carlo). The invariant-mass distribution for the p $\pi^-$ (and $\bar{p}$ $\pi^+$) combinations for this region is shown in Figure 6.7. A clear signal for $\Lambda^0$(and $\overline{\Lambda}^0$) is seen. The $\Lambda^0$ (and $\overline{\Lambda}^0$) signal region is defined as the mass interval from 1.105 GeV/c$^2$ to 1.125 GeV/c$^2$, and there are 479 candidates. A fit to this distribution that included a gaussian function, and a fourth order polynomial function for the background, yields 220 ± 23 (*stat*) background subtracted events, and a $\Lambda^0$(and $\overline{\Lambda}^0$) mass of (1.116 ± 0.0004) GeV/c$^2$, with a FWHM mass resolution of 8.0 MeV/c$^2$. This value of the $\Lambda^0$(and $\overline{\Lambda}^0$) mass is in agreement with the world average value from the particle data group [12].

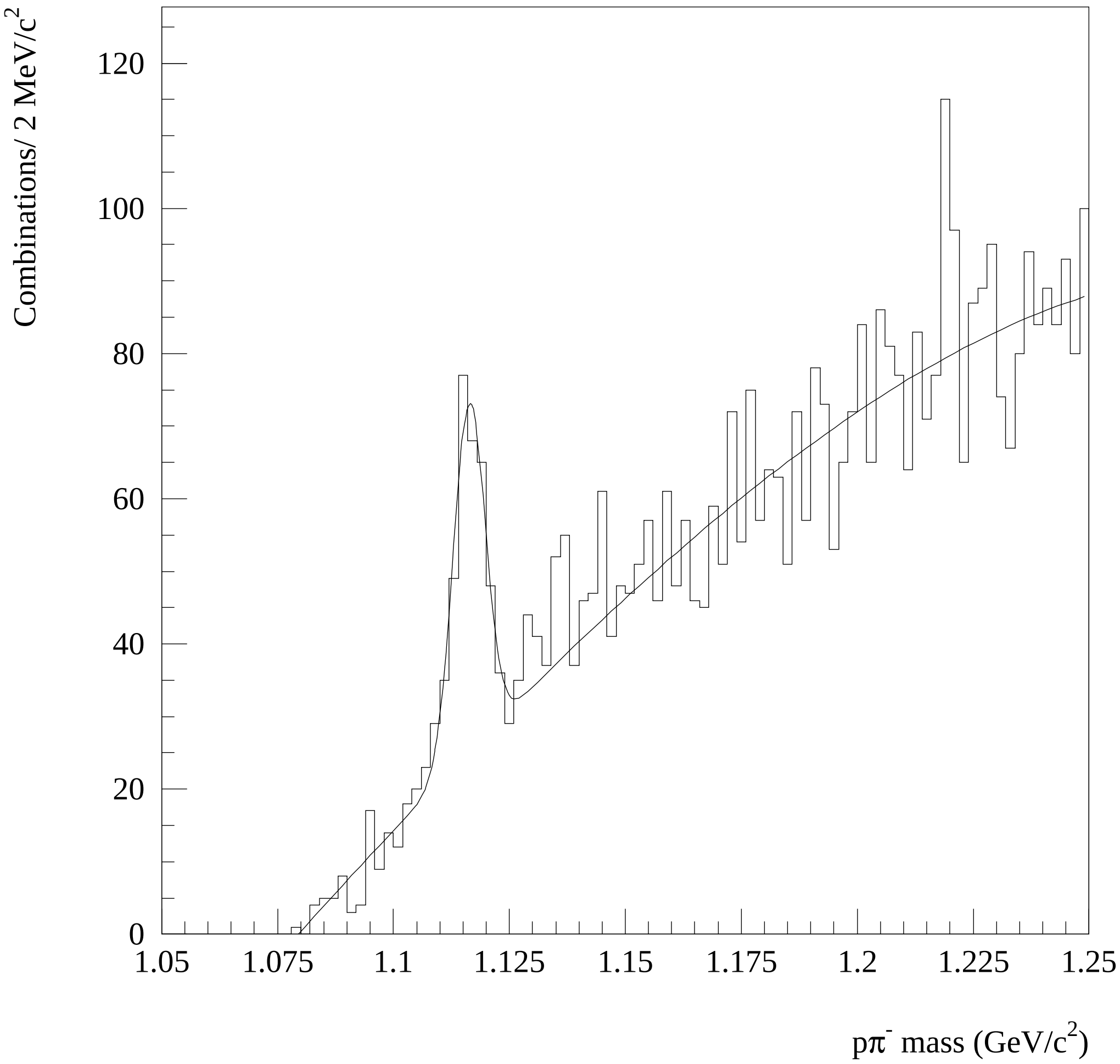

Figure 6.7. Invariant-mass distribution using only PWC tracks for p $\pi^-$ (and $\bar{p}$ $\pi^+$) combinations for tracks intercepting in the region upstream of the dipole magnet. The solid line is a fit to the data.

### 6.1.3 Reconstruction inside the dipole magnet region

For those $\Lambda^0$s that decayed inside the dipole magnet the reconstruction algorithm is similar as for those that decay upstream of the magnet. Again, the reconstruction algorithm begins by finding the (y, z) coordinates of the $\Lambda^0$ decay vertex. The y and z coordinates are determined by the intersection of the proton and pion PWC tracks in the y-z plane, see Figure 6.8. The determination of the x coordinate is the same as before. Knowing the y and z coordinates and with a given x coordinated for the $\Lambda^0$ decay vertex, the momenta of the proton and pion are estimated by measuring the radii of their unique circular trajectories through the dipole magnet projected onto the x-z plane. These circular trajectories are such that they include the point (x, z). Also at the downstream end of the dipole magnet (z=

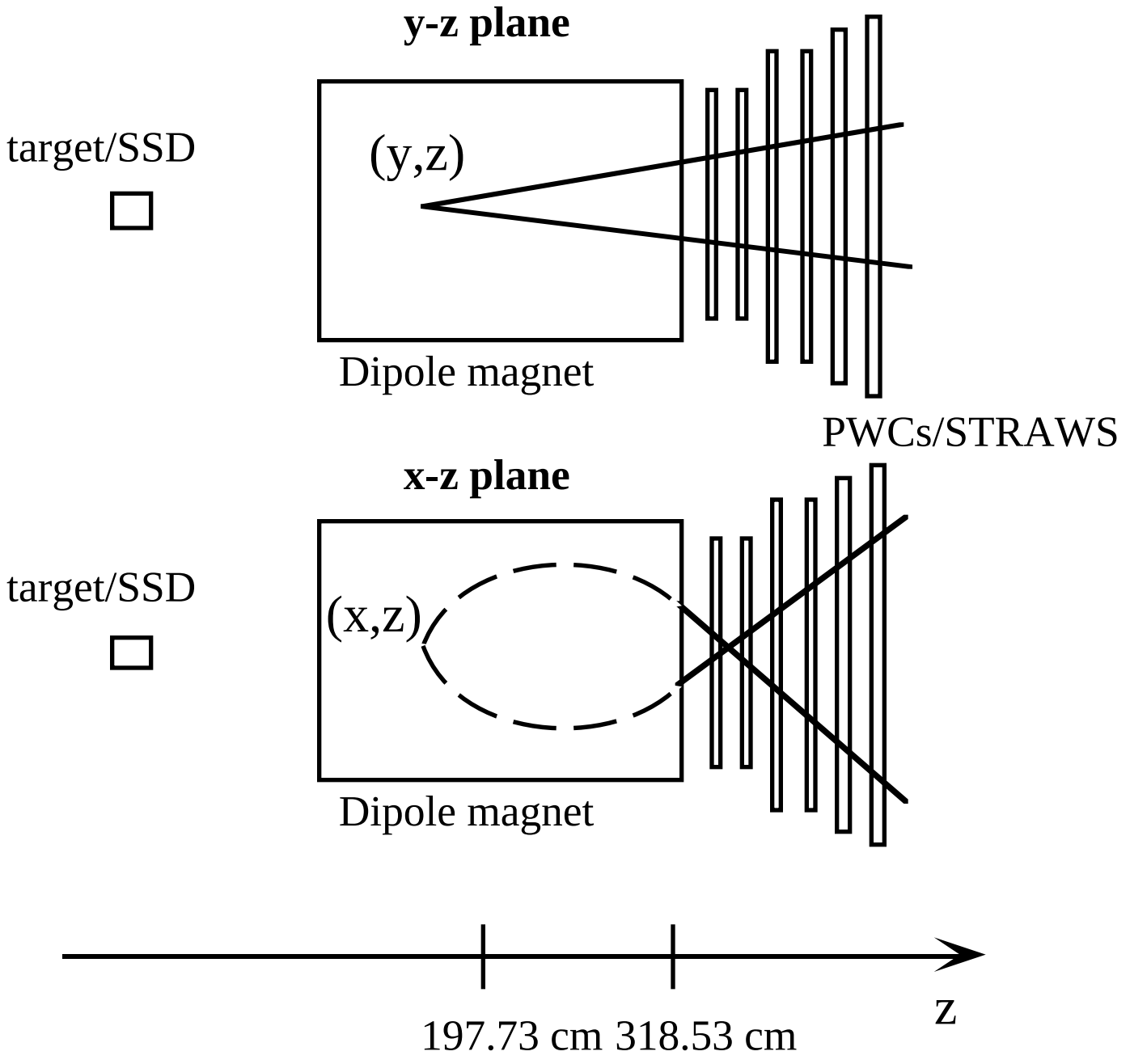

Figure 6.8. Illustration of the proton and pion PWC tracks, for those $\Lambda^0$ s that decayed in the region inside the dipole magnet. The solid lines are the reconstructed PWC tracks.

318.53 cm) they match the slopes of the reconstructed PWC tracks (see Figure 6.8). See also Appendix C for the detailed calculation of the momentum vectors.

To find the actual x coordinate of the $\Lambda^0$ decay vertex, the assumed position of the x coordinate is scanned across the x-axis for the value of x such that the relative $p_T$ is a minimum (see Figure 6.5 for the defenition of relative $p_T$). Real $\Lambda^0$ decays should have, in principle, minimum relative $p_T$ equal to zero; thus, cutting on this quantity removes undesired background to $\Lambda^0$ decays originating from accidental intersections of tracks. Figure 6.9 shows the distribution of minimum relative $p_T$ for proton and pion tracks intersecting inside the dipole magnet. Again, according to Monte-Carlo studies, 99 % of reconstructed $\Lambda^0$s (and $\overline{\Lambda}^0$s) have minimum relative $p_T$ less than 0.4. Thus, a cut on the minimum relative $p_T$ to be less than 0.4 is imposed on the data.

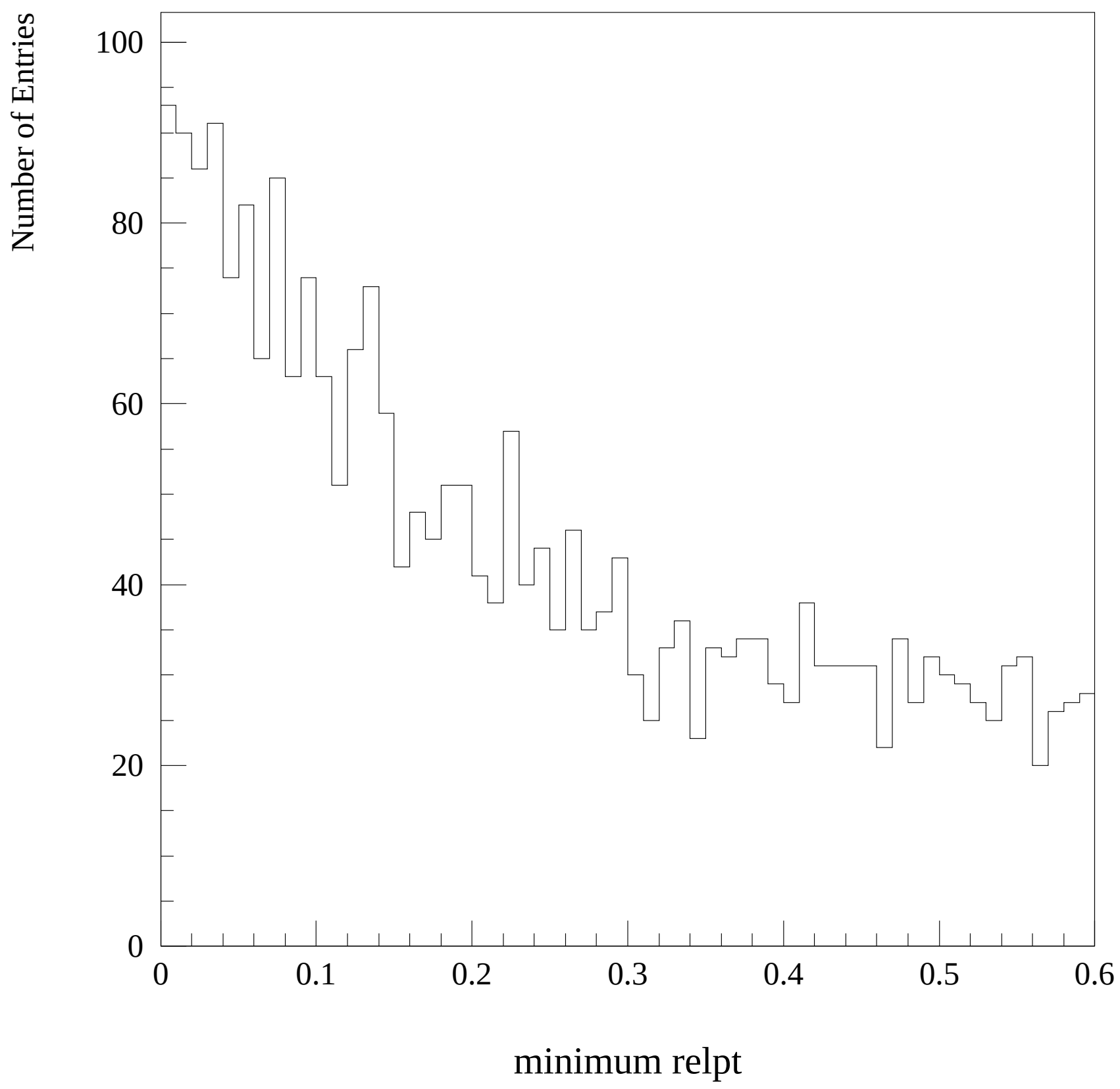

Figure 6.9. Distribution of minimum relative $p_T$ for tracks intersecting inside the dipole magnet.

To further reduce the background to $\Lambda^0$, the proton and pion PWC tracks are required to not be linked to any SSD track. The impact parameter of the proton $\delta_p$ relative to the $\Lambda^0$ production vertex is required to be less than 1.5 cm. The invariant-mass distribution for the proton-pion combinations that intersect this region is shown in Figure 6.10. A clear signal for $\Lambda^0$(and $\overline{\Lambda}^0$) is seen. The $\Lambda^0$ (and $\overline{\Lambda}^0$) signal region is defined as the mass interval from 1.10 GeV/$c^2$ to 1.13 GeV/$c^2$, and there are 582 candidates. A fit to this distribution that included a gaussian function, and a fourth order polynomial function for the background, yields 346 ± 26 (*stat*) background subtracted events, and a $\Lambda^0$(and $\overline{\Lambda}^0$) mass of (1.115 ± 0.0006) GeV/$c^2$ with a FWHM mass resolution of 14 MeV/$c^2$. This value of the $\Lambda^0$ (and $\overline{\Lambda}^0$) mass is in agreement with the world average value from the particle data group [12].

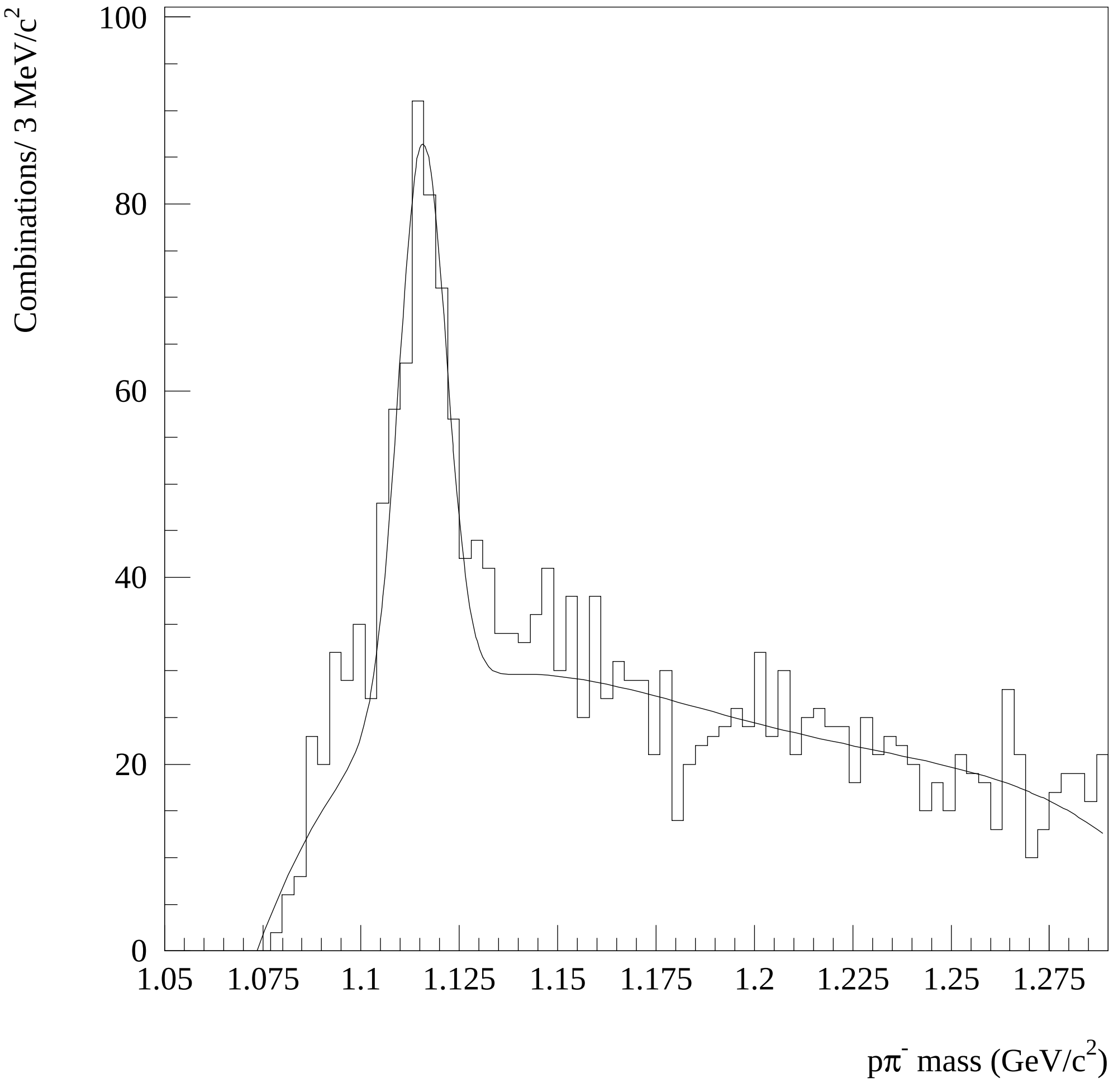

Figure 6.10. Invariant-mass distribution using only PWC tracks for p $\pi^-$ (and $\bar{p}$ $\pi^+$) combinations for tracks intercepting in the region inside the dipole magnet. The solid line is a fit to the data.

## 6.2 The $K^0_s \rightarrow \pi^+ \pi^-$ signal

Since the $K^0_s$ meson has a lifetime of the same order of magnitude ($10^{-10}$ s) as the $\Lambda^0$, the $K^0_s \rightarrow \pi^+ \pi^-$ decays have the same topology as the $\Lambda^0 \rightarrow p\ \pi^-$ decays. Thus, the $K^0_s \rightarrow \pi^+ \pi^-$ decay becomes a substantial background to the $\Lambda^0 \rightarrow p\ \pi^-$ decays. However, the $K^0_s$ will be used to cross-check the reconstruction algorithms and to clean the $\Lambda^0$ (and $\overline{\Lambda}^0$) samples. This is needed in order to avoid ambiguities when searching for the $\Lambda_b \rightarrow J/\psi\ \Lambda^0$ decay (and charge conjugate).

### 6.2.1 The $K^0_s \rightarrow \pi^+ \pi^-$ signal in the region near the target/SSDs

For decays in this region, the two pion SSD-PWC linked tracks are required to form a vertex (distance of closest approach < 50 μm). In order to resolve the $K^0_s$ vertex from the primary vertex the $K^0_s$ decay vertex should be at least 0.2 cm downstream of the J/ψ decay vertex (the assumed production point of the $K^0_s$). Again, according to Monte-Carlo, an impact parameter of a track to the primary vertex of less than 200 μm means the track is associated with the primary vertex. Thus, the impact parameter of each pion $\delta_\pi$ relative to the $K^0_s$ production point must be greater than 200 μm. Also according to Monte-Carlo the impact parameter of the reconstructed $K^0_s$ relative to its production point must be less than 120 μm. A clear signal for the $K^0_s$ is observed at the nominal $K^0_s$ mass of 0.497 GeV/c, as seen in Figure 6.11.

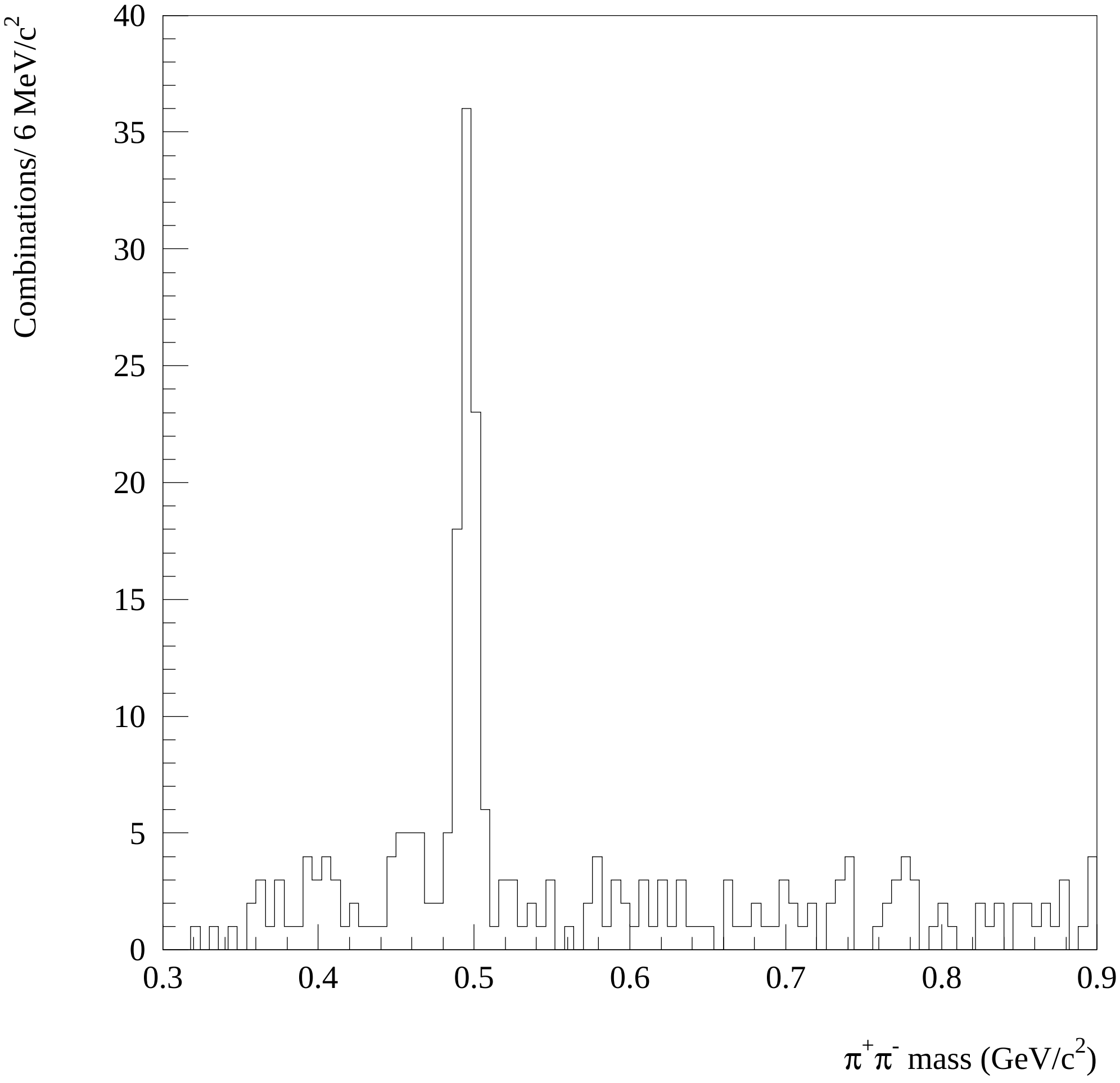

Figure 6.11. Invariant-mass distribution using only SSD-PWC tracks for $\pi^+$ $\pi^-$ track combinations intercepting in the target/SSDs region.

### 6.2.2 The $K^0_s \rightarrow \pi^+ \pi^-$ signal in the regions upstream and inside the dipole magnet

In both of these regions the requirements are that the pion PWC tracks are not linked to any SSD track, and that the minimum relative $p_T$ is less than 0.4, as in the $\Lambda^0$ case. However, unlike in the $\Lambda^0 \rightarrow p\,\pi^-$ decay where the proton tends to follow the direction of its parent $\Lambda^0$, the pions from the $K^0_s$ decay have no prefered direction. Thus, no impact parameter is imposed on the pions. To enhance the $K^0_s$ signal its transverse momentum is required to be greater than 0.4 GeV/c. These cuts resulted in clear signals in both regions for the $K^0_s$ at the nominal mass of 0.497 GeV/C. See Figure 6.12 and Figure 6.13.

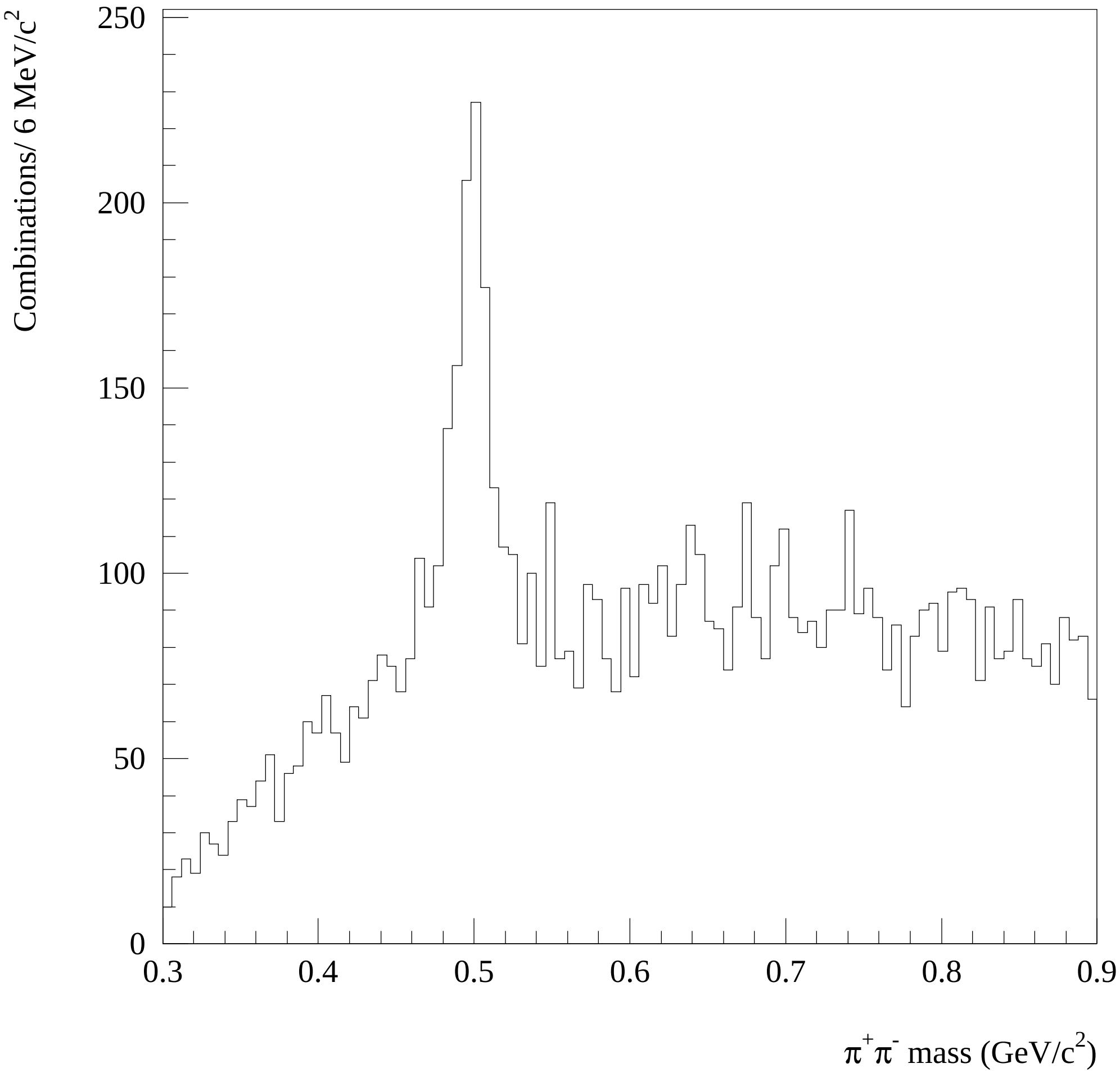

Figure 6.12. Invariant-mass distribution using only PWC tracks for $\pi^+ \pi^-$ combinations for tracks intercepting in the region upstream of the dipole magnet.

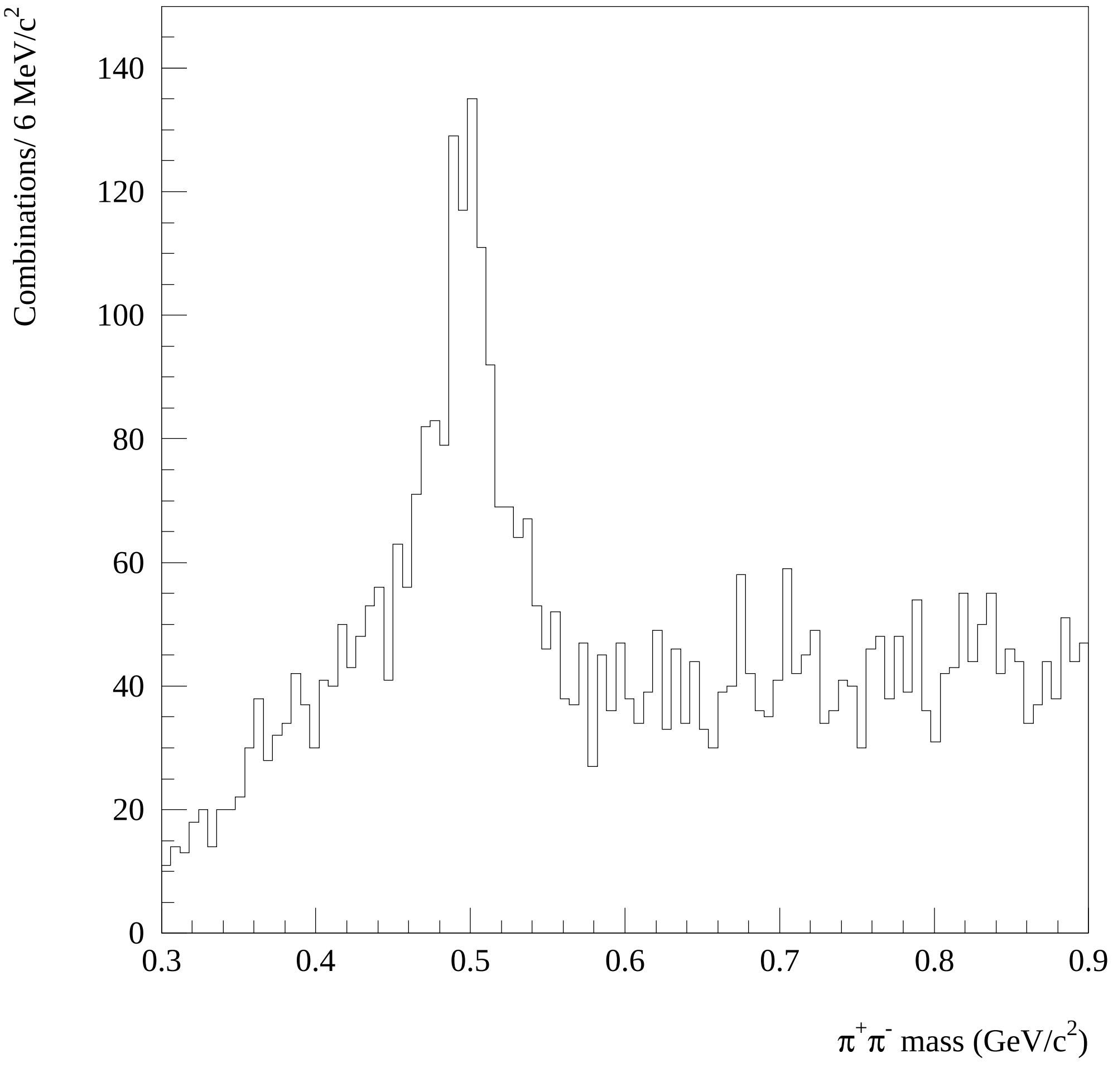

Figure 6.13. Invariant-mass distribution using only PWC tracks for $\pi^+\ \pi^-$ combinations for tracks intercepting in the region inside the dipole magnet.

### 6.2.3 Distinguishing the $\Lambda^0 \rightarrow p\,\pi^-$ (and charge conjugate) decays from the $K^0_s \rightarrow \pi^+\,\pi^-$ decays

The $K^0_s \rightarrow \pi^+ \pi^-$ and $\Lambda^0 \rightarrow p\,\pi^-$ (or $\overline{\Lambda}^0 \rightarrow \overline{p}\,\pi^+$) decays are difficult to distinguished them from one another without particle identification. One can, however, try to distinguish the two decays by plotting the Podolanski-Armenteros plot [51], defined as the plot of the magnitude of the $p_T$ of either decay daughter particle relative to the direction of flight of the parent particle versus the asymmetry in of longitudinal momentum (in the laboratory frame) of the daughter particles, $\left(p_L^+ - p_L^-\right)/\left(p_L^+ + p_L^-\right)$. It is easy to show that there is a kinematically allowed region bounded by:

$$\frac{\left(\frac{p_L^+ - p_L^-}{p_L^+ + p_L^-} - \frac{E^{*+} - E^{*-}}{E^{*+} + E^{*-}}\right)^2}{\left(\frac{2p^*}{E^{*+} + E^{*-}}\right)^2} + \frac{p_T^2}{p^{*2}} = 1;$$

where $E^{*+}$, $E^{*-}$, and $p^*$ are the energy and momenta of the daughter particles in the center-of-mass frame, and $p_{Tmax} = p^*$. Figure 6.14 shows the Podolanski-Armenteros boundaries for the $K^0_s \rightarrow \pi^+ \pi^-$, $\Lambda^0 \rightarrow p\,\pi^-$, and $\overline{\Lambda}^0 \rightarrow \overline{p}\,\pi^+$ decays, and Figure 6.15 shows the Podolanski-Armenteros distribution of the $\Lambda^0$ (and $\overline{\Lambda}^0$) candidates decaying upstream the dipole magnet. These include all the combinations between 1.105 GeV/$c^2$ and 1.125 GeV/$c^2$ in Figure 6.7. A cut on the $p_T$ of the decaying particles relative to the direction of flight of the parent particle to be less then 0.18 GeV/c is imposed to reject possible $K^0_s \rightarrow \pi^+ \pi^-$ events. This value was suggested from Monte-Carlo since 100 % of the $\Lambda^0 \rightarrow p\,\pi^-$, and $\overline{\Lambda}^0 \rightarrow \overline{p}\,\pi^+$ decays are retained. The $\Lambda^0$ (and $\overline{\Lambda}^0$) candidates were reduced to 10, 471, and 578 events, in the target/SSDs, upstream of the dipole magnet, and inside the dipole magnet regions, respectively.

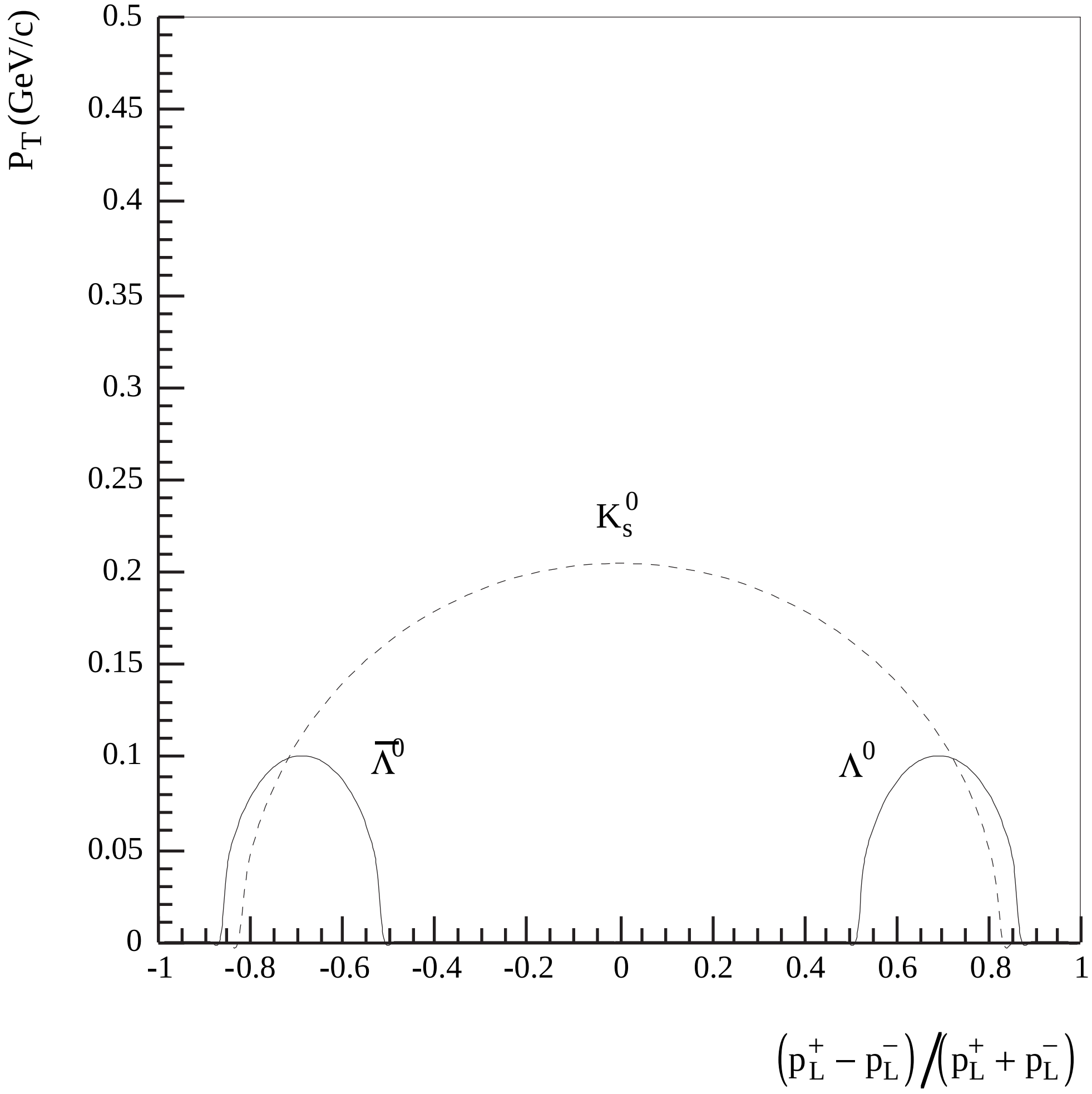

Figure 6.14. Podolanski-Armenteros plot for the $K^0{}_s \rightarrow \pi^+ \pi^-$, $\Lambda^0 \rightarrow p\, \pi^-$, and $\overline{\Lambda}^0 \rightarrow \overline{p}\, \pi^+$ decays.

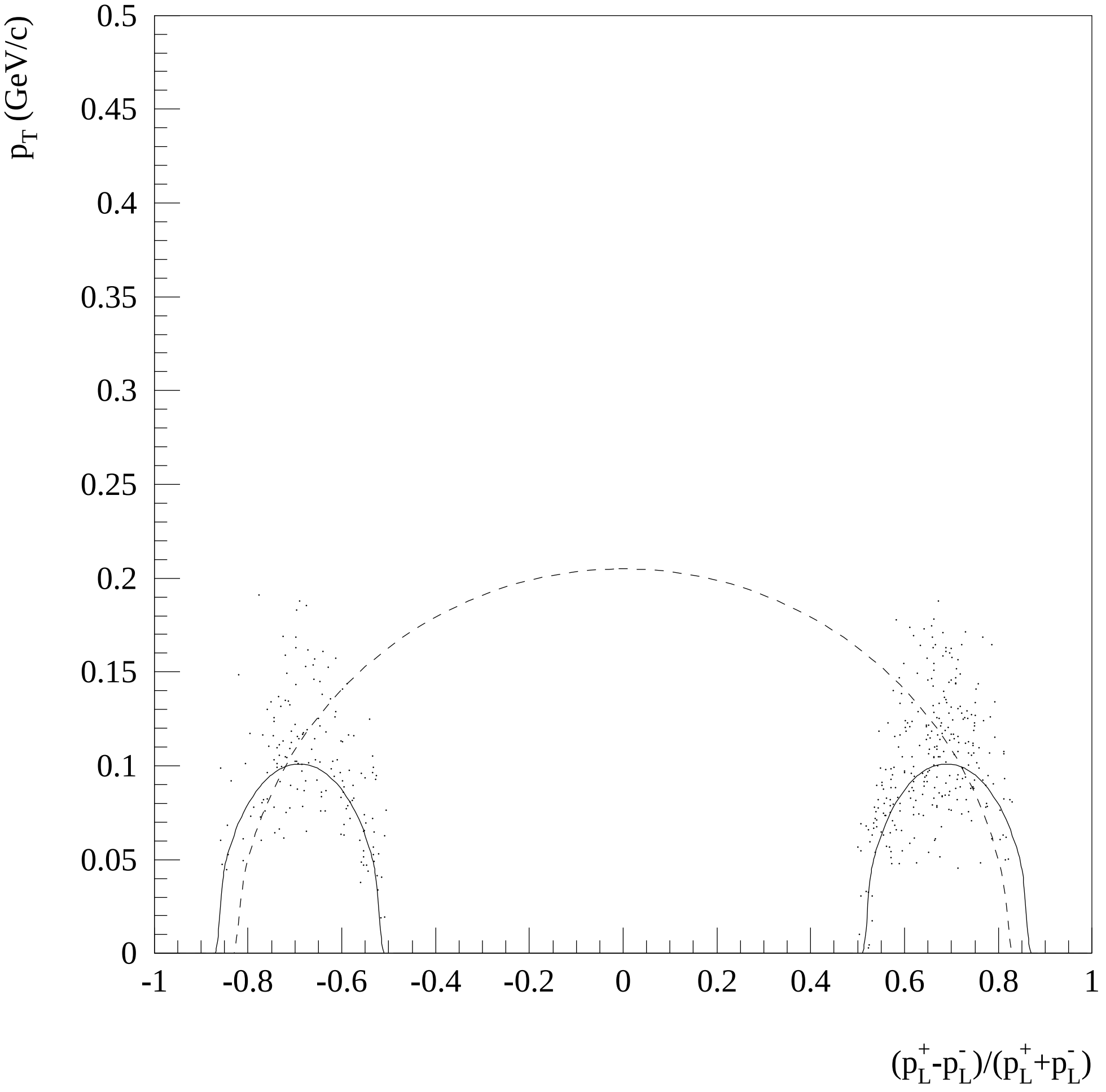

Figure 6.15. Podolanski-Armenteros plot for the $\Lambda^0$ (and $\overline{\Lambda}^0$) candidates decaying upstream the dipole magnet. These included all the combinations between 1.105 GeV/$c^2$ and 1.125 GeV/$c^2$ in Figure 6.7. The lines are the kinematical boundaries for the $K^0_s \rightarrow \pi^+ \pi^-$, $\Lambda^0 \rightarrow p\, \pi^-$, and $\overline{\Lambda}^0 \rightarrow \bar{p}\, \pi^+$ decays (see previous Figure).

However, even after applying the cut described in the previous paragraph there is still $K^0_s \rightarrow \pi^+ \pi^-$ contamination from the outer sides of the $\Lambda^0$ and $\overline{\Lambda}^0$ boundary regions. Figure 6.16 shows the mass spectrum of p $\pi^-$ and $\bar{p}\pi^+$ combinations that passed the Podolanski-Armenteros cut from the region upstream of the magnet, when both tracks are assigned pion masses. The solid line is the expected distribution obtained from Monte-Carlo, normalized to have the same number of events as the data in the in the mass interval between 0.3 GeV/$c^2$ and 0.45 GeV/$c^2$. A similar plot is shown in Figure 6.17 for those events in the region inside the magnet.

There is a clear $K^0_s \rightarrow \pi^+ \pi^-$ contamination. Because of this $K^0_s \rightarrow \pi^+ \pi^-$ background, any $\Lambda^0$ (or $\overline{\Lambda}^0$) candidate that has an invariant-mass consistent with that of the $K^0_s$ under the assumption of both tracks being pions, is excluded from the $\Lambda^0$ (and $\overline{\Lambda}^0$) sample. The $K^0_s$ mass range in each of the regions is: (0.48 - 0.51) GeV/$c^2$, (0.475 - 0.525) GeV/$c^2$, and (0.45 - 0.55 GeV/$c^2$), for the target/SSD, upstream of the magnet, and inside the magnet, respectively. This cut retained 90% (9 combinations), 67% (318 combinations), and 49% (285 combinations), in each region respectively. The efficiency of this cut according to Monte-Carlo is 90%, 67%, and 49%, for each region, respectively.

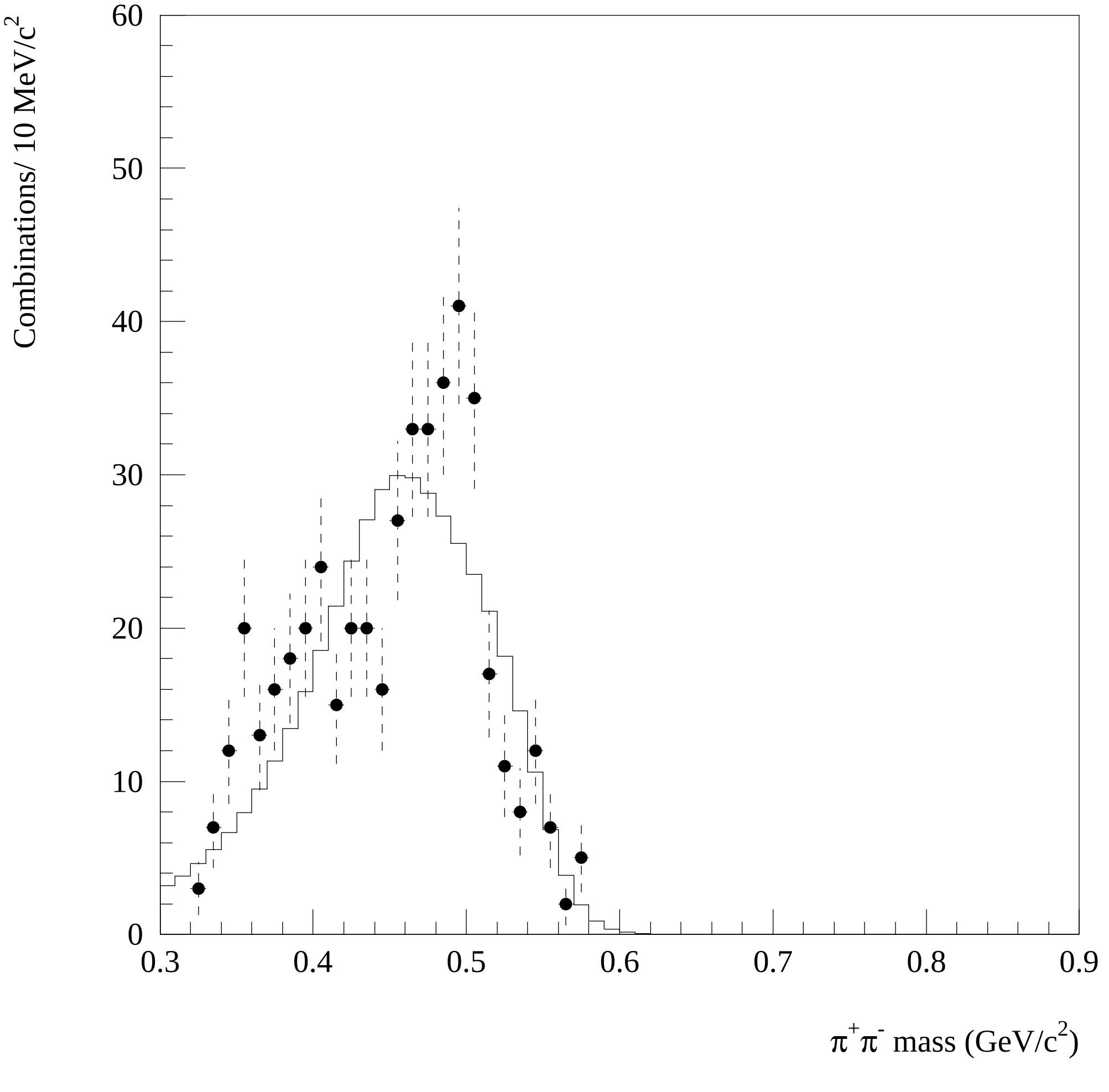

Figure 6.16. Invariant-mass spectrum of p $\pi^-$ and $\bar{p}\pi^+$ combinations that passed the criteria of being a $\Lambda^0$ decaying in the region upstream of dipole magnet when both tracks are assigned pion masses. The solid line is the expected distribution obtained from Monte-Carlo and the dots are from data.

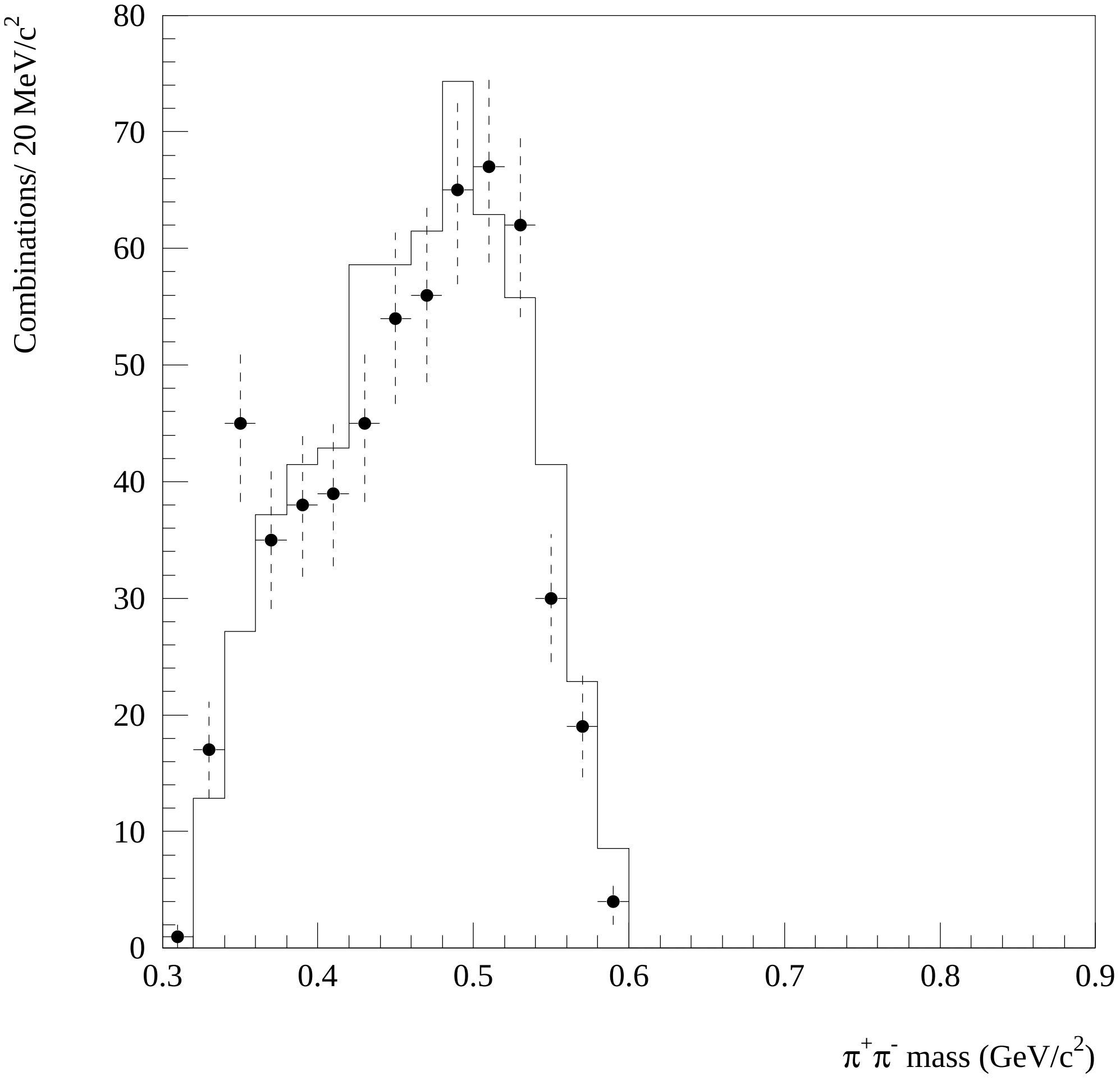

Figure 6.17. Invariant-mass spectrum of p $\pi^-$ and $\bar{p}\pi^+$ combinations that passed the criteria of being a $\Lambda^0$ decaying in the region inside of dipole magnet when both tracks are assigned pion masses. The solid line is the expected distribution obtained from Monte-Carlo and the dots are from data.

# CHAPTER 7

# THE $\Lambda_b \rightarrow J/\psi\ \Lambda^0$ DECAY CHANNEL

This chapter includes a description of the $\Lambda_b \rightarrow J/\psi\ \Lambda^0$ Monte-Carlo simulation, as well as the search method and the results from the search for $\Lambda_b \rightarrow J/\psi\ \Lambda^0$ events in our experiment.

## 7.1 The $\Lambda_b \rightarrow J/\psi\ \Lambda^0$ Monte-Carlo simulation

A Monte-Carlo simulation of $\Lambda_b \rightarrow J/\psi\ \Lambda^0$ events was used to tune the reconstruction program to search for such events in our data. It was also used to measure the acceptance of our detector and the efficiency of the reconstruction program for this type of events. The Monte-Carlo was a GEANT3-based full detector simulation [52].

The Monte-Carlo simulation generates a $b\bar{b}$ pair according to the next-to-leading order (NLO) calculations of Mango, Nason, and Rodolfi (MNR), and it includes the following parameters: the mass factorization scale $Q = m_b/2$, the $\Lambda_{QCD} = \Lambda_5 = 204$ MeV, and the MRS235 and SMRS parton distribution fuctions for the nucleon and pion, respectively [21]. The two-dimensional distribution of $x_F$ vs. $p_T^2$ for the generated b quarks is shown in Figure 7.1, and the projections on $x_F$ and $p_T^2$ are shown in Figure 7.2. The $\Lambda_b$ s are assumed to have the same momentum as their parent b quarks. They are assigned a mean lifetime of 1.07 $10^{-12}$ s, and a mass of 5.641 GeV/$c^2$. Both of these values are taken from the PDG book [12]. In each event, one of the $\Lambda_b$ s in chosen at random and forced to decay into J/$\psi$ $\Lambda^0$, with J/$\psi \rightarrow \mu^+\mu^-$ and $\Lambda^0 \rightarrow p\ \pi^-$, or the charge conjugate reactions in the case of a $\overline{\Lambda}_b$. It is assumed that the $\Lambda_b$ s are unpolaraized and that their decay is isotropic in the center-of-mass frame. The other b hadron was forced to decay into a kaon plus a randomly chosen number of pions ( up to a maximum of 5). With the remainder of the energy, hadrons in the underlying event are generated subject to energy, momentum and charge conservation, according to a longitudinal phase space. See reference [53] for a detailed description of the model.

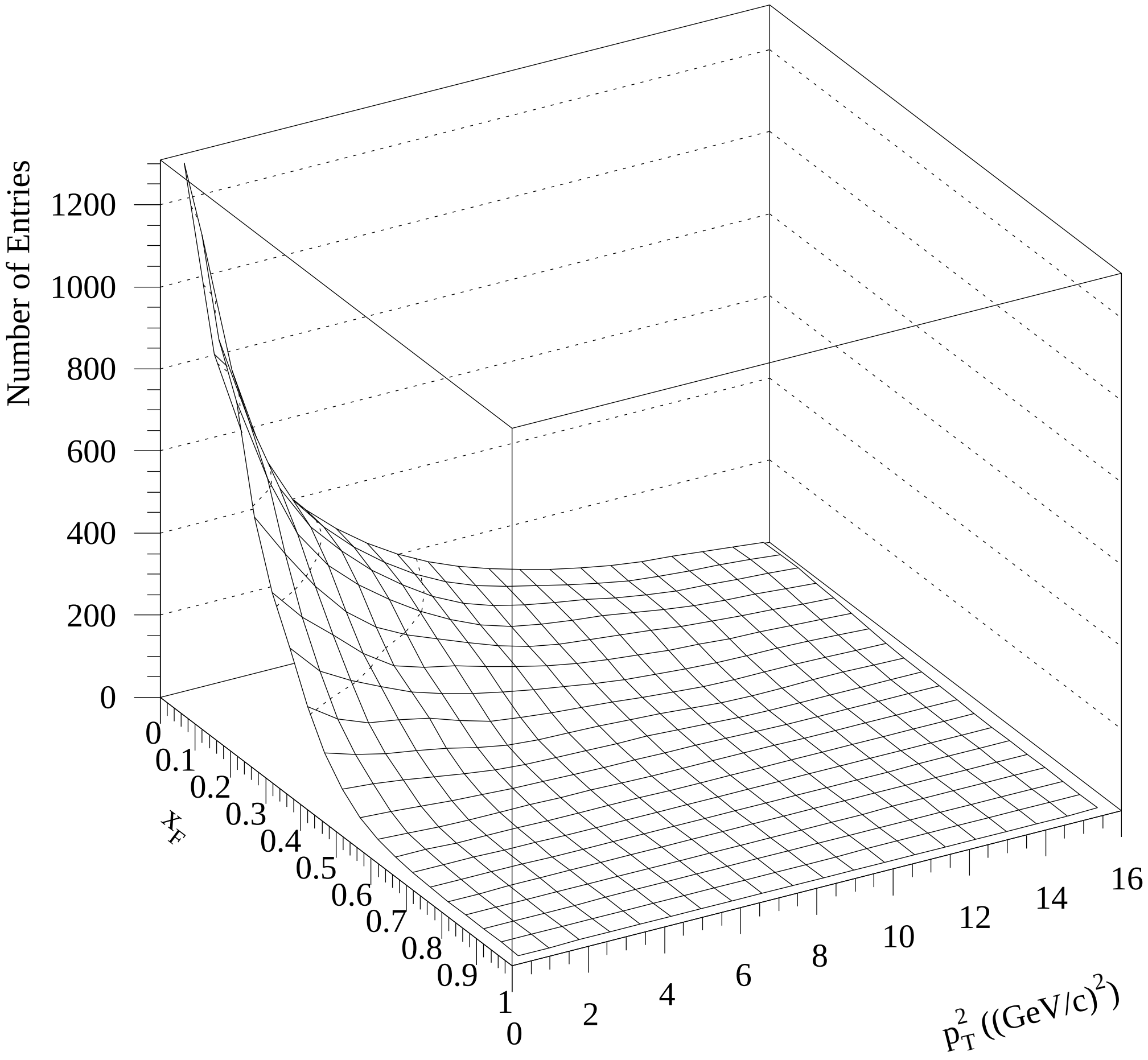

Figure 7.1. Generated $x_F$ vs. $p_T^2$ distribution of the b quark from Monte-Carlo events.

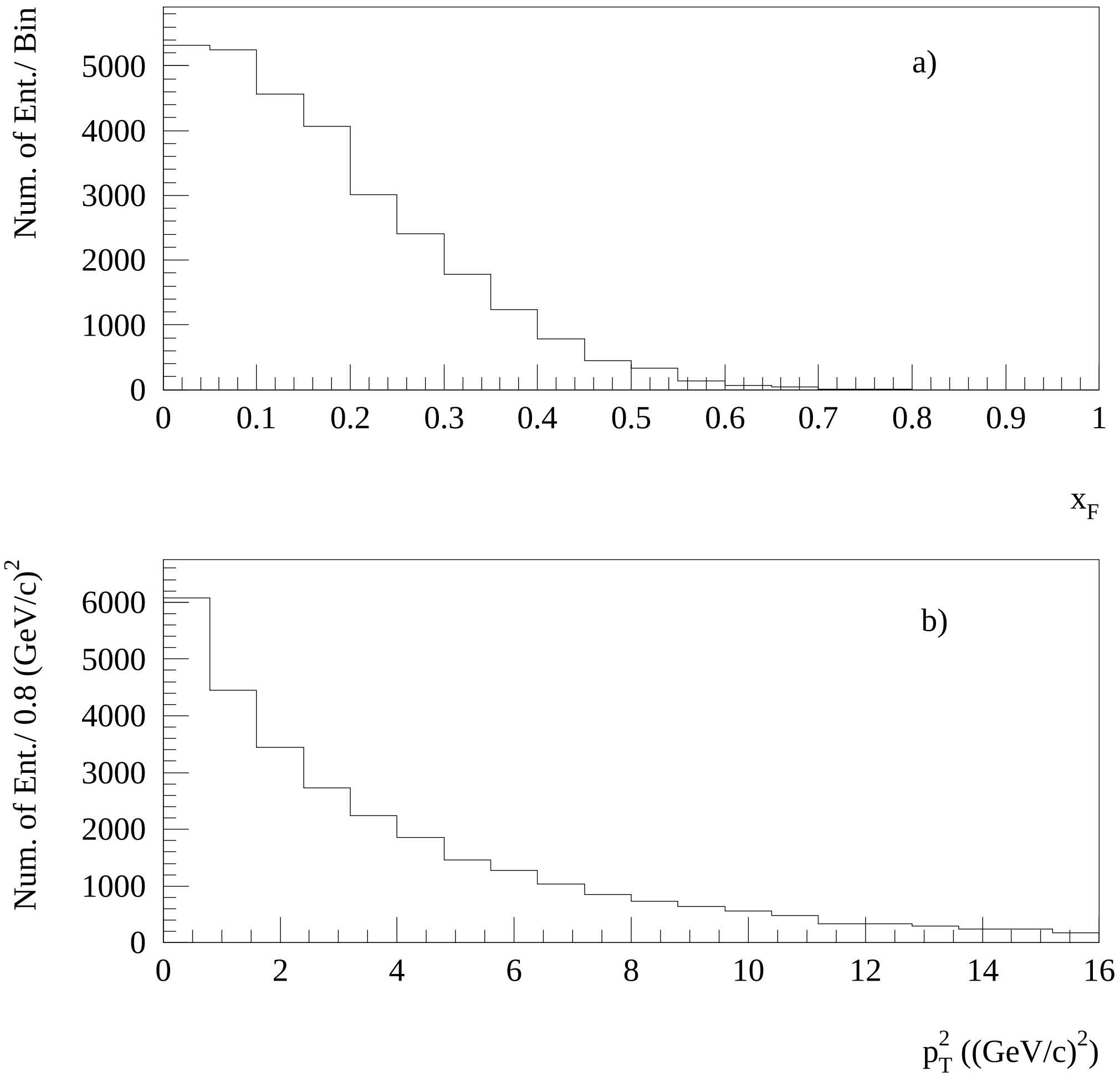

Figure 7.2. (a) Generated $x_F$ distribution of the b quark from Monte-Carlo events (projection from the two-dimesional distribution); and (b) generated $p_T^2$ distribution of the b quark from Monte-Carlo events (projection from the two-dimesional distribution).

All the events in the generated file were run through the detector geometry by GEANT. The resultant two-dimensional dimuon acceptance as a function of $x_F$ and $p_T^2$ from $\Lambda_b \rightarrow J/\psi\, \Lambda^0$ events is shown in Figure 7.3. Our detector only accepted dimuons in the $x_F$> 0 region. The integrated acceptance for these J/ψ s is 0.18 in the region $x_F > 0$.

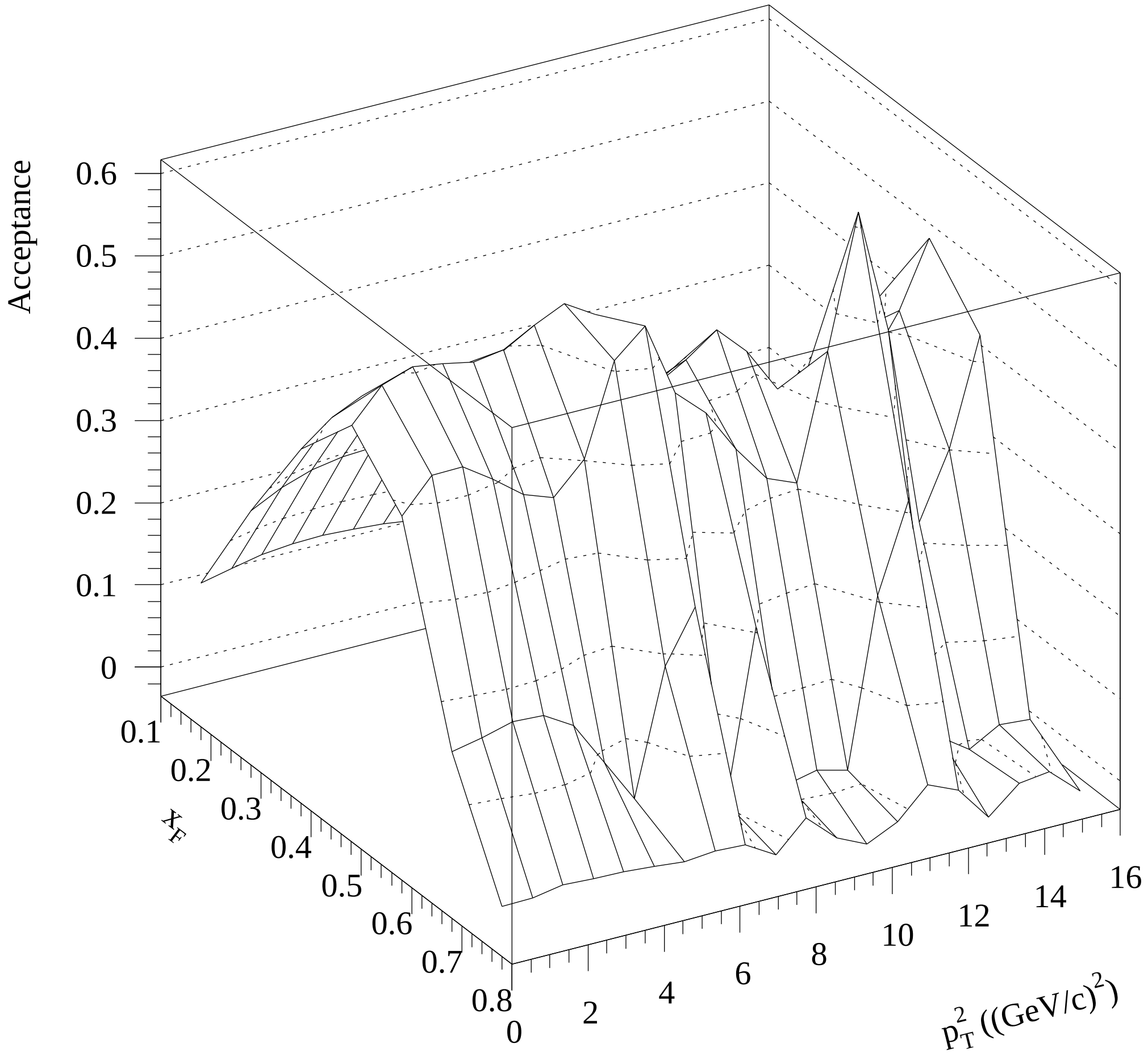

Figure 7.3. Two-dimensional dimuon acceptance as a function of $x_F$ and $p_T^2$ from $\Lambda_b \rightarrow J/\psi\, \Lambda^0$ events.

The events that had a dimuon accepted by the apparatus were digitized as spectrometer hits with appropriate detector noise and efficiency to produce a fake raw-data file. This file was then run through the same reconstruction and analysis program as that used for the data. Figure 7.4 shows a comparison of the track multiplicity in the Monte-Carlo with that in the data. Figure 7.5 and Figure 7.6 compare the momenta and $p_T$ of hadron tracks from Monte-Carlo and data.

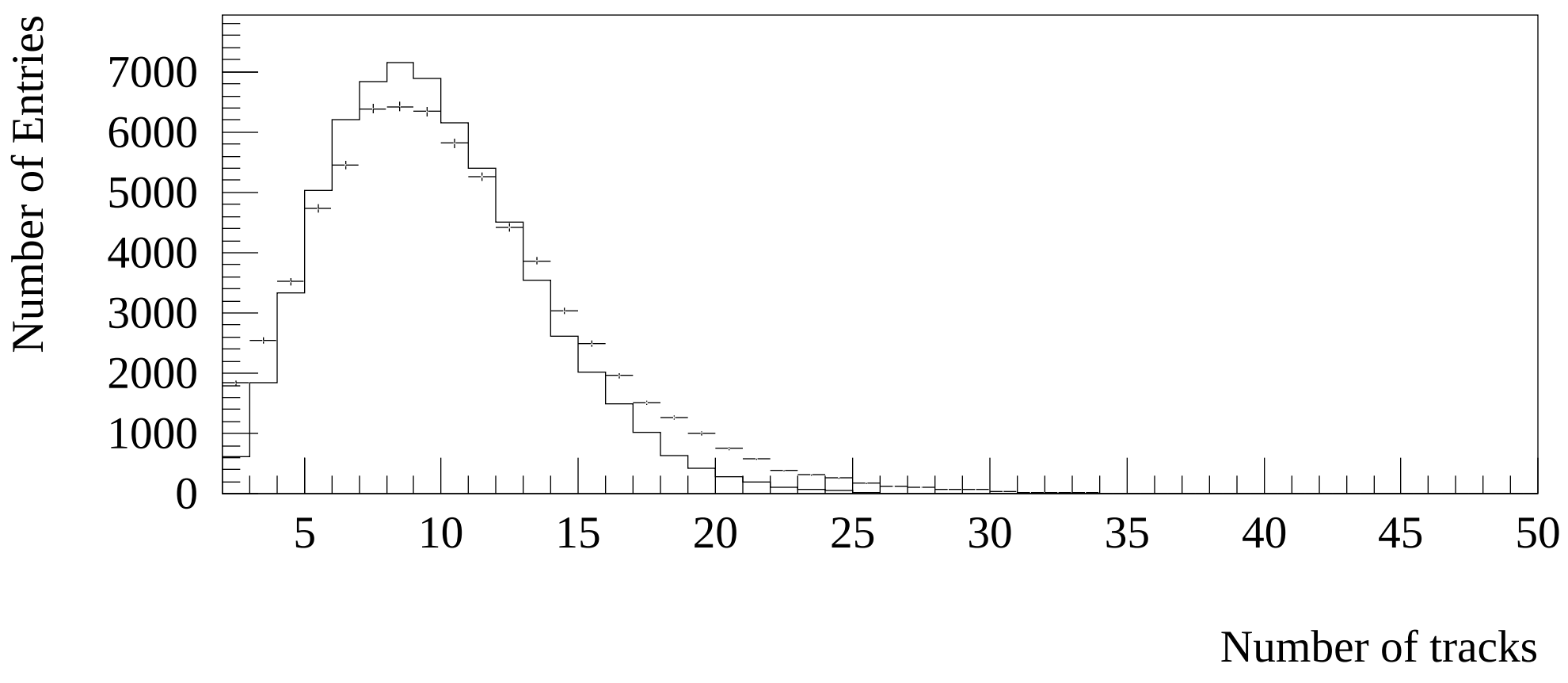

Figure 7.4. Monte-Carlo and data distributions of the reconstructed track multiplicity. The solid line represents the Monte-Carlo and the dots the data.

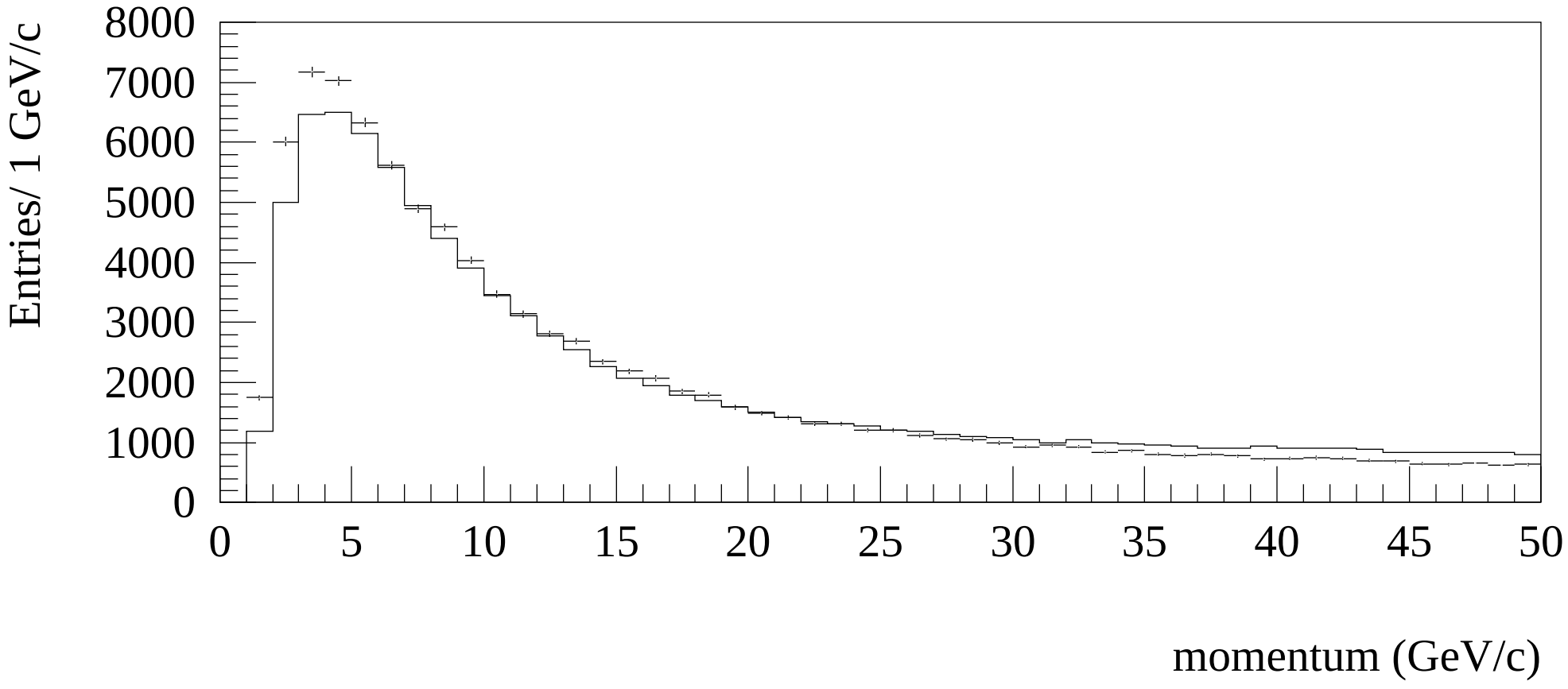

Figure 7.5. Monte-Carlo and data distributions of the momenta of charged hadron tracks. The solid line represents the Monte-Carlo and the dots the data.

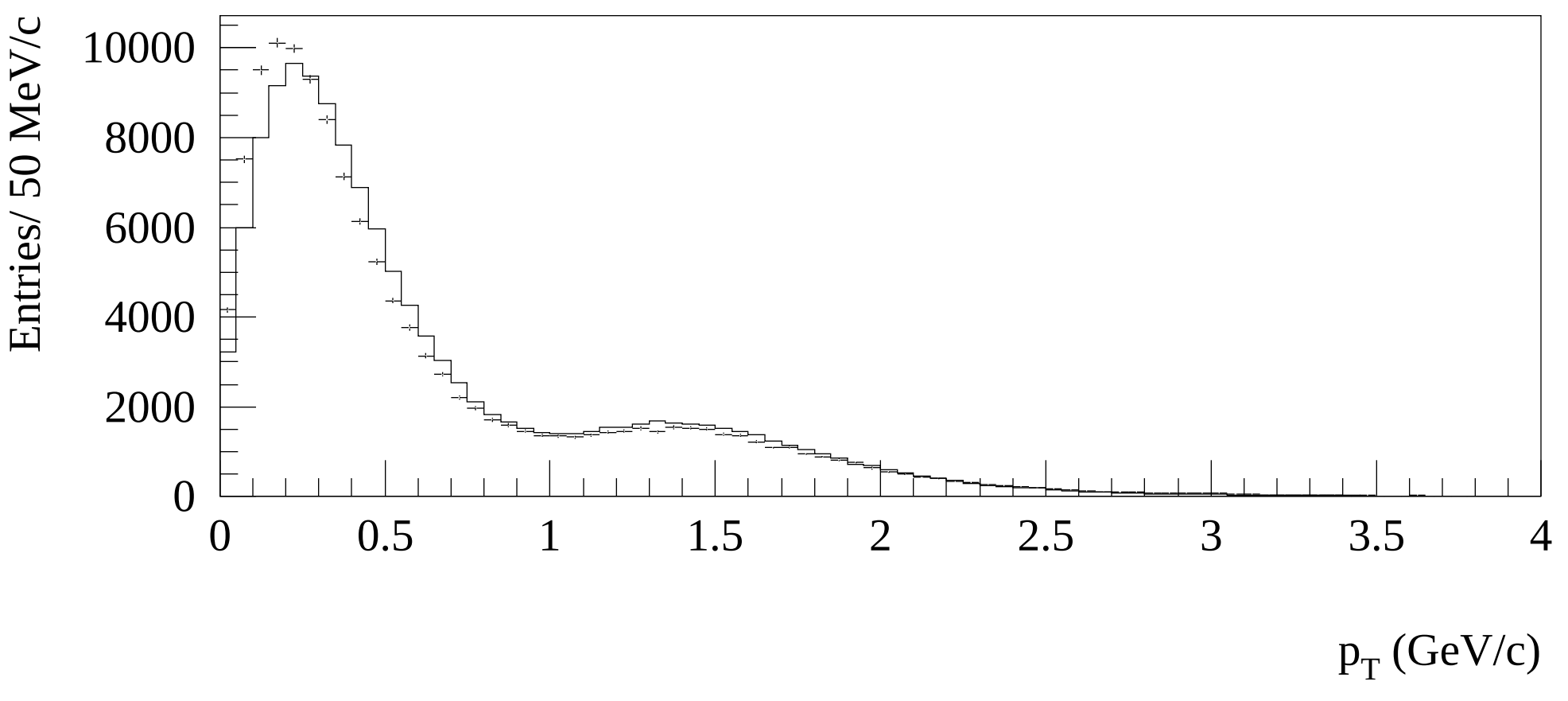

Figure 7.6. Monte-Carlo and data distributions of the $p_T$ of charged hadron tracks. The solid line represents the Monte-Carlo and the dots the data.

Figure 7.7 shows the reconstructed background subtracted signals for the J/ψ from Monte-Carlo and data. Figure 7.8, Figure 7.9, and Figure 7.10 show the reconstructed background subtracted signals of the $\Lambda^0$ (and $\overline{\Lambda}^0$) in the regions near the target/SSDs, upstream of the dipole magnet, and inside the dipole magnet, respectively. As seen, the resolution of the reconstructed J/ψ mass and $\Lambda^0$ (and $\overline{\Lambda}^0$) from Monte-Carlo are in good agreement with those from the data.

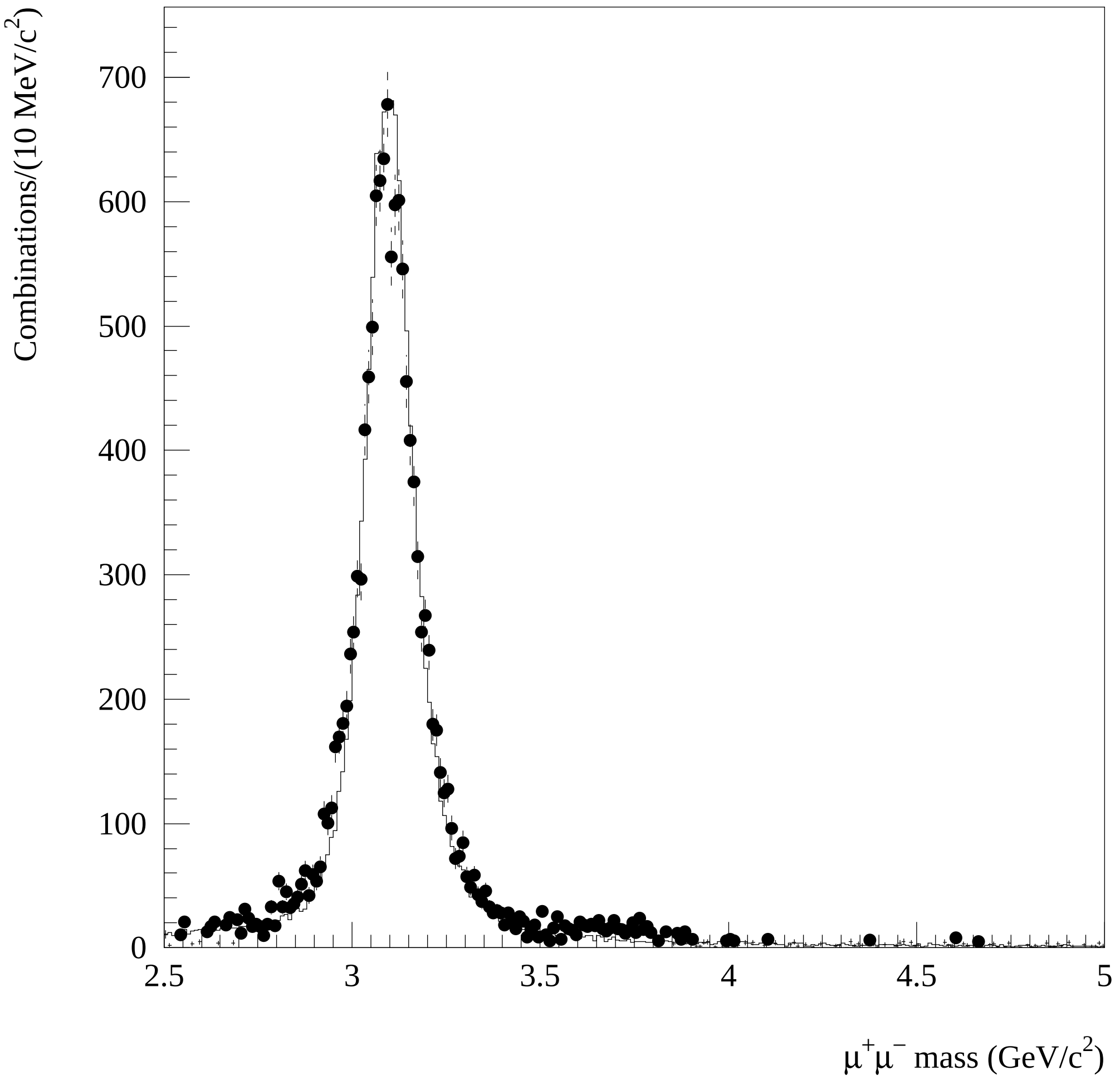

Figure 7.7. The reconstructed J/ψ signal after background subtraction, the solid line is from Monte-Carlo and the dots are from data. The Monte-Carlo is normalized to have the same number of events in the signal region as the data.

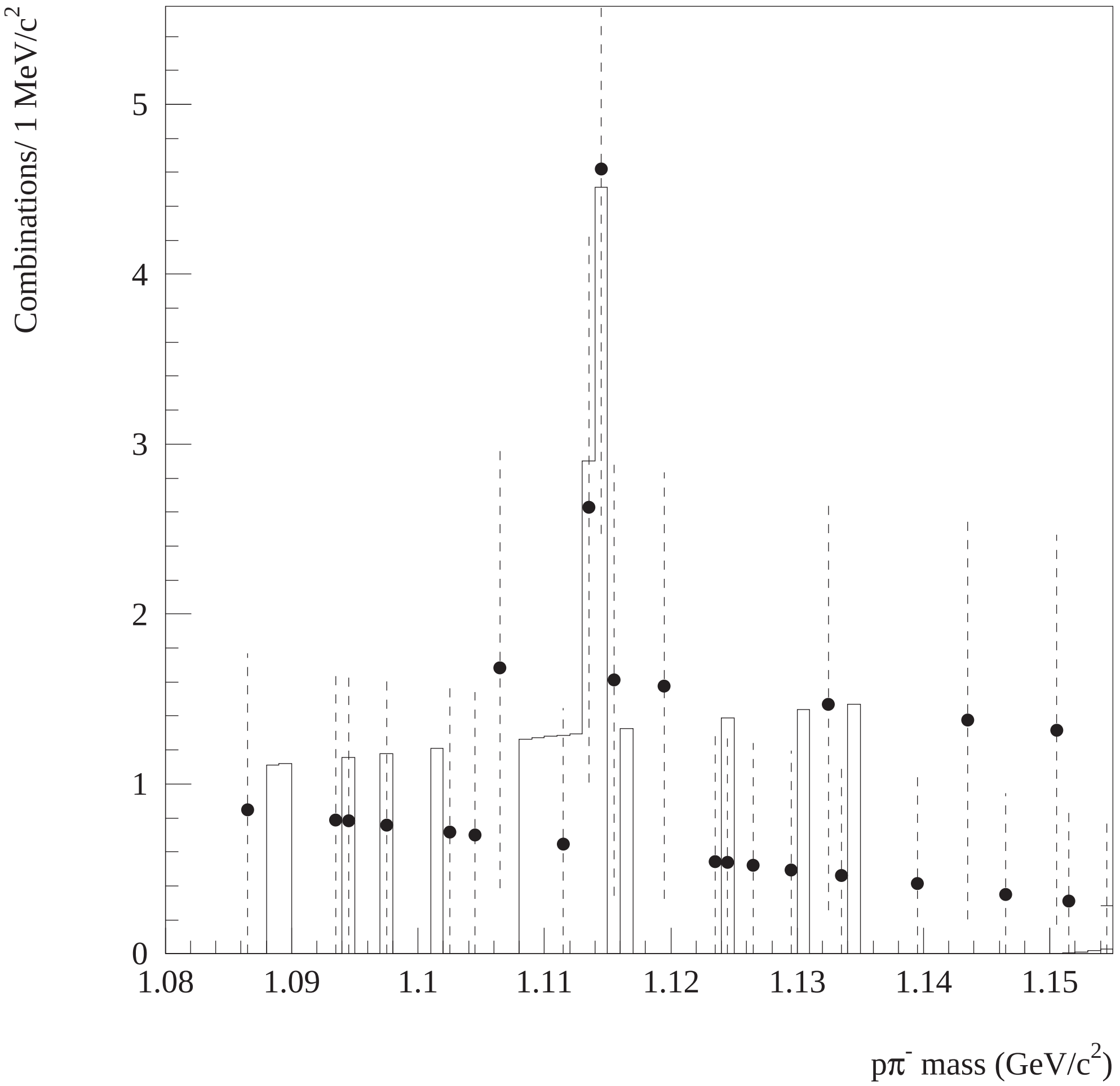

Figure 7.8. The reconstructed $\Lambda^0$ signal after background subtraction for those $\Lambda^0$ s that decayed in the target/SSD region. The solid line is from Monte-Carlo and the dots are from data. The Monte-Carlo is normalized to have the same number of events in the signal region as the data.

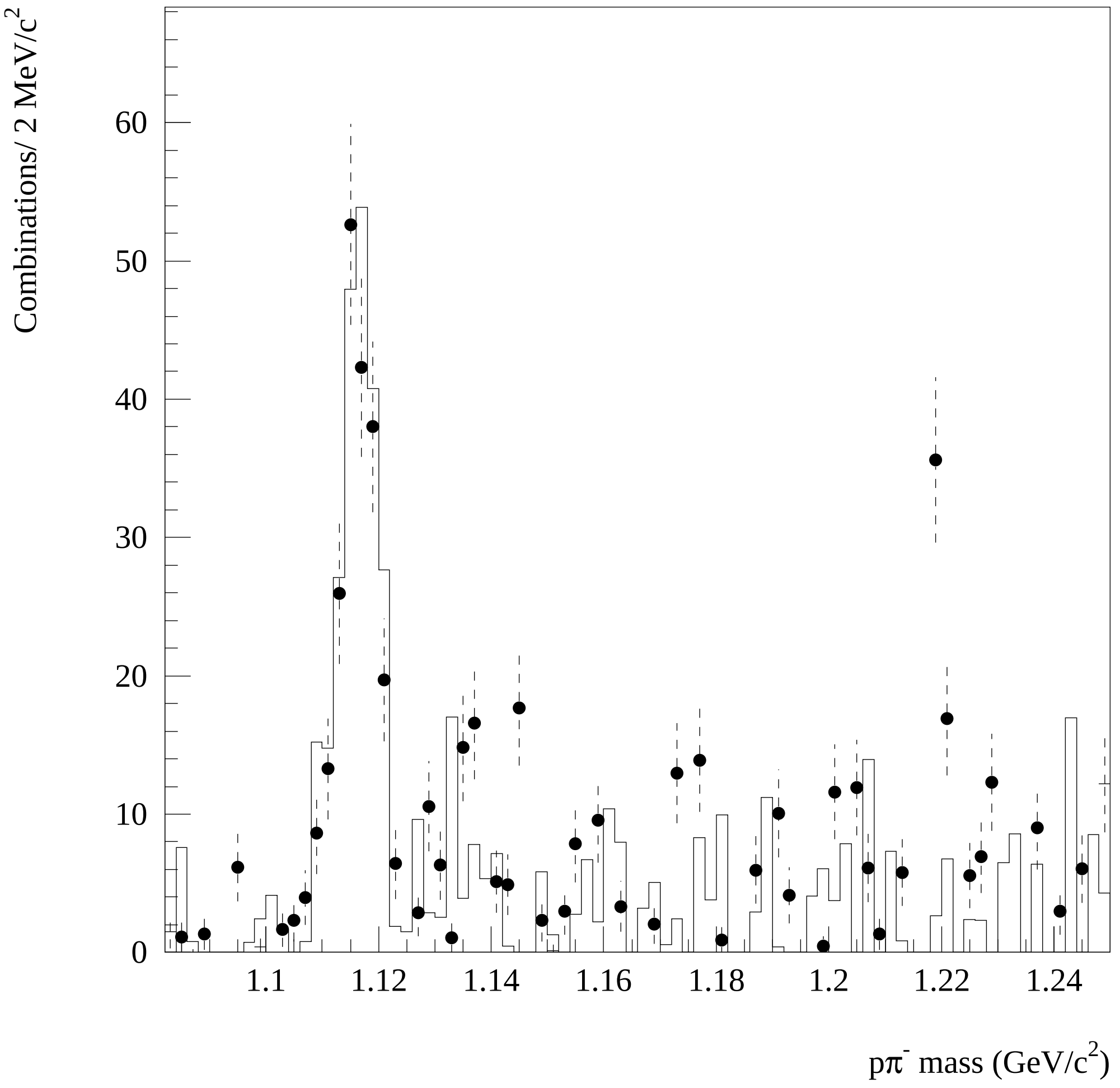

Figure 7.9. The reconstructed $\Lambda^0$ signal after background subtraction for those $\Lambda^0$s that decayed upstream of the dipole magnet. The solid line is from Monte-Carlo and the dots are from data. The Monte-Carlo is normalized to have the same number of events in the signal region as the data.

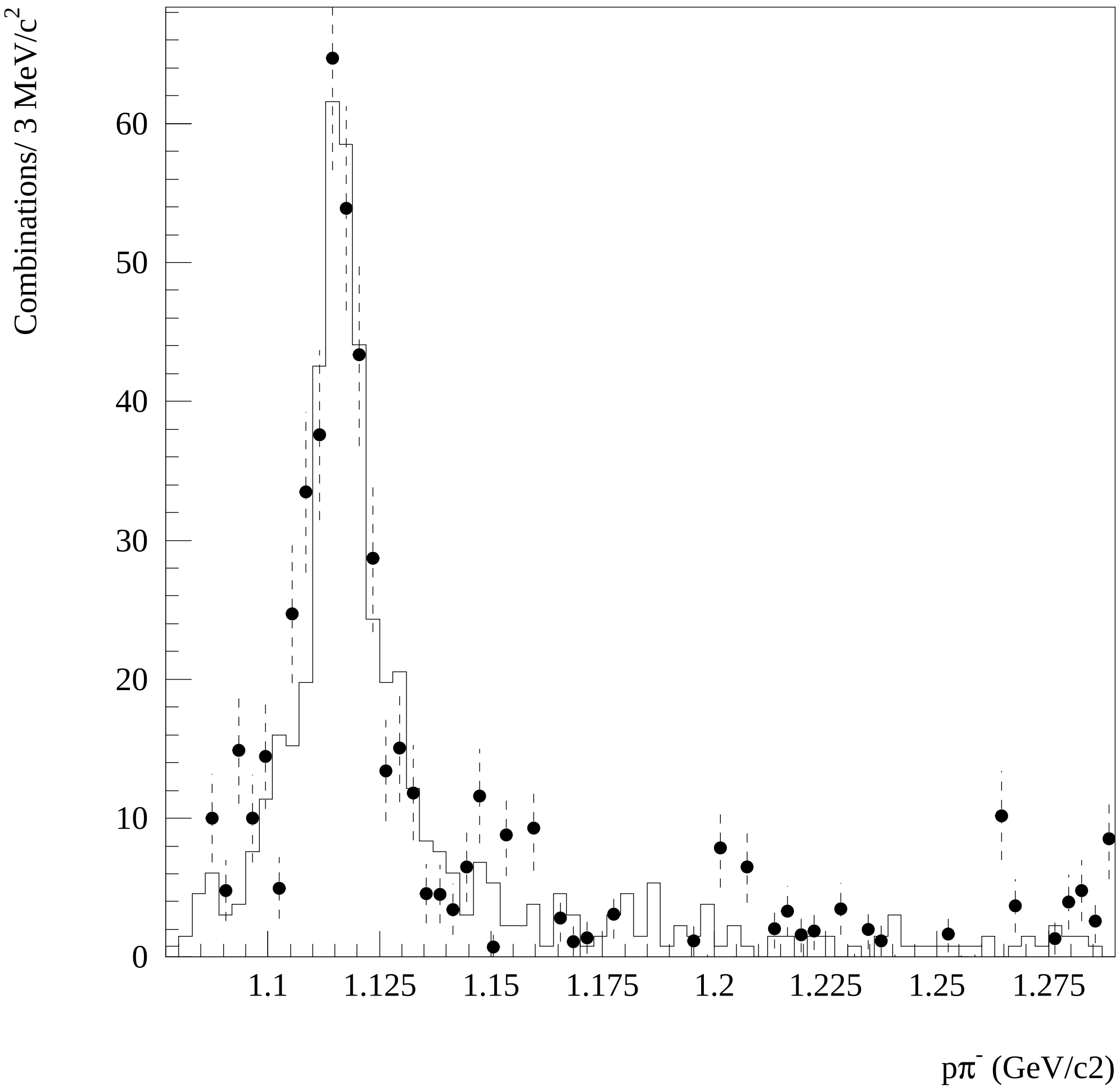

Figure 7.10. The reconstructed $\Lambda^0$ signal after background subtraction for those $\Lambda^0$s that decayed inside the dipole magnet. The solid line is from Monte-Carlo and the dots are from data. The Monte-Carlo is normalized to have the same number of events in the signal region as the data.

## 7.2 The $\Lambda_b \to J/\psi\ \Lambda^0$ search method and results

The topology of the $\Lambda_b \to J/\psi\ \Lambda^0$ decay is illustrated in Figure 7.11. Combinatorial background to the $\Lambda_b$ signal comes from three sources. The first one is from p $\pi^-$ and $\mu^+\mu^-$ backgrounds underneath the $\Lambda^0$ and J/ψ signal regions, respectively, (see Figure 5.1, Figure 6.3, Figure 6.7, and Figure 6.10). The second is from real J/ψ s combined with real $\Lambda^0$ s in the same event to give a fake $\Lambda_b$. The third is by wrongly assigning a proton mass to one of the pion tracks so that a $K^0_s \to \pi^+\ \pi^-$ could look like a $\Lambda^0 \to p\ \pi^-$ (or $\overline{\Lambda}^0 \to \overline{p}\ \pi^+$), since the experiment did not have particle identification capabilities.

To search for the $\Lambda_b \to J/\psi\ \Lambda^0$ decay, the (9 + 318 + 285 ) $\Lambda^0$ (and $\overline{\Lambda}^0$) candidates that survived the $K^0_s$ mass cut (see section 6.2.3) were required to have their transverse momenta greater than 0.8 GeV/c. This cut was imposed to filter out $\Lambda^0$s produced in the primary interactions. Since $\Lambda^0$s originated from $\Lambda_b$s have an average transverse momenta of 1.3 GeV/c whereas the directly produced $\Lambda^0$s had on average transverse momenta of 0.6 GeV/c (see Figure 7.12). This cut reduced the data to 2, 86, and 61 $\Lambda^0$ (and $\overline{\Lambda}^0$) candidates, in the target/SSDs, upstream of the dipole magnet, and inside the dipole magnet regions, respectively. According to Monte-Carlo this cut was 67 %, 75 % and 85 % efficient in the three regions, respectively.

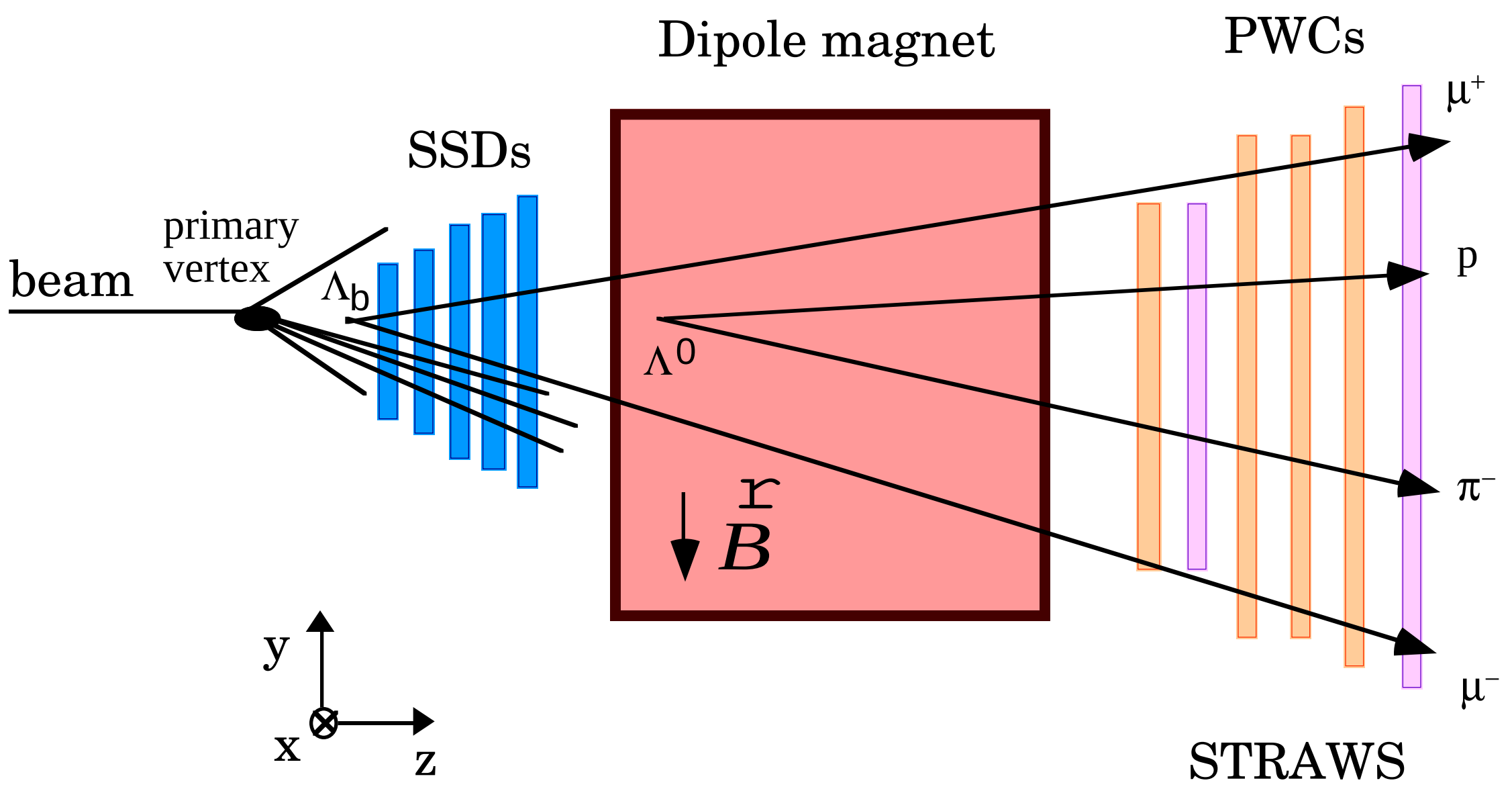

Figure 7.11. The topology of the $\Lambda_b \to J/\psi\ \Lambda^0$ decay.

The $\Lambda^0$ (and $\overline{\Lambda}^0$) candidates that survived the transverse monenta cut were then checked to see if a J/ψ candidate that passed the dimuon refit (see chapter 5) and did not have its vertex upstream of the primary vertex, was present in the event. This criteria reduced the $\Lambda^0$ (and $\overline{\Lambda}^0$) data sample to (0 + 7 + 8). Table IX shows a summary of the number of $\Lambda^0$ (and $\overline{\Lambda}^0$) candidates that survived each cut employed to reduce to this number of candidates.

Table IX Summary of the number of $\Lambda^0$ (and $\overline{\Lambda}^0$) candidates that survived each cut employed to select the sample used to search for the $\Lambda_b \rightarrow J/\psi\, \Lambda^0$ decay.

| Region | Initial sample | Podolanski-Armenteros Cut | $K^0_s$ mass Cut | $\Lambda^0$-$p_T$ Cut | J/ψ requirement |
|---|---|---|---|---|---|
| I | 10 | 10 | 9 | 2 | 0 |
| II | 479 | 471 | 318 | 86 | 7 |
| III | 582 | 578 | 285 | 61 | 8 |

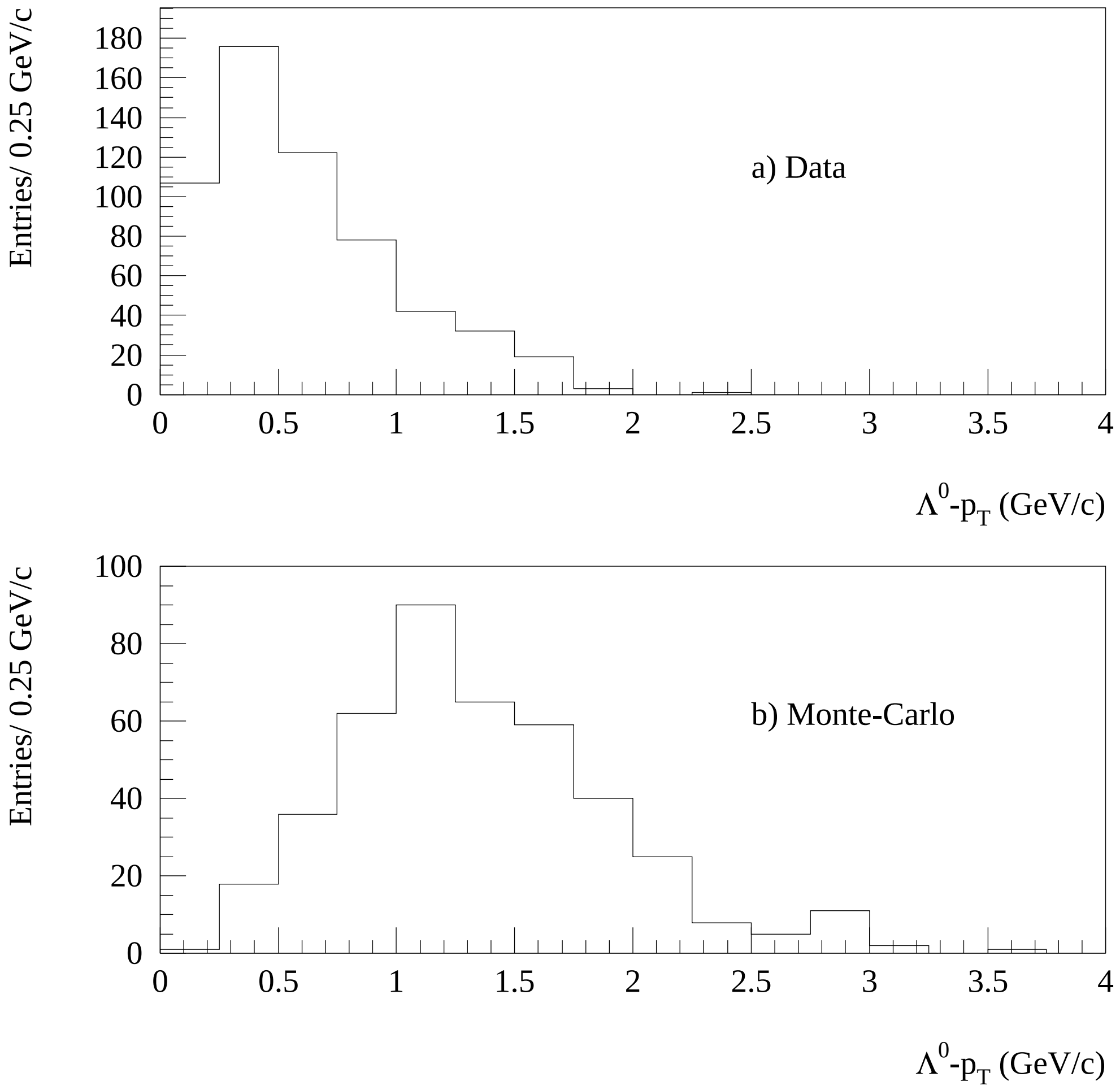

Figure 7.12. (a) $\Lambda^0$-$p_T$ distribution from data, and (b) from the $\Lambda_b \rightarrow J/\psi\ \Lambda^0$ Monte-Carlo.

The invariant-mass for the (0 + 7 + 8) $J/\psi\ \Lambda^0$ combinations summed over the three reconstruction regions is shown in Figure 7.13 (a), assuming the cuts described above. The distribution of invariant-mass of $\Lambda_b$ s reconstructed from Monte-Carlo simulation is shown in Figure 7.13 (b). This shows the $\Lambda_b$ signal region to be in the range 5.4 GeV/$c^2$ to 5.9 GeV/$c^2$. It is easy to see from Figure 7.13 (a) that there are two events in this mass range. Thus, there are two $\Lambda_b \rightarrow J/\psi\ \Lambda^0$ candidates. To estimate the shape of the combinatorial background to this distribution, a "wrong-frame" $J/\psi\ \Lambda^0$ background was produced by combining real $J/\psi$ s with real $\Lambda^0$ s (and $\overline{\Lambda}^0$ s) from different events and applying the same set of cuts as for the signal. The distribution of invariant mass from "wrong-frame" events is shown in Figure 7.13 (c). It is clear that the signal region is not in the same region where the wrong-frame background peaks.

In principle with the data shown so far one can also ask about the $B^0 \rightarrow J/\psi\ K^0$ decay. However, due to the smaller Br($B^0 \rightarrow J/\psi\ K^0$) (= (7.5 ± 2.1) x $10^{-4}$ [12]), compared to Br($\Lambda_b \rightarrow J/\psi\ \Lambda^0$) (= 1.8 x $10^{-2}$ [7]), and F($b \rightarrow K^0$ ) (≈ 0.2) compared to F($b \rightarrow \Lambda_b$ ) (≈ 0.1), and considering we can only observe $K^0{}_s$ and not $K^0{}_L$, the expected number of $B^0 \rightarrow J/\psi\ K^0$ events is about 1/24 of those expected from $\Lambda_b \rightarrow J/\psi\ \Lambda^0$ decays[1]. Thus, we do not expect to see this decay in our experiment.

[1] See chapter 8 for the expected number of $\Lambda_b \rightarrow J/\psi\ \Lambda^0$ decays.

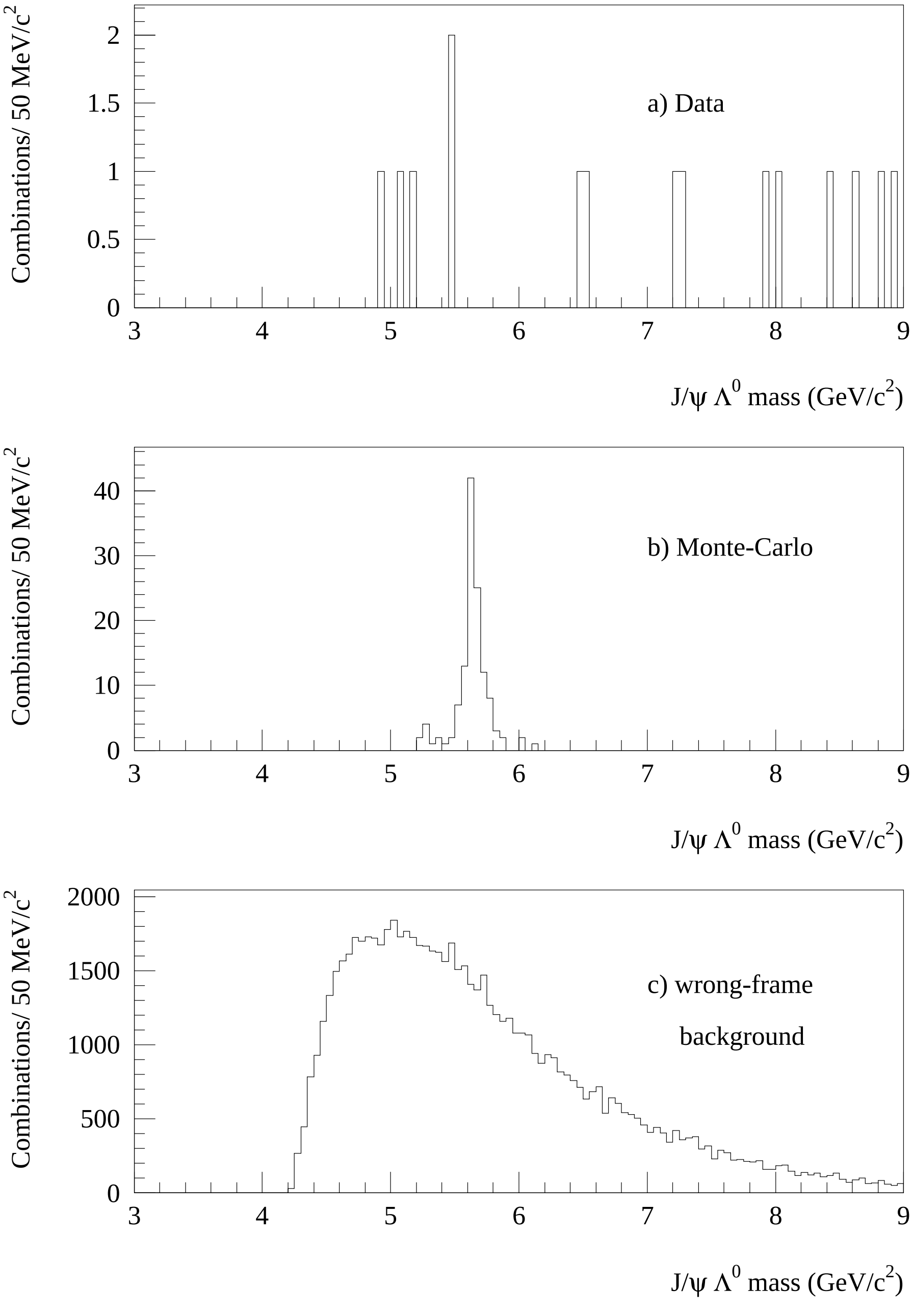

Figure 7.13. (a) The invariant mass for the J/ψ $\Lambda^0$ combinations; (b) the invariant mass for the J/ψ $\Lambda^0$ combinations from Monte-Carlo; (c) J/ψ $\Lambda^0$ wrong frame background.

# CHAPTER 8

# EVALUATION OF THE $\Lambda_b \rightarrow J/\psi\ \Lambda^0$ PRODUCTION RATE

This chapter gives a detailed explanation of the detection efficiencies for the J/ψs and $\Lambda^0$ s, and describes the procedure used to determine an upper limit to the $\Lambda_b \rightarrow J/\psi\ \Lambda^0$ production rate, $F(\Lambda_b) * Br(\Lambda_b \rightarrow J/\psi\ \Lambda^0)$. The upper limit is calculated based on two events found in the $\Lambda_b$ region (see section 7.2). It is then recalculated twice using a modified set of cuts.

## 8.1 Acceptances and efficiencies

The acceptances and efficiencies to reconstruct each signal were measured using the Monte-Carlo summation described in section 7.1. The acceptance for $J/\psi \rightarrow \mu^+ \mu^-$ with $x_F > 0$ from $\Lambda_b$ events is 18 % (see section 7.1). The reconstruction efficiency is 59 % and it breaks down in the following way:

i. The efficiency of linking each muon throughout the entire detector is 86 %.

ii. The invariant mass cut of 2.85 GeV/$c^2$ < $M_{\mu\mu}$ < 3.35 GeV/$c^2$ retains 93% of the J/ψs.

iii. The dimuon refit to the J/ψ explained in chapter 5 and appendix A is 92 % efficient.

iv. The requirement that the be vertex is not upstream of the primary vertex is 93 % efficient.

Thus, the reconstruction efficiency for $J/\psi \rightarrow \mu^+ \mu^-$ from $\Lambda_b$ events is $(0.86)^2$ (0.93) (0.92) (0.93) = 0.59. The "PSI" dimuon preselection described in section 4.1.1 is 71 % efficient for J/ψs from $\Lambda_b$ s. Therefore, the total detection efficiency for J/ψ s from $\Lambda_b$s is $\varepsilon^{J/\psi\,\mathrm{from}\,\Lambda_b}$ = (0.18)(0.59)(0.71) = 0.075.

The kinematical acceptance, that is the total fraction of the $\Lambda^0$s that decay in the three regions once the J/ψ has been accepted is 63%. The individual kinematical acceptance for each region is 3 %, 25 %, and 35 %, for the target/SSD, upstream of the dipole magnet, and inside the dipole magnet regions, respectively. The geometrical acceptance times reconstruction efficiency of $\Lambda^0$ s in each region is 8 %, 16%, and 29 %, respectively. After including the $K^0_s$ mass cut which is 90 %, 67% and 49% efficient ( see section 6.2.3), and the $\Lambda^0$-$p_T$ cut on $\Lambda^0$ (or $\overline{\Lambda}^0$) which is 67 %, 75 % and 85 % efficient (see section 7.2), the geometrical acceptance times reconstruction efficiency of $\Lambda^0$ s in each region reduces to 5 %, 8 %, and 12 %, respectively. Therefore, the weighted average total acceptance times the reconstruction efficiency for the $\Lambda^0$ once the J/ψ has been accepted is $\varepsilon^{\Lambda^0}$ = (0.03) (0.05) + (0.25) (0.08) + (0.35) (0.12) = 0.064.

## 8.2 Upper limit on the $\Lambda_b$ production rate

The observed number of events in the $\Lambda_b$ signal, $N_{sig}$, is related to the exclusive branching ratio Br( $\Lambda_b \rightarrow J/\psi\ \Lambda^0$ ) through

$$N_{sig} = 2\ \varepsilon^{J/\psi\,\mathrm{from}\,\Lambda_b}\ \varepsilon^{\Lambda^0} F(\Lambda_b)\ Br(\Lambda_b \rightarrow J/\psi\ \Lambda^0)\ \sigma_{b\bar{b}}\ Br(J/\psi \rightarrow \mu^+ \mu^-)\ Br(\Lambda^0 \rightarrow p\pi^-)\ L, \quad (8.1)$$

where $\varepsilon^{J/\psi\,\mathrm{from}\,\Lambda_b}$ is the total J/ψ detection efficiency for J/ψ s originating from $\Lambda_b$ s; $\varepsilon^{\Lambda^0}$ is the total $\Lambda^0$ detection efficiency; $F(\Lambda_b)$ is the production fraction (i.e., the probability of a b quark to hadronize into a $\Lambda_b$ or into another beauty baryon decaying to $\Lambda_b$; $Br(\Lambda_b \rightarrow J/\psi\ \Lambda^0)$ is the branching ratio for $\Lambda_b \rightarrow J/\psi\ \Lambda^0$; $\sigma_{b\bar{b}}$ is the $b\bar{b}$ production cross-section; $Br(J/\psi \rightarrow \mu^+ \mu^-)$ is the branching ratio for $J/\psi \rightarrow \mu^+ \mu^-$; $Br(\Lambda^0 \rightarrow p\pi^-)$ is the branching ratio for $\Lambda^0 \rightarrow p\pi^-$ and $L$ is the integrated luminosity. The prompt J/ψ signal from our beryllium target is used to compute $L$, this way there is a reduction of systematic uncertainties arising from efficiency corrections due to dead-time, muon halo, pretrigger and the dimuon trigger processor. $L$ is written as:

$$L = \frac{N_{J/\psi}}{\varepsilon^{J/\psi}\,\sigma_{J/\psi}\,Br\left(J/\psi \rightarrow \mu^+\mu^-\right)}, \quad (8.2)$$

where $\varepsilon^{J/\psi}$ is the total detection efficiency for prompt J/ψ s. In reference [11] we report, $N_{J/\psi}$ = 9,800 ± 130 on Be with $x_F$> 0.1. This corresponds to $\sigma_{J/\psi}\ Br(J/\psi \rightarrow \mu^+ \mu^-)/A$ = 9.2 ± 1.2 nb/nucleon for $x_F$> 0.1 on Be [11]. The acceptance for these J/ψ s with $x_F$ > 0.1 is 43%, the reconstruction efficiency is 64 % [11], and the “PSI” dimuon preselection described in section 4.1.1 is 79 %. Thus, the total detection efficiency for prompt J/ψs is $\varepsilon^{J/\psi} = (0.43)(0.64)(0.79) = 0.217$.

The total detection efficiencies $\varepsilon^{J/\psi \, \mathrm{from} \, \Lambda_b}$, $\varepsilon^{\Lambda^0}$, and $\varepsilon^{J/\psi}$ are listed in Table X. The $Br(J/\psi \rightarrow \mu^+ \mu^-) = 5.97$ %, and $Br(\Lambda^0 \rightarrow p\, \pi^-) = 63.9$ % [12]. Also E672/E706 measured $\sigma_{b\bar{b}}$ to be 47 ± 19 (*stat*) ± 14 (*sys*) nb/nucleon for $x_F > 0.0$ [11]. From equations 8.1 and 8.2:

$$F(\Lambda_b) * Br(\Lambda_b \rightarrow J/\psi\, \Lambda^0) = \frac{N_{sig}\, \varepsilon^{J/\psi}\, \sigma_{J/\psi}\, Br(J/\psi \rightarrow \mu^+ \mu^-)}{N_{J/\psi}\, 2\, \varepsilon^{J/\psi \, \mathrm{from} \, \Lambda_b}\, \varepsilon^{\Lambda^0}\, \sigma_{b\bar{b}}\, Br(J/\psi \rightarrow \mu^+ \mu^-)\, Br(\Lambda^0 \rightarrow p\, \pi^-)} \quad . \qquad (8.3)$$

To compute an upper limit to $F(\Lambda_b) * Br(\Lambda_b \rightarrow J/\psi\, \Lambda^0)$, the 2 events in the $\Lambda_b$ signal region (see section 7.2) are treated as signal with zero background. According to Poisson statistics the maximum number of signal events that the 2 events can statistically fluctuate up to is 5.3 events at the 90 % C.L[1]. Therefore

$$F(\Lambda_b) * Br(\Lambda_b \rightarrow J/\psi\, \Lambda^0) < 6.2 \times 10^{-2} \text{ at 90 \% C.L.}$$

This does not contradict the value measured by UA1 of $F(\Lambda_b) * Br(\Lambda_b \rightarrow J/\psi\, \Lambda^0) = 1.8 \pm 1.1 \times 10^{-3}$ [7]. Using the UA1 measurement and the set of cuts described in the previous chapters we expect to find 0.2 $\Lambda_b \rightarrow J/\psi\, \Lambda^0$ events. To better understand and measure the $F(\Lambda_b) * Br(\Lambda_b \rightarrow J/\psi\, \Lambda^0)$ upper limit, this number was recalculated twice under a modified set of cuts than those used above.

---

[1] See ref. [12] page 1279

Table X. The total detection efficiencies for the different particles, which are the products the kinematical and geometrical acceptance, reconstruction efficiencie and preselection efficiencie.

| | Acceptance (%) | Reconstruction efficiency (%) | Preselection efficiency (%) | Total detection efficiency (%) |
|---|---|---|---|---|
| $J/\psi$ s from $\Lambda_b$s | 18 | 59 | 71 | $\varepsilon^{J/\psi \text{ from } \Lambda_b} = 7.5$ |
| prompt $J/\psi$s | 43 | 64 | 79 | $\varepsilon^{J/\psi} = 21.7$ |
| $\Lambda^0$(and) $\overline{\Lambda}^0$ | 6.4[1] | | | $\varepsilon^{\Lambda^0} = 6.4$ |

The first set was made without including the $K^0_s$ mass cut mentioned in section 6.2.3. Without this cut the $\Lambda^0$ total detection efficiency $\varepsilon^{\Lambda^0}$ = (0.03) (0.05) + (0.25) (0.12) + (0.35) (0.25) = 0.119 (see section 8.1 for the definition of each factor). Consequently removing this cut, the $\Lambda^0$ total detection efficiency increases approximately by a factor of 2 and therefore increases the expected number of $\Lambda_b$ signal events to about 0.4 events, maximizing the signal. Without the $K^0_s$ mass cut, the $\Lambda^0$ (or $\overline{\Lambda}^0$) data sample contains 10, 471, and 578 candidates in the target/SSD, upstream of the dipole magnet, and inside the dipole magnet regions, respectively (see section 6.2.3 for all other cuts applied). Applying the cut $\Lambda^0$-$p_T$ to be greater than 0.8 GeV/c reduces the data sample to 2, 122, and 162 candidates in the three regions, respectively (see section 7.2 for the motivation of this cut). After searching each event in this data sample for a $J/\psi$ candidate that passed the dimuon refit (see chapter 5) and did not have its vertex upstream of the primary vertex, the $\Lambda^0$ (and $\overline{\Lambda}^0$) data sample reduced to (1 + 22 + 25) events. The invariant-mass for these 48 $J/\psi$ $\Lambda^0$ combinations is shown in Figure 8.1 with the solid line. The dashed line represents

1 This value is the weighted average acceptance times reconstruction efficiency for the $\Lambda^0$(and $\overline{\Lambda}^0$) in the three reconstruction regions combined, once the $J/\psi$ has been accepted.

wrong-frame background normalized to have the same number of events in the region between 6.5 GeV/$c^2$ and 9.0 GeV/$c^2$ as the data. Recall that the $\Lambda_b$ signal region is the mass interval between 5.4 GeV/$c^2$ and 5.9 GeV/$c^2$. There are 7 $\Lambda_b$ candidates in this signal region (number of events observed), and 12 background events in the wrong-frame background (number of expected background events). According to Poisson statistics, the maximum number of signal events that 7 observed events with an expected background of 12 events, can fluctuate up to is 4.2 events at the 90 % C.L[1]. Using equation 8.3 were every value is the same expect for $\varepsilon^{\Lambda^0}$ which is now 0.119, one finds that

$$F(\Lambda_b)\ Br(\Lambda_b \rightarrow J/\psi\ \Lambda^0) < 3.1 \times 10^{-2} \text{ at 90 \% C.L.}$$

The second set of cuts also does not included the $K^0_s$ mass cut, but in addition the J/ψs are required to originate from a vertex downstream of the primary vertex in the event. In this set of cuts the background is reduced to a minimum. The reconstruction of J/ψ s with $x_F > 0$ from $\Lambda_b$ events is now $(0.86)^2$ (0.93) (0.92) (0.60) = 0.38 (see section 8.1 for the definition of each factor). Thus, $\varepsilon^{J/\psi \text{ from } \Lambda_b}$ = (0.18)(0.38)(0.71) = 0.048. Giving an expected number of event of about 0.2. Requiring that the J/ψ s emerge from secondary vertices, reduces J/ψ $\Lambda^0$ data sample to (1 + 1 + 2) events. The requirements for a J/ψ to be originated from a secondary vertex are the same as those used in the J/ψ selection process for our $b\bar{b}$ cross-section calculation in [11]. (i) The primary vertex in the event is required to have at least 3 SSD-PWC linked tracks associated with the vertex. (ii) The J/ψ vertex is required to be at least 2.5 mm downstream of the primary vertex, and (iii) a significance greater than 3 is required for both longitudinal and transverse separations between the primary and secondary vertex, with the significance defined as the separation divided by the combined uncertainty. The invariant-mass for the (1 + 1 + 2) $J/\psi\ \Lambda^0$ combinations is shown in Figure 8.2. There are no events in the $\Lambda_b$ signal region. For the purpose of computing an upper limit to $F(\Lambda_b) * Br(\Lambda_b \rightarrow J/\psi\ \Lambda^0)$, the signal is zero and the is zero background. According to Poisson statistics, the maximum number of signal events that zero signal and zero background can statistically fluctuate up to is 2.3 events at the 90 % C.L [1]. Using equation 8.3 as before but with $\varepsilon^{J/\psi \text{ from } \Lambda_b}$ = 0.038, one gets

$$F(\Lambda_b) * Br(\Lambda_b \rightarrow J/\psi\ \Lambda^0) < 3.2 \times 10^{-2} \text{ at 90 \% C.L.}$$

---

1 See ref. [12] page 1279

1 See ref. [12] page 1279

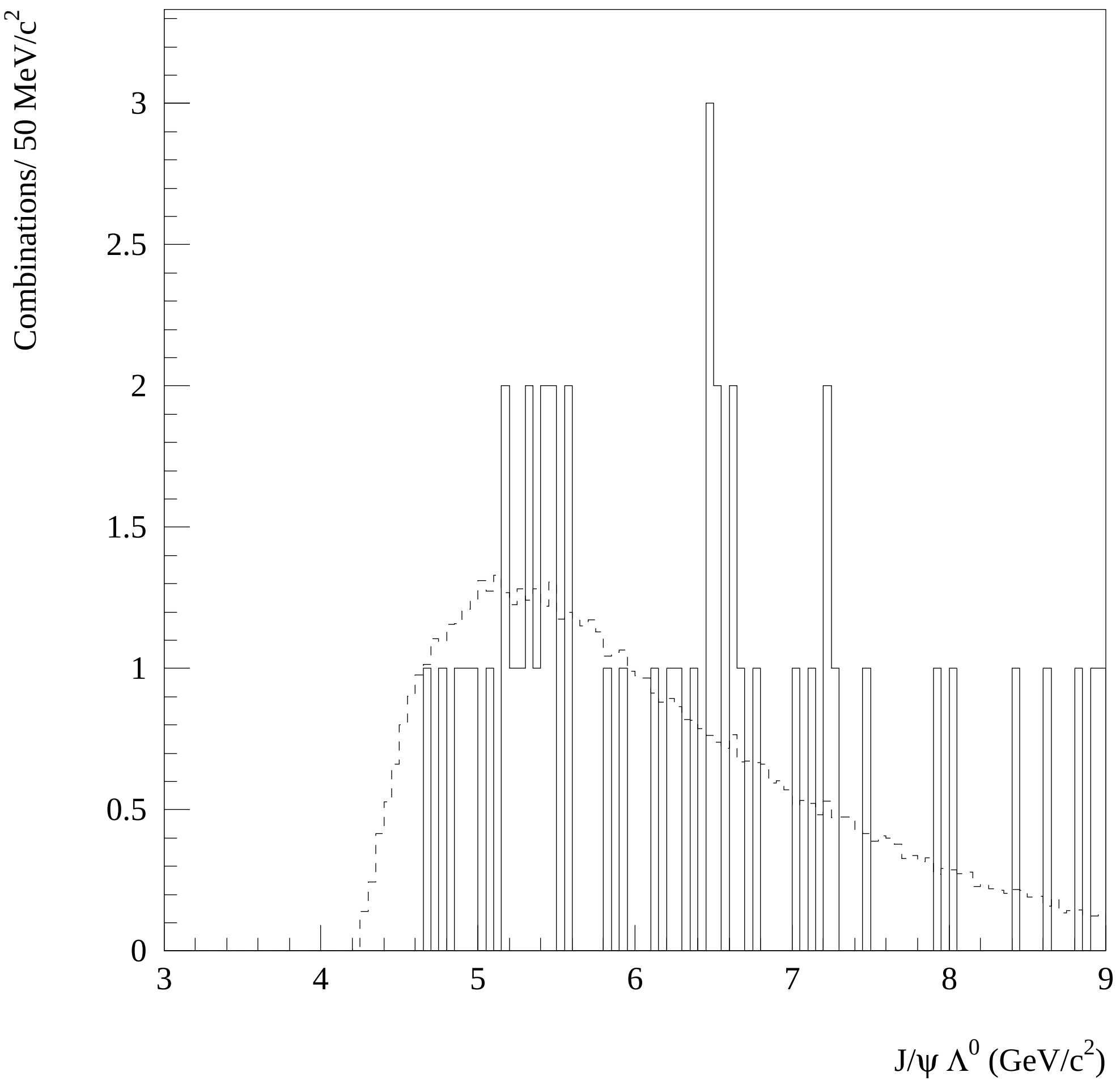

Figure 8.1. The invariant mass distribution (solid) for the $J/\psi\ \Lambda^0$ combinations, without using the $K^0_s$ mass cut, and the wrong-frame background (dashed).

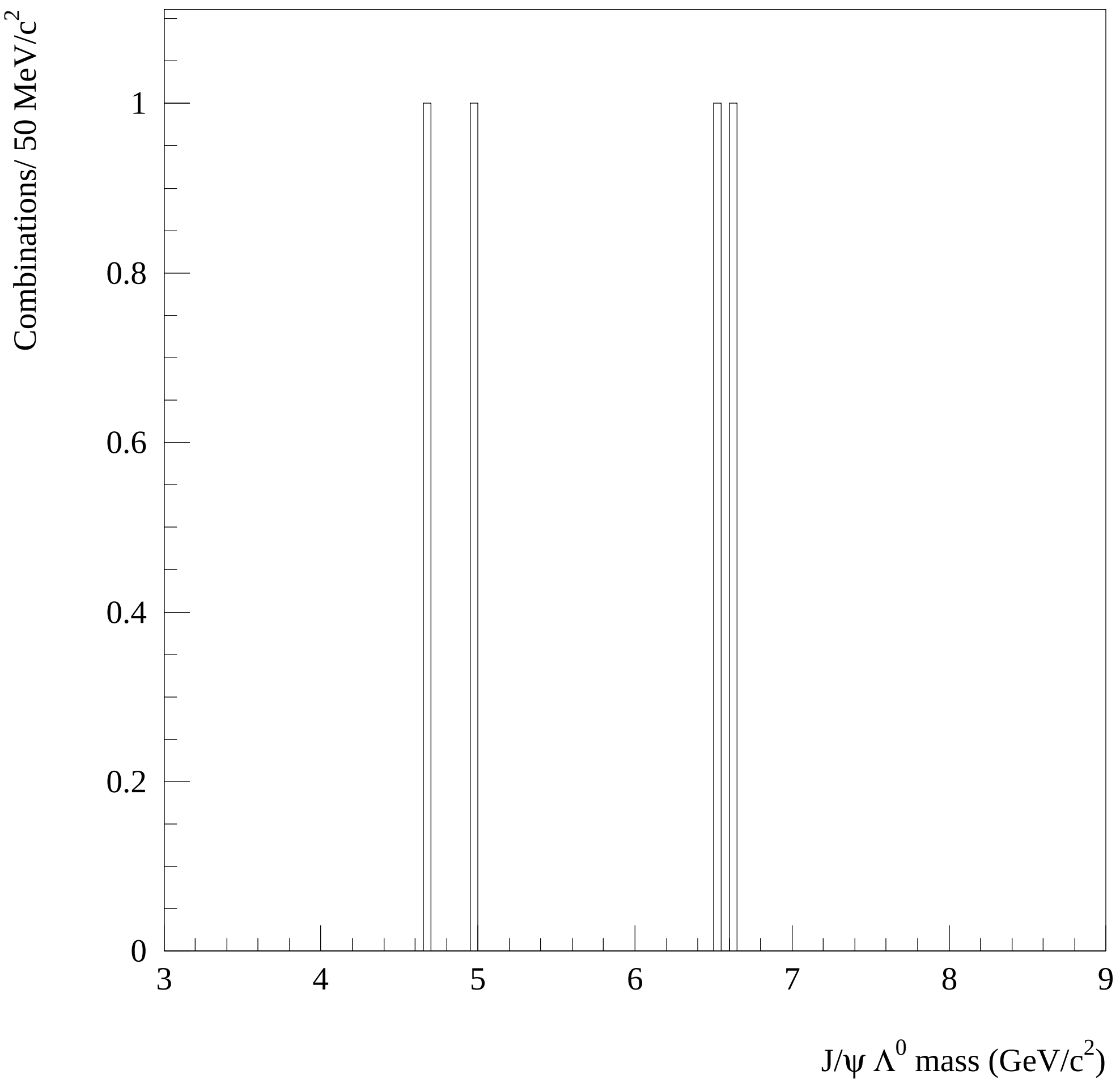

Figure 8.2. The invariant-mass distribution for the J/ψ $\Lambda^0$ combinations, without using the $K^0_s$ mass cut, and requiring that the J/ψ s originate from secondary vertices.

# CHAPTER 9

# CONCLUSIONS

An extensive analysis was performed on 8.0 $pb^{-1}$ of dimuon data produced in $\pi^{-}$ A collisions at 515 GeV/c to search for $\Lambda_b$ events in the decay channel $\Lambda_b \rightarrow J/\psi\ \Lambda^0$, with $J/\psi \rightarrow \mu^+ \mu^-$ and $\Lambda^0 \rightarrow p\ \pi^-$ (and for the conjugate reactions). A refit to the muon tracks from $J/\psi$ decays was performed with the mass constraint of two-body decay and with the constraint that both muon tracks intersect at a common point. The $\Lambda^0$ s were identified by their characteristic decay, giving the larger fraction of their momenta to the protons. Using an iterating algorithm, $\Lambda^0$ s were reconstructed in three regions of the E672/E706 spectrometer, near the target/SSDs, between the SSDs and the dipole magnet, and inside the dipole magnet. A total of 575 ± 35 $\Lambda^0$ (and $\overline{\Lambda}^0$) background subtracted events were reconstructed. A $K^0_s \rightarrow \pi^+ \pi^-$ signal was also reconstructed in all three regions using the same technique. The $K^0_s$ mass signal was used to cross-check the reconstruction algorithm and to clean the $\Lambda^0$ (and $\overline{\Lambda}^0$) data sample, giving (9 + 318 + 285) $\Lambda^0$ (and $\overline{\Lambda}^0$) candidates.

To search for the $\Lambda_b \rightarrow J/\psi\ \Lambda^0$ (and charge conjugate reaction), $J/\psi$ s that passed the muon refit were combined with clean $\Lambda^0$ s (or $\overline{\Lambda}^0$s) when they existed in the same event, giving a total of (0 + 7 + 8) $J/\psi\ \Lambda^0$ (and charge conjugate) combinations. The results show 2 events in the $\Lambda_b$ mass region. Considering the two $\Lambda_b$ events as signal with zero background, an upper limit to $F(\Lambda_b) * Br(\Lambda_b \rightarrow J/\psi\ \Lambda^0)$ was found to be less than 6.2 $x10^{-2}$ at 90 % C.L. An upper limit was also calculated, without using the $K^0_s$ mass cut for the $\Lambda^0$ s (and $\overline{\Lambda}^0$s), then, also requiring that the $J/\psi$ s originate from secondary vertices, giving that $F(\Lambda_b) * Br(\Lambda_b \rightarrow J/\psi\ \Lambda^0) < 3.1\ x10^{-2}$ at 90 % C.L and $F(\Lambda_b) * Br(\Lambda_b \rightarrow J/\psi\ \Lambda^0) < 3.2\ x10^{-2}$ at 90 % C.L, respectively. The upper limits should be compared to $(1.8 \pm 1.1)\ x\ 10^{-3}$ measured by UA1 [7].

This study was performed using a Monte-Carlo simulation for $b\overline{b}$ production from next-to-leading order (NLO) calculations of Mango, Nason, and Rodolfi (MNR), which included the mass factorization scale $Q = m_b/2$, the $\Lambda_{QCD} = \Lambda_5 = 204$ MeV, and the MRS235 and SMRS parton distribution functions for the nucleon and pion, respectively [21]. In addition It was assumed that the $\Lambda_b$ s are unpolaraized, and no fragmentation was included.

In conclusion E672 did not have the sufficient sensitivity to contradict the UA1 measurement, nor give a lower limit than the ones established by ALEPH ($< 0.4\ x\ 10^{-3}$ at 90 % C.L.), OPAL ($< 1.5\ x10^{-3}$90 % C.L.), and CDF ($< 0.5\ x\ 10^{-3}$ at 90 % C.L.), (see Refs. [8] and [9]). However, ALEPH and OPAL both used $e^+e^-$ interactions, UA1 and CDF used $p\overline{p}$ interactions , and CDF had a limited sensitivity to reconstruct $\Lambda^0$ s (and $\overline{\Lambda}^0$s) that have a $\pi^-$ (and $\pi^+$) with $p_T < 0.4$ GeV/c [9]. Here we used $\pi^-$ A collisions and reconstructed $\Lambda^0$ s (and $\overline{\Lambda}^0$s) with the $p_T$ of the $\pi^-$ (and $\pi^+$) between 0 GeV/c and 0.4

GeV/c. Thus, our search for the $\Lambda_b \rightarrow J/\psi\ \Lambda^0$ (and charge conjugate reaction) in complementary to that of the others mentioned above.

# APPENDIX A

## FIT TO THE J/$\psi$

The muons that form the decaying J/$\psi$ are fitted using a Least-Squares method (explained in detail below) with the constraint that the two muons intersect at a common point in space and that the invariant mass of the combined pair is equal to 3.097 GeV/$c^2$ [1]. The $\chi^2$ of the fit is used to reduce the combinatorial background when there are more than two muon candidates in a given event. The fit improves the resolution of the measured momenta of the muons by a factor of 2 and vertex position by 15%.

### General least-squares estimation with constraints

The mathematical formulation of the iterative procedure will be derived without making any reference to any special physical problem.

[1] This is the value published by [12].

## The iteration procedure

Let $\vec{M} = \{m_1, m_2, ..., m_N\}$ be a vector of N measurable variables, which have initial measurements $\vec{M}^0 = \{m_1^0, m_2^0, ..., m_N^0\}$, with errors contained in the covariance matrix $E(\vec{M}^o)$. In addition, let $\vec{U} = \{u_1, u_2, ..., u_J\}$ be a set of J unmeasurable variables. The $\vec{M}$ variables and the $\vec{U}$ variables are related and have to satisfy a set of K constraint equations

$$f_k(m_1, m_2, ..., m_N, u_1, u_2, ..., u_J) = 0 \;, \quad k = 1, 2, ..., K \;.$$

According to the Least-Squares Principle, the best estimates of the $\vec{M}$ and $\vec{U}$ variables are those for which

$$\chi^2\left(\vec{M}, \vec{M}^o\right) = \left(\vec{M}^o - \vec{M}\right)^T E^{-1}\left(\vec{M}^o\right)\left(\vec{M}^o - \vec{M}\right) = \text{minimum} \text{ , and} \qquad \text{(A.1)}$$

$$\vec{f}\left(\vec{M}, \vec{U}\right) = \vec{0},$$

where the superscript T indicates the transpose of the matrix.

The problem will now be solved using the Lagrange multiplier method. Introducing the K component vector $\vec{\lambda} = \{\lambda_1, \lambda_2, ..., \lambda_K\}$ of Lagrangian multipliers, the problem can be rephrased by requiring

$$\chi^2\left(\vec{M}, \vec{U}, \vec{\lambda}\right) = \left(\vec{M}^o - \vec{M}\right)^T E^{-1}\left(\vec{M}^o\right)\left(\vec{M}^o - \vec{M}\right) + 2\vec{\lambda}^T \vec{f}\left(\vec{M}, \vec{U}\right) = \text{minimum} \;. \qquad \text{(A.2)}$$

There are now a total N+J+K variables. The values of $\vec{M}, \vec{U}$ and $\vec{\lambda}$ that make the $\chi^2$ minimum, must also satisfy the following set of equations

$$\begin{aligned} \nabla_M \chi^2 &= -2\,E^{-1}\left(\vec{M}^o\right)\left(\vec{M}^o - \vec{M}\right) + 2\,F_M^T\,\vec{\lambda} = \vec{0}, && \text{(N equations)} \\ \nabla_U \chi^2 &= 2\,F_U^T\,\vec{\lambda} = \vec{0}, && \text{(J equations)} \\ \nabla_\lambda \chi^2 &= 2\,\vec{f}\left(\vec{M}, \vec{U}\right) = \vec{0}\,, && \text{(K equations)} \end{aligned} \qquad \text{(A.3)}$$

where the matrices $F_M, F_U$ of dimensions KxN and KxJ respectively are defined by

$$\left(F_M\right)_{k\,i} \equiv \frac{\partial f_k}{\partial m_i}\,,\ \left(F_U\right)_{k\,j} \equiv \frac{\partial f_k}{\partial u_j}.$$

Simplifying equations (A.3) gives

$$-E^{-1}\left(\vec{M}^0\right)\left(\vec{M}^0 - \vec{M}\right) + F_M^T\vec{\lambda} = \vec{0}, \qquad \text{(A.4a)}$$

$$F_U^T\,\vec{\lambda} = \vec{0}, \qquad \text{(A.4b)}$$

$$\vec{f}\left(\vec{M}, \vec{U}\right) = \vec{0}\,. \qquad \text{(A.4c)}$$

The solution to the set of equations (A.4) for the N+J+K variables must be, in general, found by iterations producing successively better approximations. To do this, suppose that the n-th iteration has been performed and it is still necessary to find a better solution. For the n-th iteration the approximate solution is given by the values of $\vec{M}^n, \vec{U}^n, \vec{\lambda}^n$, corresponding to the function value $(\chi^2)^n$. By performing a Taylor expansion of the constraint equations (A.4c) about the point $(\vec{M}^n, \vec{U}^n)$ one obtains:

$$f_k^n + \sum_{i=1}^{N} \left( \frac{\partial f_k}{\partial m_i} \right)^n (m_i^{n+1} - m_i^n) - \sum_{j=1}^{J} \left( \frac{\partial f_k}{\partial u_i} \right)^n (u_j^{n+1} - u_j^n) + \ldots = 0 \ , \ \ k = 1, 2, \ldots, K.$$

Neglecting all terms of second or higher order, this equation can be written as

$$\vec{f}^n + F_M^n \left( \vec{M}^{n+1} - \vec{M}^n \right) + F_U^n \left( \vec{U}^{n+1} - \vec{U}^n \right) = \vec{0} \ , \tag{A.5}$$

where all superscripts n indicate that the $\vec{f}^n, F_M^n, F_U^n$ are to be evaluated at the point $(\vec{M}^n, \vec{U}^n)$. The other two equations in A.4, a and b, are then,

$$E^{-1} \left( \vec{M}^{n+1} - \vec{M}^0 \right) + \left( F_M^T \right)^n \vec{\lambda}^{n+1} = \vec{0}, \tag{A.6a}$$

$$\left( F_U^T \right)^n \vec{\lambda}^{n+1} = \vec{0} \ . \tag{A.6b}$$

These two equations, together with the expanded constraint equations (A.5) make it possible to express all variables of the (n+1)-th iteration in terms of quantities of the preceding iteration. These solutions are

$$\begin{aligned} \vec{U}^{n+1} &= \vec{U}^n - \left( F_U^T \ S^{-1} \ F_U \right)^{-1} \ F_U^T \ S^{-1} \ \vec{r} \ , \\ \vec{\lambda}^{n+1} &= S^{-1} \left[ \vec{r} + F_U \left( \vec{U}^{n+1} - \vec{U}^n \right) \right], \\ \vec{M}^{n+1} &= \vec{M}^0 - E \ F_M^T \ \vec{\lambda}^{n+1} \ , \end{aligned} \tag{A.7}$$

where

$$\vec{r} = \vec{f}^{\,n} + F_M^n\left(\vec{M}^0 - \vec{M}^n\right),$$
$$S \equiv F_M^n\ E\ \left(F_M^T\right)^n .$$

In equations (A.7) the matrices $F_M, F_U, S$, and the vector $\vec{r}$ are evaluated at the point $(\vec{M}^n, \vec{U}^n)$. The value of $\vec{U}^0$ can be found from the set of the constraint equation, setting $\vec{f}\left(\vec{M}^0, \vec{U}^0\right) = 0$ and solving for $\vec{U}^0$. With the new values for $\vec{M}^{n+1}, \vec{U}^{n+1}$ and $\vec{\lambda}^{n+1}$ one can calculate the value of the function $\left(\chi^2\right)^{n+1}$ for the (n+1)-th iteration and compare it with the previous value of $\left(\chi^2\right)^n$, by using

$$\left(\chi^2\right)^{n+1} = \left(\vec{\lambda}^{n+1}\right)^T S\ \vec{\lambda}^{n+1} + 2\left(\vec{\lambda}^{n+1}\right)^T \vec{f}^{\,n+1}, \qquad \text{(A.8)}$$

where the matrix S is evaluated for the n-iteration. Once a satisfactory solution is found the iterative procedure is stopped. This is when the value of the $\chi^2$ and the values of the $\vec{M}$ and $\vec{U}$ vectors are converging [54].

## Kinematical analysis of the $J/\psi \to \mu^+\mu^-$ decay

The following section applies the previous formulation to the $J/\psi \to \mu^+\mu^-$ decays as seen in the MWEST spectrometer. Let us assume that the two muon tracks have been measured and that the track reconstruction program has provided for each muon track a first approximation for the kinematical variables: 1/p ( inverse momentum); $m_x$ (track slope in the x-z plane ); $m_y$ (track slope in the y-z plane ); $b_x$ (intercept of the track with the x-axis in the x-z plane); and $b_y$ (intercept of the track with the y-axis in the y-z plane); as well as a covariance matrix (E) for these values. Since the decay vertex of the $J/\psi$ (x, y, z) is unspecified and the magnitude and direction of its momentum is unknown, the problem in this case involves six unmeasurable variables

$$\vec{U} = \left\{ \frac{1}{p_{J/\psi}}, m_{x\,J/\psi}, m_{y\,J/\psi}, x, y, z \right\},$$

and ten measurable variables for both muon tracks,

$$\vec{M} = \left\{ \frac{1}{p_{\mu^+}}, m_{x\mu^+}, m_{y\mu^+}, b_{x\mu^+}, b_{y\mu^+}, \frac{1}{p_{\mu^-}}, m_{x\mu^-}, m_{y\mu^-}, b_{x\mu^-}, b_{y\mu^-} \right\}.$$

The algebraic constraint equations are: the three equations describing momentum conservation; the equation from energy conservation; and the geometrical line-equations of the tracks in the x-z and y-z planes. This gives a total of eight equations:

$$f_1 = -\frac{m_{x\,J/\psi}}{\frac{1}{p_{J/\psi}}\sqrt{m_{x\,J/\psi}^2 + m_{y\,J/\psi}^2 + 1}} + \frac{m_{x\mu^+}}{\frac{1}{p_{\mu^+}}\sqrt{m_{x\mu^+}^2 + m_{y\mu^+}^2 + 1}} + \frac{m_{x\mu^-}}{\frac{1}{p_{\mu^-}}\sqrt{m_{x\mu^-}^2 + m_{y\mu^-}^2 + 1}} \quad ;$$

$$f_2 = -\frac{m_{y\,J/\psi}}{\frac{1}{p_{J/\psi}}\sqrt{m_{x\,J/\psi}^2 + m_{y\,J/\psi}^2 + 1}} + \frac{m_{y\mu^+}}{\frac{1}{p_{\mu^+}}\sqrt{m_{x\mu^+}^2 + m_{y\mu^+}^2 + 1}} + \frac{m_{y\mu^-}}{\frac{1}{p_{\mu^-}}\sqrt{m_{x\mu^-}^2 + m_{y\mu^-}^2 + 1}} \quad ;$$

$$f_3 = -\frac{1}{\frac{1}{p_{J/\psi}}\sqrt{m_{x\,J/\psi}^2 + m_{y\,J/\psi}^2 + 1}} + \frac{1}{\frac{1}{p_{\mu^+}}\sqrt{m_{x\mu^+}^2 + m_{x\mu^+}^2 + 1}} + \frac{1}{\frac{1}{p_{\mu^-}}\sqrt{m_{x\mu^-}^2 + m_{x\mu^-}^2 + 1}} \quad ;$$

$$f_4 = -\sqrt{\frac{1}{\left(\frac{1}{p_{J/\psi}}\right)^2} + m_{J/\psi}^2} + \sqrt{\frac{1}{\left(\frac{1}{p_{\mu^+}}\right)^2} + m_{\mu}^2} + \sqrt{\frac{1}{\left(\frac{1}{p_{\mu^-}}\right)^2} + m_{\mu}^2} \quad ;$$

$$f_5 = -x + z\, m_{x\mu^+} + b_{x\mu^+} \quad ;$$

$$f_6 = -x + z\, m_{x\mu^-} + b_{x\mu-} \quad ;$$
$$f_7 = -y + z\, m_{y\mu^+} + b_{y\mu^+} \quad ;$$
$$f_8 = -y + z\, m_{y\mu^-} + b_{y\mu^-} \quad ;$$

where $m_{J/\psi}$ and $m_\mu$ are the masses of the J/ψ and muon, respectively. Since the problem involves 8 constraint equations and 6 unmeasurable variables this is then a 2c-fit.

## Results from the fit

In a given event, all the unlike-sign muon pairs with fully linked SSD-PWC-MUON tracks, and with an invariant-mass in the mass interval between 2.85 GeV/c$^2$ to 3.35 GeV/c$^2$, and having a distance of closest approach between the two muon tracks less then 50 μm, were fitted with the above technique. For convergence, the values of the $\chi^2$, $\vec{M}$, and $\vec{U}$ variables were required to be within 0.001 of the preceding values of each variable, respectively. In each event only the dimuon with the lowest $\chi^2$ was kept. Figure A. 1 shows the distribution of the $\chi^2$ per degree of freedom of the fit. A cut on the $\chi^2$/d.o.f less than 5.0 was established. Before the fit, there were a total of 13,053 muon pairs. After the fit 12,340 dimuons survived the fit requirements. These were identified as J/ψ s.

From the $\Lambda_b \to J/\psi\, \Lambda^0$ Monte-Carlo, the reconstruction efficiency of the initial track finding for J/ψ s from $\Lambda_b$ s was found to be 49.8 %, and by using the refit procedure this efficiency was improved to 59 %. In Figure A. 2 the dashed line shows the residuals of the J/ψ momentum between the generated value from $\Lambda_b \to J/\psi\, \Lambda^0$ Monte-Carlo and that reconstructed by the initial track finding. The solid line shows the residuals of the momentum after the refit. Figure A. 3 shows the residuals of the z-coordinate of the J/ψ decay vertex between the generated value from Monte-Carlo and the reconstructed, the dashed line for the initial track finding and the solid line for the refit. As one can see, refitting the J/ψ resulted in an improvement of its momentum measurement resolution by a factor of 2, and in the vertex resolution by 15%.

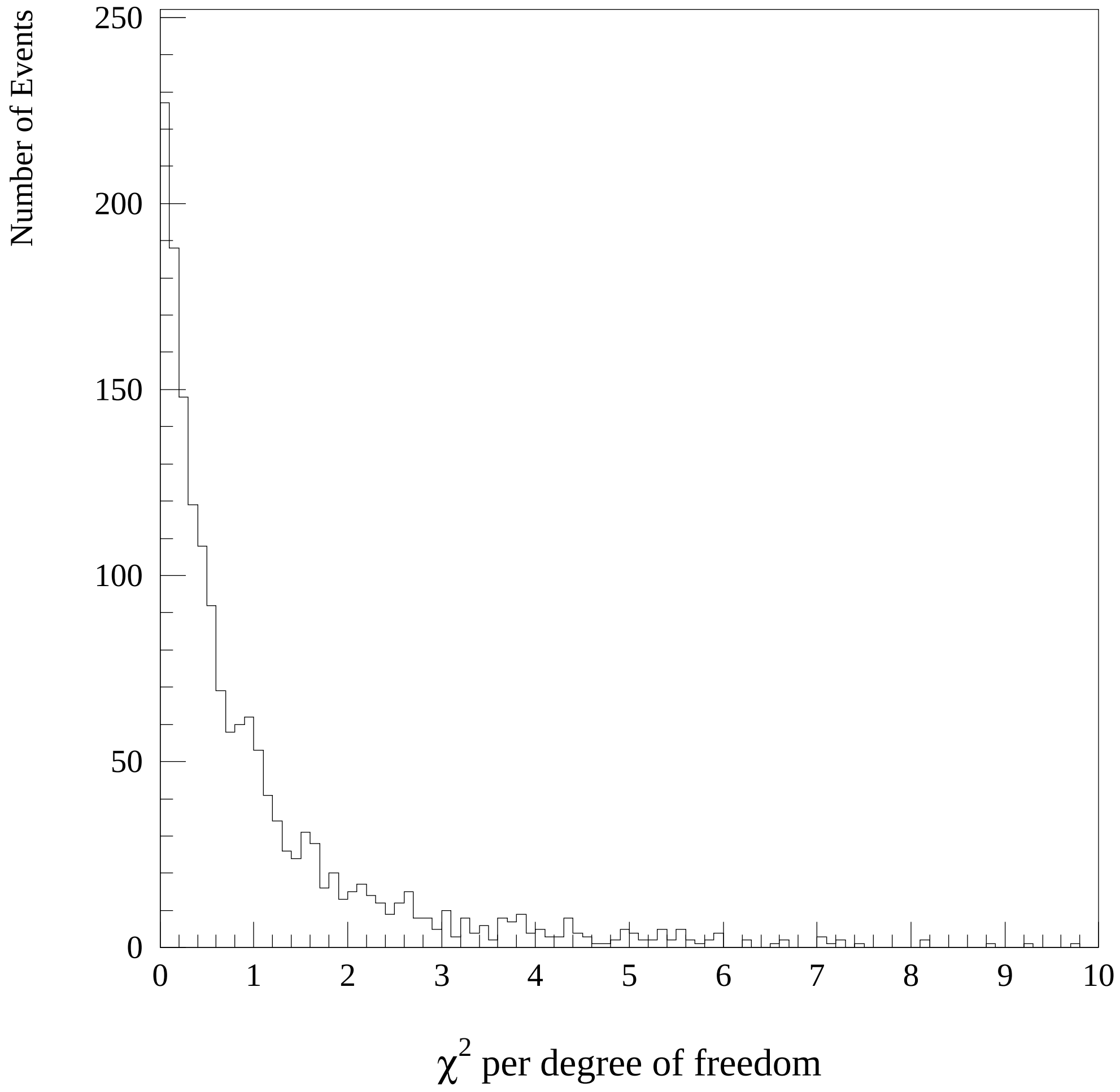

Figure A. 1. The $\chi^2$ per degree of freedom distribution from the J/ψ refit.

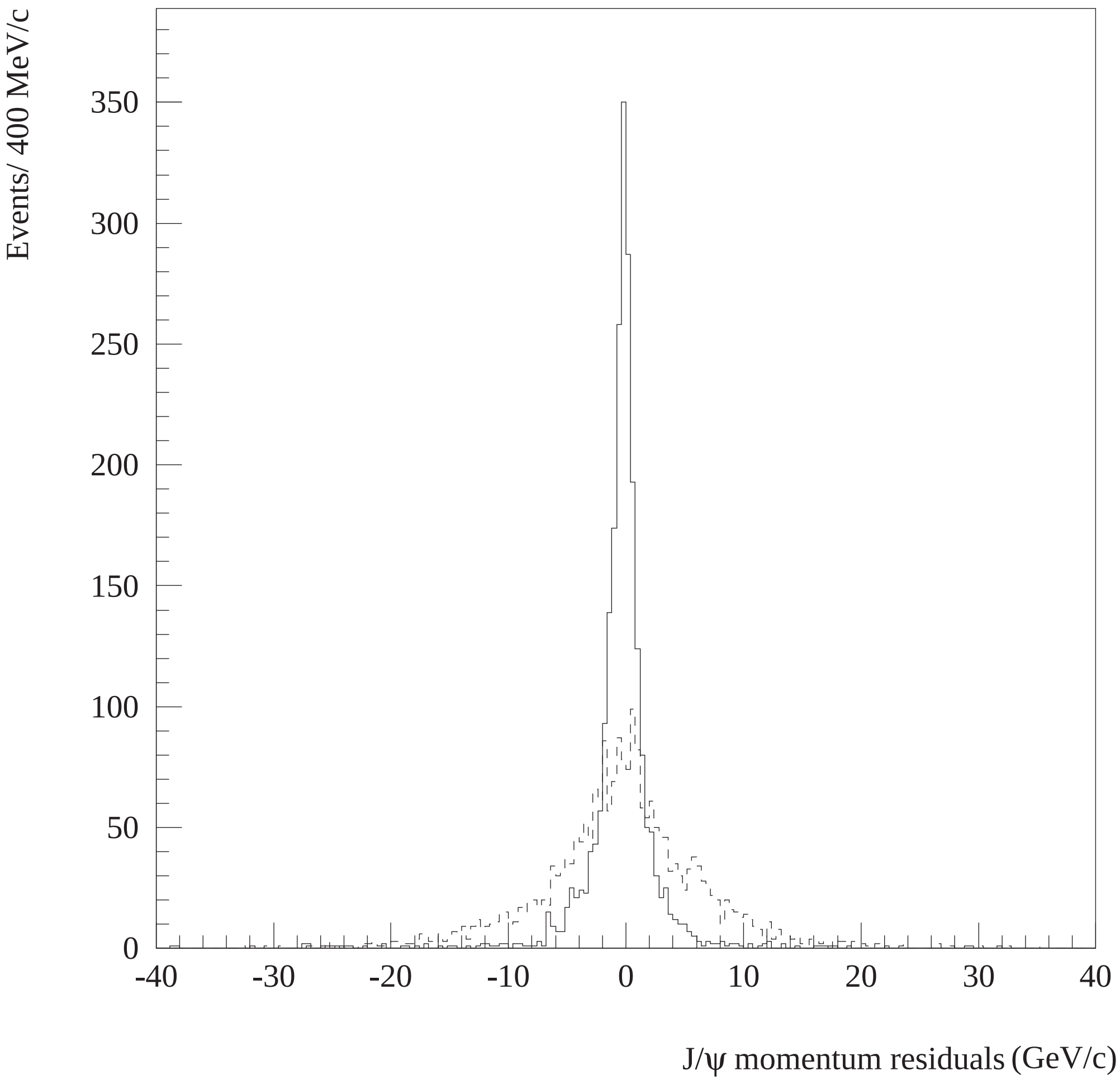

Figure A. 2. Residuals of the J/ψ momentum between the generated momentum value in the $\Lambda_b \rightarrow J/\psi\ \Lambda^0$ Monte-Carlo and the reconstructed value. The dashed line is for the values reconstructed by the initial track finding. The solid line shows the residuals of the momentum after the refit.

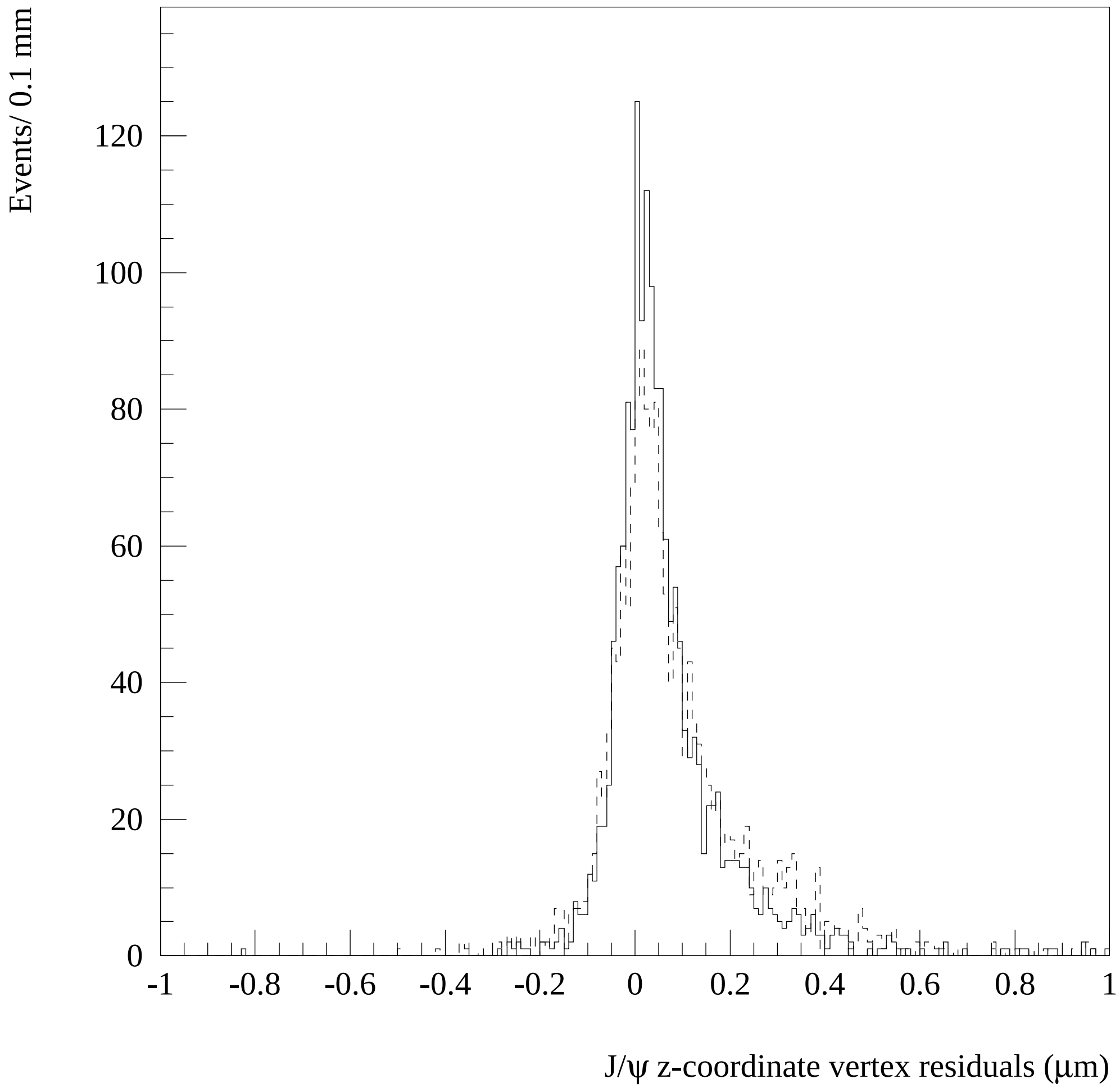

Figure A. 3. Residuals of the z-coordinate of the decay vertex between the value generated in the Monte-Carlo and the value reconstructed. The dashed line is for the values found by the initial track finding and the solid is for those after the refit.

# APPENDIX B

## MOMENTUM MEASUREMENT IN THE REGION UPSTREAM OF THE DIPOLE MAGNET

Assuming the decay vertex of the $\Lambda^0$ (x, y, z) and the parameters of the PWC track $m_{xD}$ (x-slope), $m_{yD}$ (y-slope), $b_x$ (x-intercept) $b_y$ (y-intercept) are known, and that the magnetic field is uniform throughout the dipole magnet, $\vec{B} = -|\vec{B}|\hat{j}$. The momentum vector is estimated in the following way: First, the coordinates of the PWC track at the center of the magnet are computed,

$$z_{center} = 197.73 \text{ cm},$$

$$x_{center} = z_{center} + b_x,$$

Then, to estimate the bend of the trajectory of particle (proton or pion), the x-slope $m_{xU}$ of the imaginary line in the x-z plane from (x, z) to ($x_{center}$, $z_{center}$) is computed

$$m_{xU} = \frac{x - x_{center}}{z - z_{center}},$$

see Figure B. 1. $M_{xU}$ is the slope of the proton or pion trajectory at the $\Lambda^0$ decay vertex. Since the magnetic field is along the negative y-axis (see section 2.5.2) the electric charge associated with the track is then,

$$Q = \frac{m_{xD} - m_{xU}}{\left|m_{xD} - m_{xU}\right|}.$$

The magnitude of the momentum of the track in the x-z plane is:

$$P_{xz} = \frac{\left|\vec{P}_{T_{kick}}\right|}{\left|\frac{m_{xD}}{\sqrt{1+m_{xD}^2}} - \frac{m_{xU}}{\sqrt{1+m_{xU}^2}}\right|},$$

where $\left|\vec{P}_{T_{kick}}\right| = 0.3\left|\vec{B}\right|L$, and $\left|\vec{B}\right|$is the magnetic field (in Tesla), and L (in meters) is the length of the field along the z-axis. From $P_{xz}$ one can calculate the z component of the momentum:

$$P_z = \frac{P_{xz}}{\sqrt{1+m_{xU}^2}};$$

and from this the x and y components:

$$P_x = P_z\, m_{xU};$$

and

$$P_y = P_z\, m_{yD}.$$

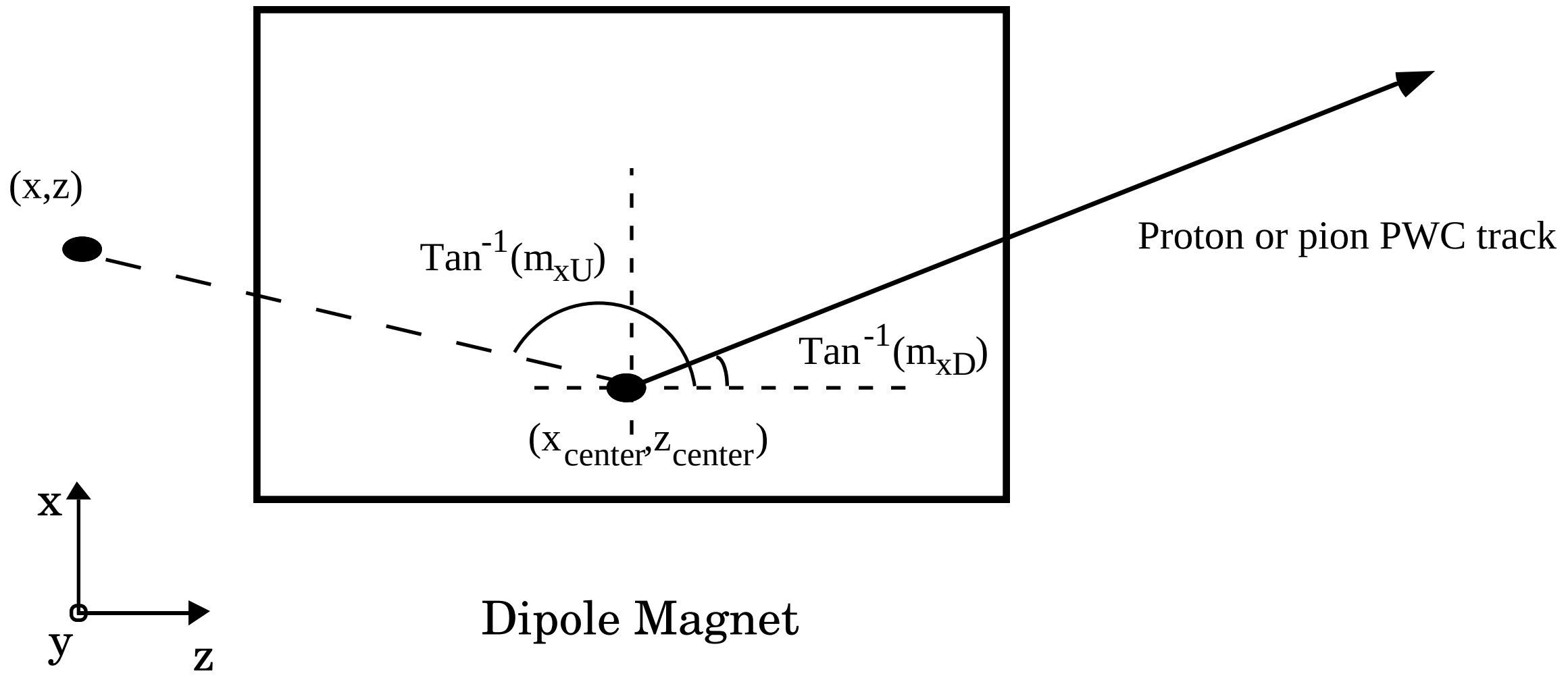

Figure B. 1. The thin lens approximation is used to compute the momenta of the proton and pion tracks of those $\Lambda^0$ s that decayed upstream of the dipole magnet.

## APPENDIX C

# MOMENTUM MEASUREMENT IN THE REGION INSIDE THE DIPOLE MAGNET

Assuming that the decay vertex of the $\Lambda^0$ (x,y,z), and the parameters of the PWC track $m_{xD}$ (x-slope), $m_{yD}$ (y-slope), $b_x$ (x-intercept), $b_y$ (y-intercept) are known, and that the magnetic field is uniform throughout the dipole magnet, $\vec{B} = -|\vec{B}|\hat{j}$. The magnitude of the momentum is determined by measuring the radius R of the circular trajectory described by the proton or pion particle inside the magnetic field. R is related to the momentum of the particle in the x-z plane by, $|\vec{P}_{xz}| = 0.3|\vec{B}|R$, where $|\vec{B}|$ is measured in Tesla, and R in meters. The momentum vector is estimated in the following way: Labeling the center of the circular trajectory as ($x_o$, $z_o$), see Figure C. 1, it is easy to see that,

$$(x - x_o)^2 + (z - z_o)^2 = R^2;$$

$$(x_{end} - x_o)^2 + (z_{end} - z_o)^2 = R^2; \text{ and}$$

$$m_{xD} \cdot \frac{x_{end} - x_o}{z_{end} - z_o} = -1;$$

where, $z_{end}$ = 318.53 cm, and $x_{end} = m_{xD} + b_x$.

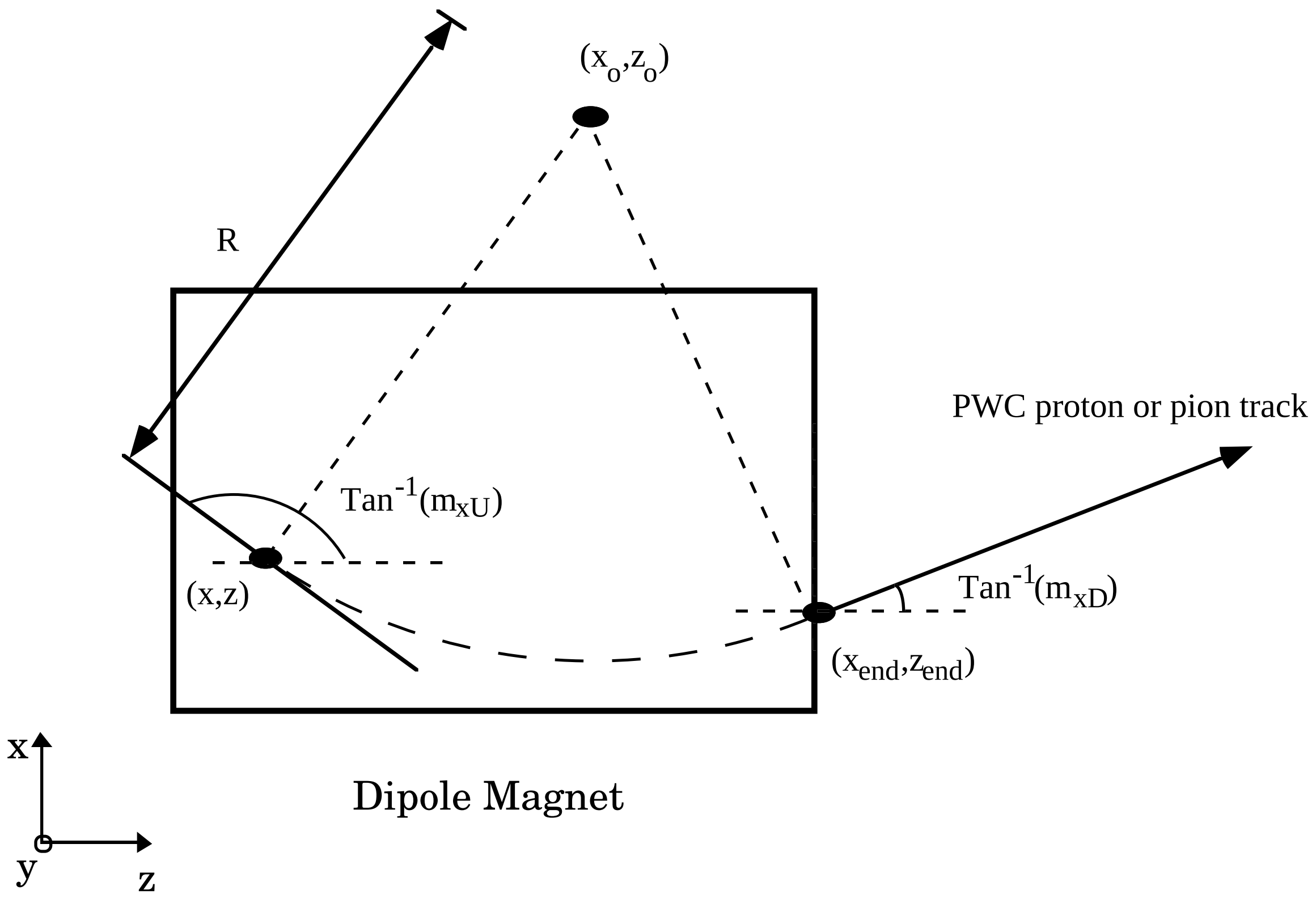

Figure C. 1. Circular trajectory described by of a proton or pion of a $\Lambda^0$ that decays inside the dipole magnet.

These three equations have three unknowns, $x_o$, $z_o$, and R. It is not so difficult to solve for these unknown variables and obtain

$$z_o = z_{end} - \frac{1}{2}\frac{\left(x - x_{end}\right)^2 + \left(z - z_{end}\right)^2}{\frac{\left(x - x_{end}\right)}{m_{xD}} - \left(z - z_{end}\right)};$$

$$x_o = x_{end} + \frac{1}{m_{xD}}\left(z_{end} - z_o\right);\text{ and}$$

$$R = \sqrt{\left(1 + \frac{1}{m_{xD}^2}\right)\left(z_{end} - z_o\right)}.$$

Having this, $m_{xU}$ can now be computed,

$$m_{xU} = \frac{z_o - z}{x - x_o},$$

and the electric charge of the particle is found,

$$Q = \frac{m_{xD} - m_{xU}}{\left|m_{xD} - m_{xU}\right|}.$$

As mentioned previously the momentum of the particle in the x-z plane is $\left|\vec{P}_{xz}\right| = 0.3\left|\vec{B}\right|R,$ .
From $P_{xz}$ one can calculate the z component of the momentum:

$$P_z = \frac{P_{xz}}{\sqrt{1 + m_{xU}^2}};$$

and from this the x and y components:

$$P_x = P_z\, m_{xU};$$

and

$$P_y = P_z\, m_{yD}.$$

## REFERENCES

[1]. S. W. Herb *et al.*, Phys. Rev. **39** 252 (1977).

[2]. M. Basile *et al.*, Lett. Nuovo Cimento **31** 97 (1981).

[3]. D. Drijard *et al.*, Phys. Lett. **B108** 361 (1982).

[4]. M. Basile *et al.*, Nuovo Cimento **68** 289 (1982).

[5]. G. Bari *et al.*, Nuovo Cimento **A104** 1787 (1991).

[6]. M.W. Arenton *et al.*, Nucl. Phys. **B274** 707 (1986).

[7]. UA1 Collab., C. Albajar *et al.*, Phys. Lett. **B 273** 540 (1991).

[8]. OPAL Collab., R. Akers, "Search for Exclusive $\Lambda_b$ Decays with the OPAL Detector at LEP", submitted to the Lepton-Photon Symposium, Beijing, China, Aug. 10-15, 1995; ALEPH Collab., A. Bonissent, "$B_s$ and $\Lambda_b$ at LEP", Proceedings of the XXVIII Rencontres de Moriond, Les Arcs, France, Mar. 20-27, 1993, p. 413.

[9]. CDF Collab., F. Abe *et al.*, Phys. Rev. **D 47** 2639 (1993).

[10]. ALEPH Collab., D. Buskulic, "Measurement of the b Baryon Lifetime", CERN-PPE/95-65, (1995); DELPHI Collab., P. Abreu, "Lifetime and production rate of beauty baryons from Z decays", CERN-PPE/95-54,

(1995); OPAL Collab., R. Akers, "Measurement of the Average b-Baryon Lifetime and the Production Branching Ratio", CERN-PPE/95-90, (1995).

[11]. E672/E706 Collab., R. Jesik *et al*., Phys. Rev. Lett. **74**, 495 (1995).

[12]. Review of Particle Properties, Phys. Rev. **D 50** (1994).

[13]. CDF Collab., F. Abe *et al*., Phys. Rev. Lett. **74**, 2626 (1995).

[14]. D0 Collab., S. Abachi *et al*., Phys. Rev. Lett. **74**, 2632 (1995).

[15]. S. Weinberg, Rev. Mod. Phys. **52**, 515 (1980).

[16]. Halzen, F. and Martin, A., "Quarks & Leptons", New York, John Wiley & Sons, Inc., 1984.

[17]. Barger, V. and Phillips, R., " Collider Physics", New York, Addison-Wesley Publising Co., 1987.

[18]. J. Smith and W.K. Tung, "Heavy-Flavor Production", in Proceedings of the Workshop on B Physics at Hadron Accelators, Snowmass, Colorado, June 21-July 2, p.19, 1993.

[19]. P. Nason, S. Dawson, R.K. Ellis, Nucl. Phys. **B303** 607 (1988).

[20]. E.L. Berger, Phys. Rev. **D37** 1810 (1988).

[21]. M. Mangano, P. Nason, G. Rodolfi, Nucl. Phys. **B405** 507 (1993); **B327** 49 (1989); **B335** 260 (1990); **B373**, 295 (1992).

[22]. M. Catansi *et al*., Phys. Lett. **B187** 431 (1987).

[23]. P. Bordalo *et al*., Z. Phys. **C39**, 7 (1988).

[24]. K. Kodama *et al*., Phys. Lett. **B303** 359 (1993).

[25]. CDF Collab. K. Byrum, “Charmonium Production, b quark and B Meson Production and bb Correlations at CDF”, Proceedings of the XXVII International Conference on high Energy Physics, 20-27 July 1994 Glasgow Scotland UK, p. 989, ed P. J. Bussey and I. G. Knowles, Institute of Physics Publishing, Bristol and Philadelphia; D0 Collab., S. Abachi *etal*., Phys. Rev. Lett. **74**, 3548 (1995).

[26]. B. Anderson, G Gustafson, G. Ingelman, and T. Sjöstrand, Nucl. Phys. **B 197** 45 (1982).

[27]. T. Mannel and G. A. Schuler, Phys. Lett. **B279** 194 (1992).

[28]. OPAL Collab, P.D Acton *et al*., Phys. Lett. **B 281** 394 (1992).

[29]. J. L. Cortes, X. Y. Pham, and A. Tounsi, Phys. Rev. **25** 188 (1982).

[30]. J. H. Kühn and R. Rükl, Phys. Lett. **B 135** (1984) 477; J. H. Kühn, S. Nussinov and R. Rükl, Z. Phys. **C 5** 117 (1980).

[31]. Hai-Yang Cheng and B. Tseng “The $\Lambda_b \rightarrow J/\psi + \Lambda^0$ Revisited”, IP-ASTP-22-94, Nov., 1994.

[32]. E706 Collab., G. Alverson *et al*., Phys. Rev **D48** 5 (1993).

[33]. E672 Collab., V. Abramov *et al*., “ FERMILAB-Pub-91/62-E, Mar. 1991. See also R. Li, Ph. D. Thesis, Indiana University, Indiana (1993).

[34]. R. Miller “A Measurement of the E706 Beam Momentum for the 1990 Negative Runs”, E706 note no. 200 (1994).

[35]. I. Kourbannis, Ph. D. Thesis, North Eastern, Boston (1990).

[36]. S. Easo, Ph. D. Thesis, The University of Pittsburg, Pittsburg, Pennsylvania (1989).

[37]. R. Crittenden and R. Li, “ An introduction to the the Mu-B MWPC”, E672 note, (1989).

[38]. C. Yosef, Ph. D. Thesis, North Eastern, Boston (1990).

[39]. “Bison Interrupt and Gate Control”, Fermilab Computing Department, HN-3.2 (1984).

[40]. V. Sirotenko , E672 internal note.

[41]. “Vaxonline system”, Fermilab Computing Department, PN252.0 (1982).

[42]. E672 Collab., Nucl. Instr. and Meth. **A270** 99 (1988).

[43]. “2738 PCOS III Manuel”, LeCroy Research Systems Corporation (1988).

[44]. H. J. Martin, “Fits to the Muon Data:”, E672 note no. 221 (1988).

[45]. E672/E706 Collab., A. Gribushin *et al.*, “Production of J/ψ and ψ(2S) Mesons in $\pi^-$ Be Collisions at 515 GeV/c”, FERMILAB-Pub-95/298-E, Sep. 1995, submitted to Phys. Rev. D.

[46]. E672/E706 Collab., V. Koreshev *et al.*, “Production of Charmonium States in $\pi^-$ Be Collisions at 515 GeV/c”, in preparation.

[47]. H. Fritzsch, Phys. Lett. **67B** 217 (1977); M. Gluck *et al*., Phys. Rev. **D17** 2324 (1978); J.H Kuhn, Phys. Lett. **89B** 385 (1980); R. Gavi, *et al*., CERN-TH-7526/94 (1994).

[48]. R. Baier and R. Ruckl, Z Phys. **C19** 251 (1983), G.A. Schuler, CERN-TH.7170/94; M. Mangano, CERN-TH/95-190, Proceedings of the X-th Topical Workshop on Proton-Antiproton Collider Physics, Batavia, Il, May 1995.

[49]. E672/E706 Collab., F. Vaca *et al*., “Production of Vector Mesons in Hadron-A Interactions”, in preparation.

[51]. J. Podolanski and R. Armenteros, Phil. Mag 45, 13 (1954)

[52]. R. Brun, “GEANT33 Users’s Guide”, CERN DD/EE/84-1, (1984)

[53] J. Bell, Phys. Rev. **D19** 1 (1979).

[54]. A. G. Frodesen, 0. Sjeggestadt, and H. Tofte, “Probability and Statistics in Particle Physics”, Ed. Universitets forl (1979).

# VITA

## *Francisco J. Vaca*

8821 S. Escanaba
Chicago, IL 60617
(312) 933-4751

University of Illinois at Chicago
Department of Physics, M/C 273
845 W. Taylor St., Room 2263
Chicago, IL 60607
(312) 996-6751
FAX: (312) 996-9016
Internet E-Mail: francisco@fnalv.fnal.gov

### Personal

| | |
|---|---|
| Date of Birth: | April 2, 1968 |
| Place of Birth: | Zamora, Michoacán, México |
| Nationality: | US |
| Marital status: | Married, one child |
| Languages: | English and Spanish |

### Education

**Universidad Michoacana** 1985 - 1989
Morelia, Michoacán, México
*B.S. degree in Physics and Mathematics*, Dissertation: Maximization of Power in Thermodynamic and Biological cycles.

**University of Illinois at Chicago** 1989 - 1992
Chicago, IL
*M.S. degree in Physics*

**University of Illinois at Chicago** 1989 - 1995
Chicago, IL
*Ph.D. degree in Physics*, Dissertation: Search for $\Lambda_b$ in $\pi^-$ N collisions at 515 GeV.

### Academic Achievements and Distinctions

| | |
|---|---|
| 1985-1989 | Scholarship from the Universidad Michoacana, México |
| 1991 | President, Society of Physics Students, UIC Chapter |
| 1991 | Sigma Pi Sigma (Physics Honors Society), UIC Chapter, Recognition |
| 1989-1992 | Illinois Minority Graduate Incentive Program Fellowship |
| 1992 | Latino Committee at UIC, Recognition |
| 1992-1993 | Martin Luther King, Jr. Award, UIC |
| 1993 | First Prize award, International Committee for Future Accelerators India School on Instrumentation in Elementary Particle Physics (ICFA 93), Bombay, 15-27 Feb. 1993 |
| 1993-1994 | University of Illinois at Chicago Graduate Fellowship |
| 1993-1994 | National Hispanic Scholarship Foundation Scholarship |
| 1994-1995 | University of Illinois at Chicago Graduate Fellowship |
| 1994-1995 | National Hispanic Scholarship Foundation Scholarship |

## Summers Schools Attended

- Universidad Autónoma de México Molecular Vision of Matter Summer School, Cuernavaca, Morelos, México, 1-30 Aug. 1988
- International Committee for Future Accelerators India School on Instrumentation in Elementary Particle Physics (ICFA 93), Bombay, February 15-27, 1993
- Advanced Study Institute on Techniques and Concepts of High Energy Physics (ASI 94), St. Croix, USVI, June 16-27, 1994
- Lafex International School on High Energy Physics (LISHEP 95), Rio de Janeiro, February 6-22, 1995

## Professional Affiliations

| | |
|---|---|
| 1989 - present | Society of Physics Students, UIC chapter |
| 1991 - present | American Physical Society, Division of Particles and Fields |

## Professional Experience

**Universidad Michoacana** 1986- 1989
Morelia, Michoacán, México
*Teaching Assistant*, Assisted in teaching undergraduate courses (calculus, linear algebra)

**University of Illinois at Chicago** 1989 - 1994
Chicago, IL
*Teaching Assistant*,Taught undergraduate physics laboratories

**University of Illinois at Chicago** 1991 - 1994
Chicago, IL
*Research Assistant*, Research assistant with high-energy physics group working on Fermilab experiment E672

**University of Illinois at Chicago** 1994 - 1995
Chicago, IL
*Supplemental Instructor*, Engineering College, Minority Engineering Recruitment and Retention Program: tutor for undergraduate physics

**University of Illinois at Chicago** 1994
Chicago, IL
*Consultant*, Engineering College, Minority Engineering Recruitment and Retention Program: participated as the content expert in the creation of videos on undergraduate physics topics

**Columbia College Chicago** 1993 -1995
Chicago, IL
*Part-time Faculty*, teaching undergraduate physics class and laboratory

## Presentations

- "Fermilab Experiment E672", poster session, annual Department of Energy review of the Fermilab physics program, Fermilab, Mar. 29, 1994.
- "Charmonium Production and a Search for $\Lambda_b$ in Fermilab E672/E706", American Physical Society April Meeting, Washington DC, April 18-21, 1995.

## Publications

A. Gribushin *et al*., "Production of J/$\psi$ and $\psi$(2S) Mesons in $\pi^-$ Be Collisions at 515 GeV/c", FERMILAB-Pub-95/298-E, Sep. 1995, submitted to Phys. Rev. D.

R. Jesik *et al*., "Bottom Production in $\pi^-$ -Be Collision at 515 GeV/c", Phys. Rev. Lett. **74,** 459 (1995).

R. Jesik *et al*., "Hadronic Production of Beauty", Proceedings of the XXVIII Rencontres de Moriond, Les Arcs, France, Mar. 20-27, 1993, p. 385.

L. Dauwe *et al*., "Hadronic Production of B Mesons", Proceedings of the 1992 A.P.S. Division of Particles and Fields Meeting, Batavia, IL, Nov. 9-14, 1992, p. 759.

H. Mendez *et al*., "Hadronic Production of $\chi_c$ Mesons", Proceedings of the 1992 A.P.S. Division of Particles and Fields Meeting, Batavia, IL, Nov. 9-14, 1992, p. 756.

R. Jesik *et al*., "Heavy Flavor Production in $\pi^-$-A Collisions at 530 GeV/c", Proceedings of the XXVI International Conference on High Energy Physics, Dallas, TX, Aug. 5-12, 1992, ed. J. Stanford (American Institute of Physics, 1993), p. 824.

A. Zieminski *et al*., "Production of $\chi_c$ States in $\pi^-$- Nucleus Collisions at 530 GeV/c", Proceedings of the XXVI International Conference on High Energy Physics, Dallas, TX, Aug. 5-12, 1992, ed. J. Stanford (American Institute of Physics, 1993), p. 1062.

R. Li *et al*., "Hadroproduction of Vector Mesons and $\chi_c$ States", Proceedings of the XXVII Rencontres de Moriond, Les Arcs, France, Mar 21-28, 1992, p. 385.

B. Abbott *et al*., "Effects of a Hadron Irradiation on Scintillating Fibers", IEEE Trans. Nucl. Sci. **40**, 476 (1993).

## In Press

A. Gribushin *et al*., "Hadronic production of Heavy Quarks", to be published in the Proceedings of the XXIX Rencontres de Moriond, Les Arcs, France, Mar 20-27, 1994.

H. Mendez *et al*., " Hadroproduction of Charmonium States ", to be published in the Proceedings of the XXX Rencontres de Moriond, Les Arcs, France, Mar 19-26, 1995

## In Preparation

V. Koreshev *et al*., “Production of Charmonium States in $\pi^-$ Be Collisions at 515 GeV/c”.

F. Vaca *et al*., “Production of Vector Mesons in Hadron-A Interactions”.